\documentclass[preprints,article,accept,moreauthors,LaTeX]{mdpi}
\firstpage{1} 
\pubvolume{1}
\issuenum{1}
\articlenumber{0}
\pubyear{2021}
\copyrightyear{2021}
\externaleditor{Communicated by: {Francesca Loi and Tiziana Venturi} 
} 
\datereceived{28 September 2021} 
\dateaccepted{18 December 2021} 
\datepublished{22 December 2021} 
\hreflink{https://doi.org/} 
\usepackage[utf8]{inputenc}
\DeclareUnicodeCharacter{223C}{\ensuremath{}}
\DeclareUnicodeCharacter{2212}{\ensuremath{}}
\usepackage{url}
\usepackage{lscape}
\usepackage{subfigure}
\usepackage{caption}
\usepackage{aas_macros}

\Title{Finding Rare Quasars: VLA Snapshot Continuum Survey of FRI Quasar Candidates Selected from the LOFAR Two-Metre Sky Survey (LoTSS)}

\TitleCitation{Finding Rare Quasars: VLA Snapshot Continuum Survey of FRI Quasar Candidates Selected from the LOFAR Two-Metre Sky Survey (LoTSS)}

\Author{G\"ulay G\"urkan 
$^{1\,*}$\orcidA{}, Croston, J. 
$^{2}$, Hardcastle, M. J. $^{3}$, Mahatma, V. $^{1}$, Mingo B. $^{2}$ and Williams, L. W. $^{4}$}

\AuthorNames{G\"ulay G\"urkan, Judith Croston, Martin Hardcastle, Vijay Mahatma, Beatriz Mingo, Wendy Williams}

\AuthorCitation{G\"urkan, G.; Croston, J.; Hardcastle, M.; Mahatma, V.; Mingo, B.; Williams, W.}

\address{%
$^{1}$ \quad Th\"uringer Landessternwarte Tautenburg [TLS], Sternwarte 5, D-07778 Tautenburg, Germany
\\
$^{2}$ \quad School of Physical Sciences, The Open University, Walton Hall, Milton Keynes, MK7 6AA, UK
\\ 
$^{3}$ \quad Centre for Astrophysics Research, University of Hertfordshire, College Lane, Hatfield AL10 9AB, UK
\\
$^{4}$ \quad Leiden Observatory, Leiden University, P.O. Box 9513, NL-2300 RA Leiden, The Netherlands
}
\corres{Correspondence: gulaygurkan.astro@gmail.com}

\abstract{The radiative and jet power in active galactic nuclei is generated by accretion of material on to supermassive galactic-centre black holes.  For quasars, where the radiative power is by definition very high, objects with high radio luminosities form ∼10 per cent of the population, although it is not clear whether this is a stable phase. Traditionally, quasars with high radio luminosities have been thought to present jets with edge-brightened morphology (Fanaroff-Riley II - FR II) due to the limitations of previous radio surveys (i.e.,  FRIs were not observed as part of the quasar population). The LOw Frequency ARray (LOFAR) Two-metre Sky Survey (LoTSS) with its unprecedented sensitivity and resolution covering wide sky areas has enabled the  first systematic selection and investigation of quasars with core-brightened morphology (Fanaroff-Riley I - FR). We carried out a Very Large Array (VLA) snapshot survey to reveal inner structures of jets in selected quasar candidates; 15 ({25 per cent}) out of 60 sources show clear inner jet structures that are diagnostic of FRI jets and 13 quasars ({~22 per cent}) show extended structures similar to those of FRI jets. Black hole masses and Eddington ratios do not show a clear difference between FRI and FRII quasars. FRII quasars tend to have higher jet powers than FRI quasars. Our results show that the occurrence of FRI jets in powerful radiatively efficient systems is not common, probably mainly due to two factors: galaxy environment and jet power.}


\begin{document}


\section{Introduction}
\textls[-25]{Quasars are a class of Active Galactic Nuclei (AGN) that produce immense energy by the fast accretion onto supermassive black holes in the centre of massive galaxies. Based on the unification model (in which the main ingredient is the orientation effect) radio galaxies and quasars are the same objects seen at different angles. An object is called a quasar if it is viewed within a cone of half-angle approximately 45$^{\circ}$. They are common at moderate and high redshifts which make them excellent sites to probe the conditions of the early Universe, e.g., \cite{yang20}. Investigations of quasars are not only crucial in their own right, but also essential for galaxy formation and AGN feedback studies, e.g.,} \cite{gurkan141,gurkan19,mc21}. They allow us to constrain cosmological parameters, e.g., \cite{gurkan142} and to investigate the radio emission in low-luminosity (or faint) AGN at moderate/high redshifts, owing to the strong lensing effect which is just one of the powerful tools for studying AGN at moderate/high redshifts,~\citep{mckean21,hartley21}. For a recent review on radio galaxies and radio-loud quasars we refer the reader to~\citep{hc20}.

Fanaroff $\&$ Riley \cite{fr74} classified radio sources based on their radio morphologies as FRI (core-brightened) and FRII (edge-brightened) and suggested that the division between FRI and FRII is not only in radio morphology but also in radio luminosity (i.e.,  the traditional FRI/FRII break is expected to be $\sim5\times 10^{25}$ W Hz$^{-1}$ at 150 MHz). It is worth noting that this picture is changing with new sensitive radio surveys because many more FRII AGN below the traditional FRI/FRII break have been discovered [Mingo et al. in prep] \cite{miraghaei17,mingo19,macconi20,grandi21,verdul21} but it is still the case that FRI sources tend not to be found at the highest radio powers.

Among AGN, optically powerful quasars have not traditionally been associated with FRI radio morphology. The reason behind this is not based on any motivated physical model, e.g., \cite{baum-zirbel-odea95,falcke+95} but is due to a lack of these sources in the previous surveys (such as the 3C surveys, \citep{1962wref1,1983wref2}). It should be mentioned that radiatively efficient AGN with a given radiative power can have jet powers that range continuously from zero up to as much as their radiative power (\citep{mc21} and references therein), but this does not rule out FRI quasars. Since previous radio surveys revealed quasars having extended emission and only associated with FRII morphology a quasar has broadly been expected to be either a radio-quiet source (i.e.,  with no significant radio emission except, perhaps core emission) or a radio-loud source with FRII morphology. There are certainly radio-loud sources showing compact structures like compact steep-spectrum sources and gigahertz peaked sources so naturally there also needs to be a continuum of sources between them. Refs. \cite{Blundell&Lacy95,Heywood+07} discussed this bias and provided evidence that some quasars present FRI-like structures (six objects). However, radio surveys at that time were not sufficient to provide a large sample of FRI quasars only a few candidates from the dedicated efforts by \cite{Heywood+07} as the surveys were not sensitive enough to reveal any FRI structures. Recently, Ref. \cite{radcliff21} showed that some of the X-ray selected quasars do not show any radio emission (not even a core emission is detected). These results also suggest that systems with varying jet power can exist independent of AGN power. 

Hitherto, moderate to high-redshift FRIs have been poorly studied in comparison to FRII sources as FRIIs are typically brighter than FRI objects and can be detected out to high redshifts, e.g., \cite{rigby08}. 
In addition, the FRIs were found to be relatively low-power radio sources with jets that are relativistic on parsec scales, but decelerate to sub-relativistic bulk speeds on kpc scales. Quasars, on the other hand, are powerful sources amongst the AGN population. Understanding the occurrence and nature of quasars with FRI morphology on a scale of tens of kpc, how they compare to FRI radio galaxies on hundreds of kpc scales and how they fit in the interplay between AGN accretion, radio morphology and source environment will be only possible through studies based on statistically complete samples.

Systematic identification of large samples of FRI quasar {\it candidates} has only been possible with the LOw Frequency ARray (LOFAR) Two-metre Sky Survey (LoTSS, \cite{shim19}) thanks to its unprecedented capabilities in terms of high sensitivity, angular resolution and survey speed at low radio frequencies. Tier 1 of LoTSS is designed to survey the northern sky using the LOFAR High Band Antennas (HBA) at an average frequency of 144~MHz. The survey has already collected data over large areas reaching the target sensitivity ($\sim$70 $\upmu$Jy beam$^{-1}$) and angular resolution of 6 arcsec [Data Release I/DR1]~\cite{shim19}.  We \citep{gurkan19} recently selected a sample of quasars from the Sloan Digital Sky Survey (SDSS) quasar catalogue fourteenth data release [DR14Q;] \citep{myers+15} over the Hobby-Eberly Telescope Dark Energy Experiment (HETDEX) region and evaluated their radio properties using the LoTSS data set. We found that optically selected quasars have a wide range of radio continuum properties for a given accretion power and a simple bimodality is not an adequate way to describe the radio properties of quasars. In a recent study we also presented a model of the radio luminosity distribution of the quasars that assumes that every quasar displays a superposition of two sources of radio emission: AGN (jets) and star formation \citep{mc21}. The observed radio flux density distribution of the quasar sample that spans a wide range of redshift and optical luminosity is well described by the two-component model. This reinforces the idea that the jet-launching mechanism operates in all quasars but with different efficiencies, which also confirms the findings by \citep{gurkan19} and is consistent with the lack of bimodality (see for studies suggesting the opposite \citep[][]{ivezic02,white07}) observed for less powerful populations \citep[][]{mingo14,gurkan15}. Therefore, quasars with low-power kinetic output (also quasars with moderate power but in particular in rich environments) are expected to form FRI structures.

With the motivation of investigating FRI quasars we carried out a pilot project using the Jansky Very Large Array (VLA) to perform a snapshot survey of a sample of 60~quasars showing extended structures that would be classified as FRI at 6 arcsec resolution. These 60 targets were homogeneously selected from LoTSS (selecting the sample based on low-frequency radio emission also enabled us to minimise  contamination from Doppler boosting) in order to reveal the inner structures of the target objects, confirm their radio morphological classification and further evaluate their properties in a statistically meaningful manner. In this paper we present the results of the pilot study that used the recent VLA radio continuum observations. The layout of this paper is as follows. Descriptions of the sample and the data used in this work are given in Section \ref{sec2}. The primary analysis and key results are given in Section \ref{sec3}. We finish with a summary and discussion of future work in Section 4. Throughout the paper we use the most recent Planck cosmology~\citep{planck16}: $H_{0}$ = 67.7 km s$^{-1}$ Mpc$^{-1}$, $\Omega_{m}$=0.308 and $\Omega_{\Lambda}$=0.692. Spectral index $\alpha$ is defined in the sense $S\sim\nu^{-\alpha}$.

\section{DATA}
\label{sec2}
\subsection{The Sample and Optical Data}

Our quasar sample is drawn from the SDSS quasar catalogue DR14Q \citep{myers+15}, which includes all SDSS-IV/the extended Baryon Oscillation Spectroscopic Survey [eBOSS;] \citep{blanton17} objects that were spectroscopically targeted as quasar candidates and that are confirmed as quasars via a new automated procedure combined with a partial visual inspection of spectra. The DRQ14 provides a compilation of all the spectroscopically-confirmed quasars identified in the course of any of the SDSS iterations and released as part of the SDSS. The SDSS, Wide-field Survey Explorer [WISE;] \cite{wright10} and the Palomar Transient Factory [PTF;]~\cite{rau09,law09} imaging data were used to select target quasars. The CORE sample is selected homogeneously to targets quasars at z $>0.9$ based on the method that utilises optical and WISE-optical colours. There are also quasar candidates that are selected based on the photometric variability measured
from the PTF. We refer the reader to \citep{Paris18} for the details of the SDSS-DR14Q sample selection. We cross-matched the value-added catalogue of the LoTSS-DR1 \citep{williams19} over the HETDEX region with the SDSS DR14 quasar catalogue~\citep{Paris18} using a 3$''$ matching radius in order to select quasars for visual inspection. We then generated images for visual inspection, consisting of the Panoramic Survey Telescope and Rapid Response System [Pan-STARRS;] \citep{panstarrs20} image with LOFAR and the Faint Images of the Radio Sky at Twenty Centimeters [FIRST;] \citep{first95} contours overlaid. This process generated $\sim$6000~sources, all of which were inspected visually in order to classify sources as FRI and FRII. The requirement to classify a source as an FRI quasar is that it is brighter in the centre (not including the bright core component) and gets fainter as the distance from the centre increases. We identified 60 FRI quasar candidates of which the majority were convincing FRI quasars, together with a few more complex objects. It is important to emphasise that the combination of LOFAR and FIRST information is necessary for this efficient selection as FIRST picks out the central flat spectrum and compact components, whereas LOFAR is sensitive to the large-scale steep spectrum structures. Our classification has also been confirmed for some of our selected objects by a method which uses the surface brightness profile and the distance of extended structures of a radio source to its optical counterpart [LoMorph;] \cite{mingo19}, which was developed for morphological classification of LoTSS sources. Only 519 quasars meet the angular size and flux requirements of LoMorph \cite{mingo19} and 109 of them were cleanly classified as FRI or FRII by this method (32 FRI quasars and 77 FRII quasars). Only 15 objects out of the selected 60 FRI quasar candidates met the criteria that was used by \cite{mingo19} (i.e., there are 15 common objects between our target sample and their cleanly classified sample). All of these 15 sources were classified as FRI or small FRI by LoMorph. In this paper we focus on the selected 60 FRI quasar candidates and compare their properties with quasars that are cleanly classified as FRI or FRII by \cite{mingo19} but not included in our VLA snapshot survey.

The DR14Q quasar catalogue contains $i$ band absolute magnitudes ($k$-corrected to $z=2$), SDSS magnitudes and fluxes at $u$,$g$,$r$,$i$,$z$ bands together with various properties derived from these. These measurements were used in the analyses presented in this work. Additionally, we obtained extinction-corrected $i$ band absolute magnitudes $k$-corrected to $z=0$ by converting the $i$ band absolute magnitudes $k$-corrected to $z=2$ using the conversion given by \cite{Richards+6} in the same way as done by \cite{gurkan19}. Whenever available, we used black hole masses estimated using the MgII and CIV emission line widths (There are complications in the estimation of virial black hole masses using MgII and CIV lines that should be noted such as different line measurement techniques, the treatment of the narrow line component in both lines and the choice for the method of subtracting the continuum emission underneath the line. Addressing these caveats would go beyond the scope of this work), optical bolometric luminosities which are estimated using bolometric corrections by~\cite{Richards+6} using one of the quasar luminosities at 1350, 3000, 5100 \AA \, at the source redshift, and Eddington ratios published by \cite{Shen11}. Otherwise, we used estimates of the same quasar properties from \cite{Kozlowski17}. 

\subsection{LoTSS DRI Image Data}
The HETDEX Spring field (right ascension 10h 45 m 00 s to 15 h 30 m 00 s and declination 45$^{\circ}$00$'$00$''$ to 57$^{\circ}$00$'$00$''$) was observed with LOFAR as part of LoTSS DR1. This field was targeted as it is a large contiguous area at high elevation for LOFAR, whilst having a large overlap with the SDSS imaging and spectroscopic data. The creation of the radio images is described by \citep{shim19}. Radio flux densities at 144 MHz for all SDSS quasars in our sample were directly taken from the value-added LOFAR/HETDEX catalogue \citep{williams19}. These LoTSS radio flux densities at 144 MHz were used to derive the rest frame radio luminosities (using a spectral index of 0.7) of our sample sources which were used as an approximation of the kinetic energy output/jet power. 

\begin{figure}[H]
\begin{adjustwidth}{-\extralength}{0cm}
\centering 
\hspace{8em}\includegraphics[width=8. cm]{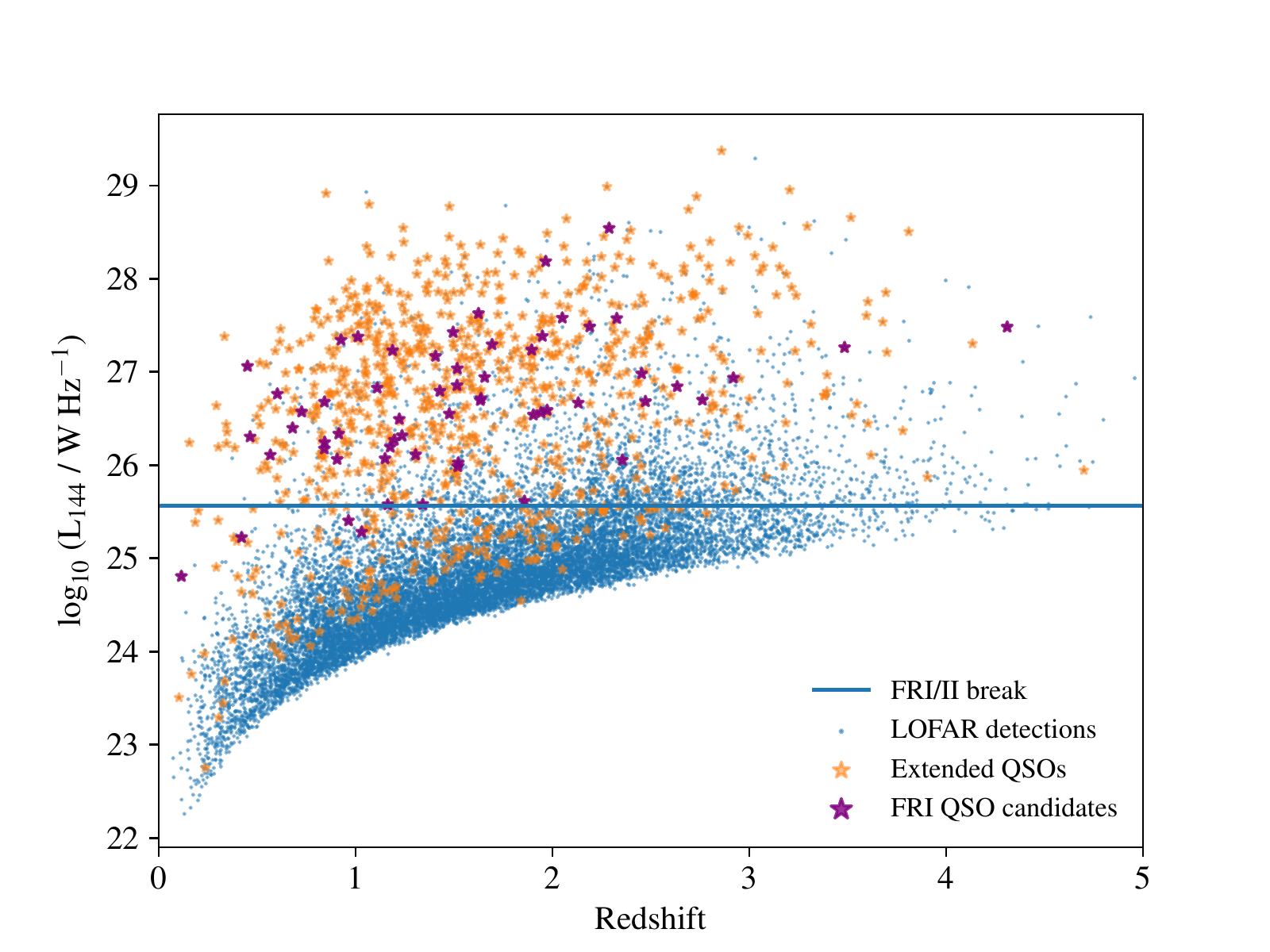}
\includegraphics[width=7.2 cm]{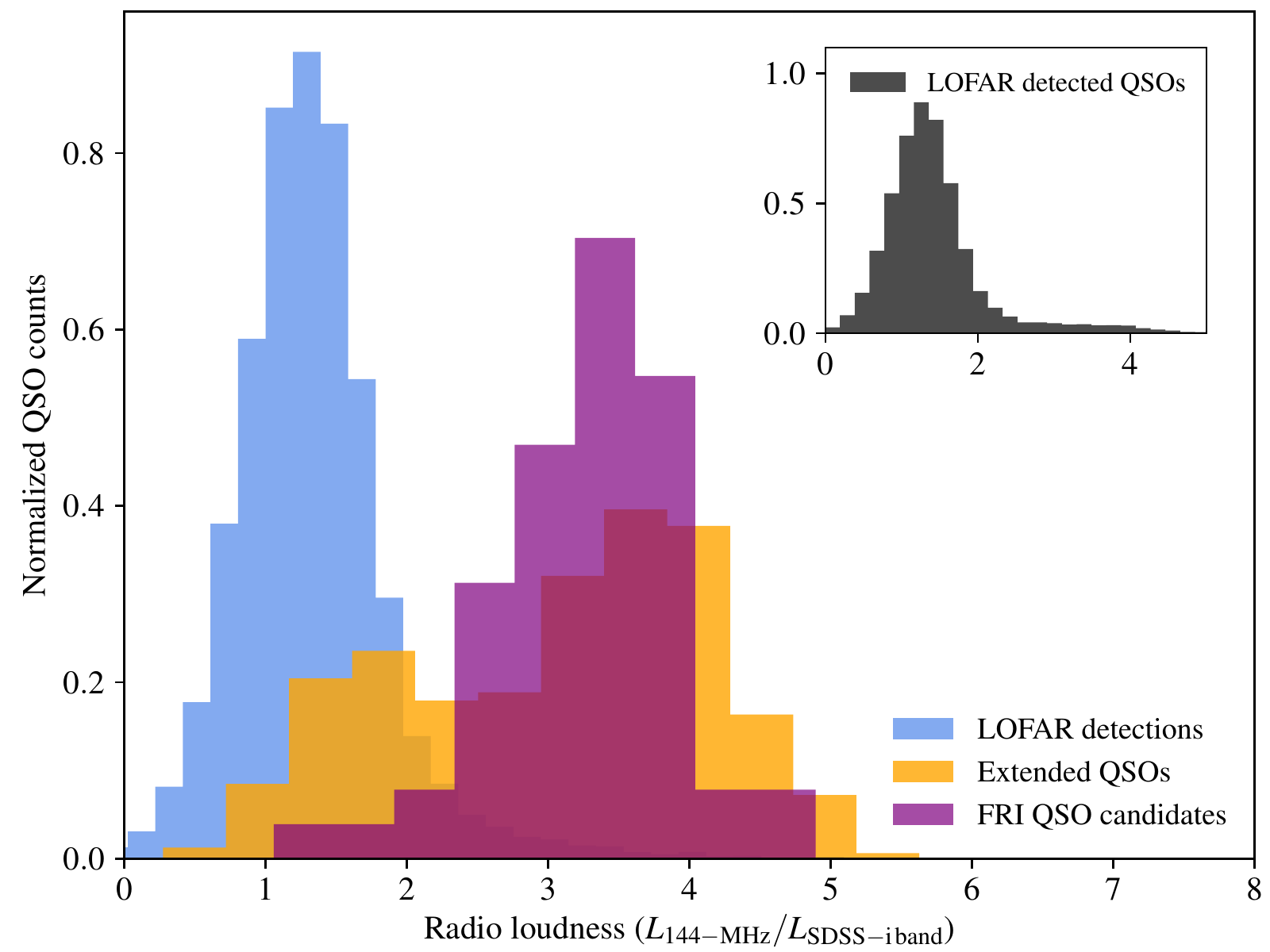}
\end{adjustwidth}
\caption{\textbf{Left-panel}: The distribution of 144-MHz luminosity of quasars as a function of their redshifts. Quasars detected in the LoTSS DR1  are shown as blue circles, sources showing extended (resolved in the LOFAR maps) emission are indicated by orange stars (these were obtained using the value-added catalogue) and our FRI quasar candidates (a subset of extended radio sources) are shown as purple stars. The blue solid line shows the traditional FRI/FRII break at 144 MHz. \textbf{Right-panel}: The radio loudness distribution of quasars (colour coded the same way as the left panel to show different classes) Inner panel shows the same histogram of all LOFAR detected quasars without dividing them into different classes (i.e., extended objects and FRI quasar candidates). \label{fig1}}
\end{figure}

On the left panel of  Figure \ref{fig1} we show the distribution of 144-MHz luminosity of the selected quasars as a function of their redshifts. Different symbols are used to show quasars detected in the LoTSS DR1 (blue circles), sources showing extended (resolved in the LOFAR maps) emission (orange stars; these were obtained using the value-added catalogue) and our FRI quasar candidates (purple stars). The blue solid line shows the traditional FRI/FRII break at 144 MHz. The radio-loudness parameter has traditionally been used to separate RQ and RL AGN, although we are not using this parameter we show the distribution of this parameter as a reference for other studies. The radio loudness distribution of quasars (colour coded the same way as the left panel to show different classes) is seen in the right panel of Figure \ref{fig1}. Here we define the radio loudness parameter for our quasar sample using the ratio of $L_{144}$ to $i$ band luminosity, as we did in \cite{gurkan19}. As can be clearly seen in both panels of Figure \ref{fig1} almost all of the FRI quasar candidates are above the FRI/FRII break (left panel) and they have a range of radio loudness parameter (covering a range of 0$-$5.5). The traditional radio-loud/radio-quiet dichotomy would predict that quasars with $R<2$ (at 144 MHz, \citep{gurkan19,morabito19}) are radio-quiet but we know that all of the objects classified as FRI or FRII in the sample have radio emission dominated by AGN.

\subsection{VLA Observations}
The 60 FRI quasar candidates were observed with the VLA on 29 September 2019 in the A-configuration (giving a resolution of around 1.3 arcsec). Observations were made in the broad-band L-band system in 8-bit mode (comprising 16 contiguous subbands of 64 MHz to cover the entire 1–2 GHz of the L-band receiver). Since we only required snapshot observations, exposure times were 10 min for each target (scheduled as two identical 5-h sessions, which allowed to improve the snapshot uv coverage by observing each source for 2 $\times$ 5 min ), reaching background rms levels of $\sim$30 $\upmu$Jy beam$^{-1}$ in the final combined images.The observations were targeted at the SDSS optical counterpart position for each~source. 

The quasar 3C 286 was used as the flux and bandpass calibrator, and since the large area of sky covered by our sources (the HETDEX-Spring region is $\sim$424 sq deg) five phase calibrators were used in total to correct for ionospheric variations over the course of ten hour observation time. The two epochs of observations were reduced separately. Prior to calibration the AOFlagger software \citep{offringa12} was used on each data set to automatically flag radio-frequency interference (RFI). The data were then reduced using the CASA VLA pipeline version 5.3.1 using CASA version 5.1.10 \citep{mcmullin7}. Various calibration tables were inspected to check the quality of calibration, and baselines displaying residual RFI or erratic phase variations were flagged manually in CASA. Images were produced by combining both epochs of observations in the $uv$-plane using the CASA image reconstruction technique CLEAN [making use of the Clarke algorithm;] \cite{clark80}.

\begin{figure}[H]
\includegraphics[width=10.cm]{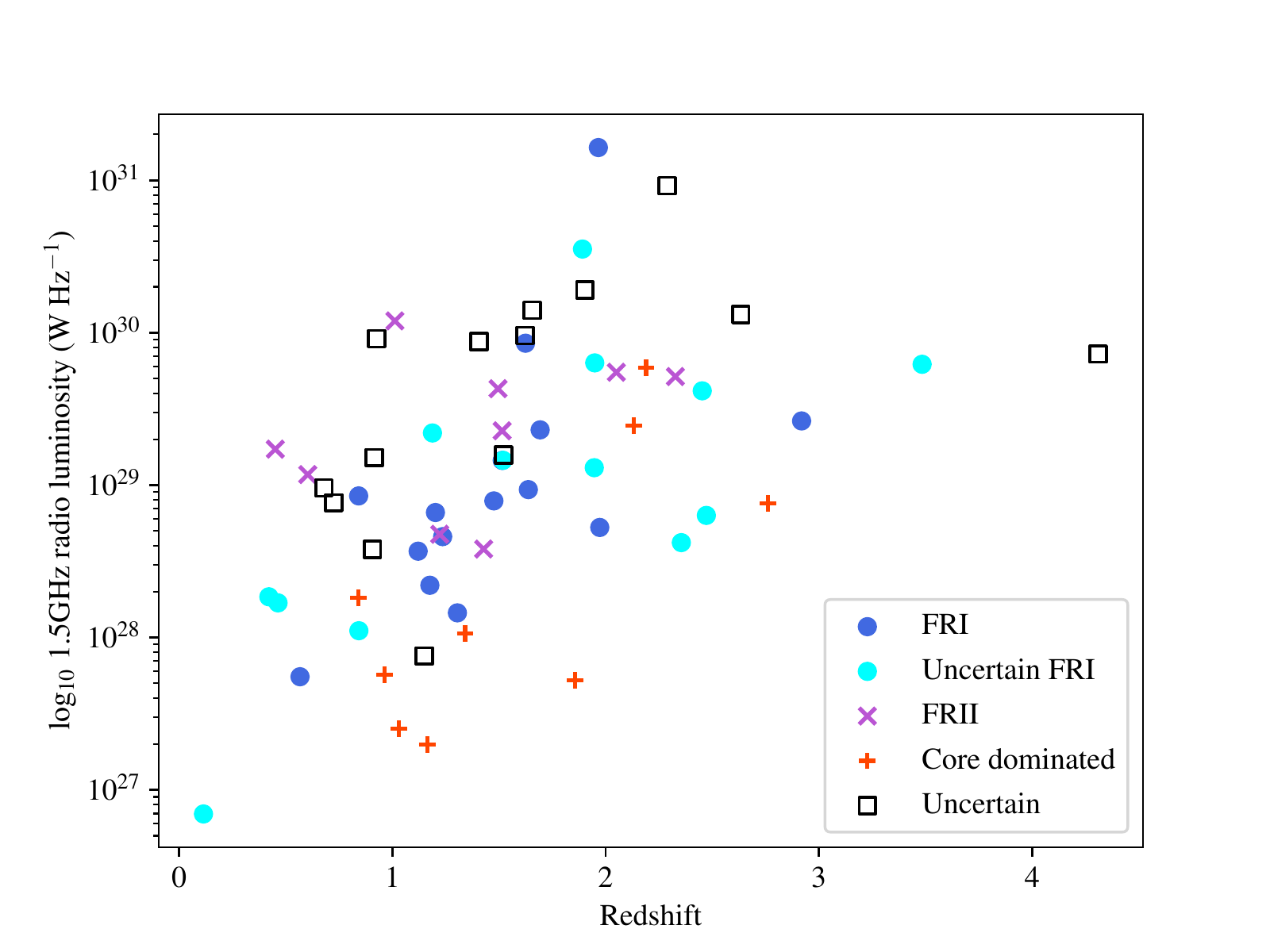}\\
\caption{The distribution of 1.5-GHz VLA luminosity of quasars as a function of their redshifts. See Section \ref{sec3.1} for the classification of sources. For a clean representation, we show FRII quasars and uncertain FRIIs using the same symbol. \label{fig0}}  
\end{figure}

Flux densities of quasars not showing any extended structure (i.e., point sources) were estimated by Gaussian fitting in CASA. For sources with extended emission we applied a mask to select pixels at least three times the off-source noise and measured the total flux (for each source separately). Errors on these flux densities were calculated by taking into account the noise and calibration errors \citep{pb17}. In Figure \ref{fig0} we show the VLA luminosity distribution of our sample sources as a function of their redshifts.

\section{Results}
\label{sec3}
\subsection{Classification Statistics of the Targets Using VLA Observations}
\label{sec3.1}
\begin{figure}[H]
\hspace{0.em}\includegraphics[width=6.8cm]{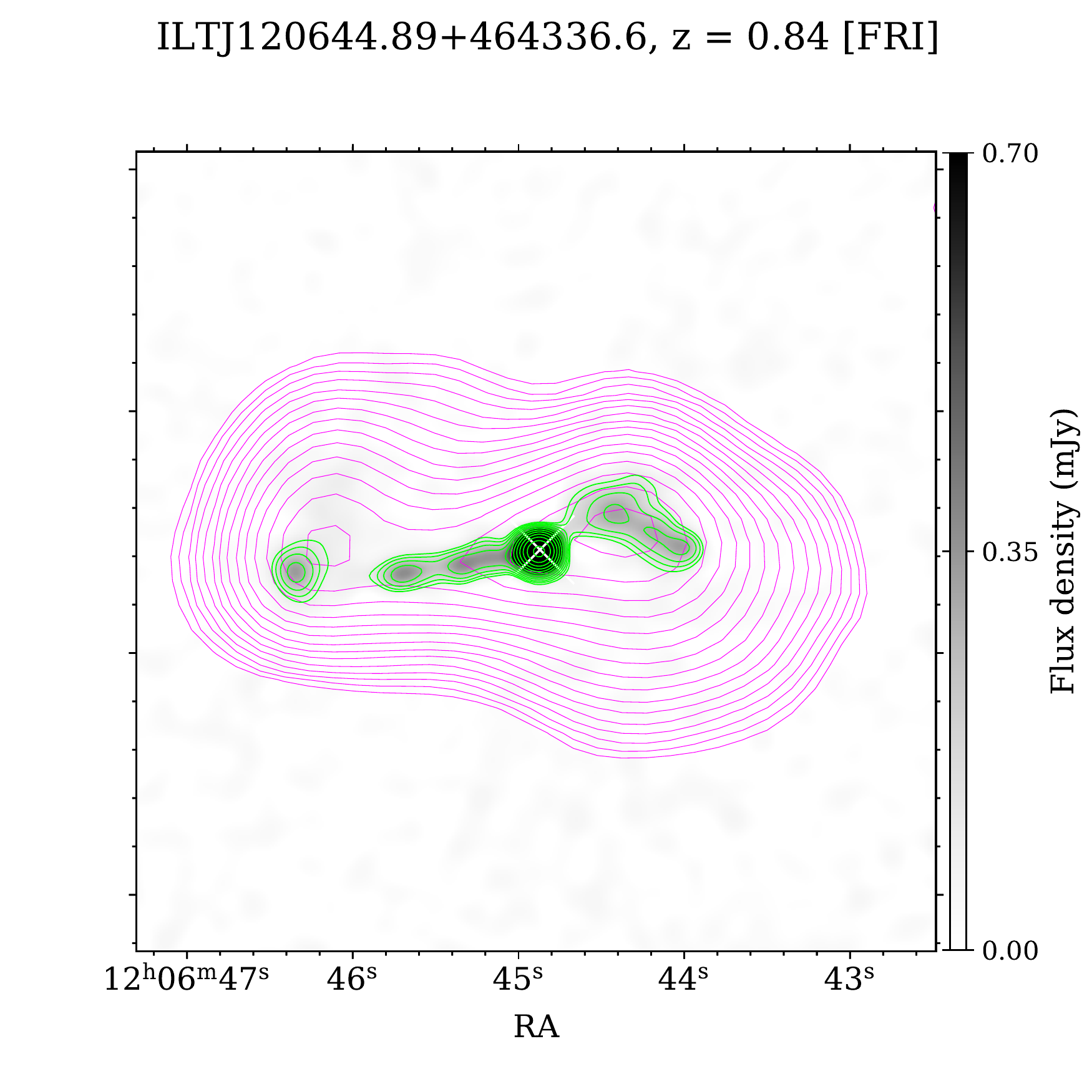}
\hspace{0.em}\includegraphics[width=6.8cm]{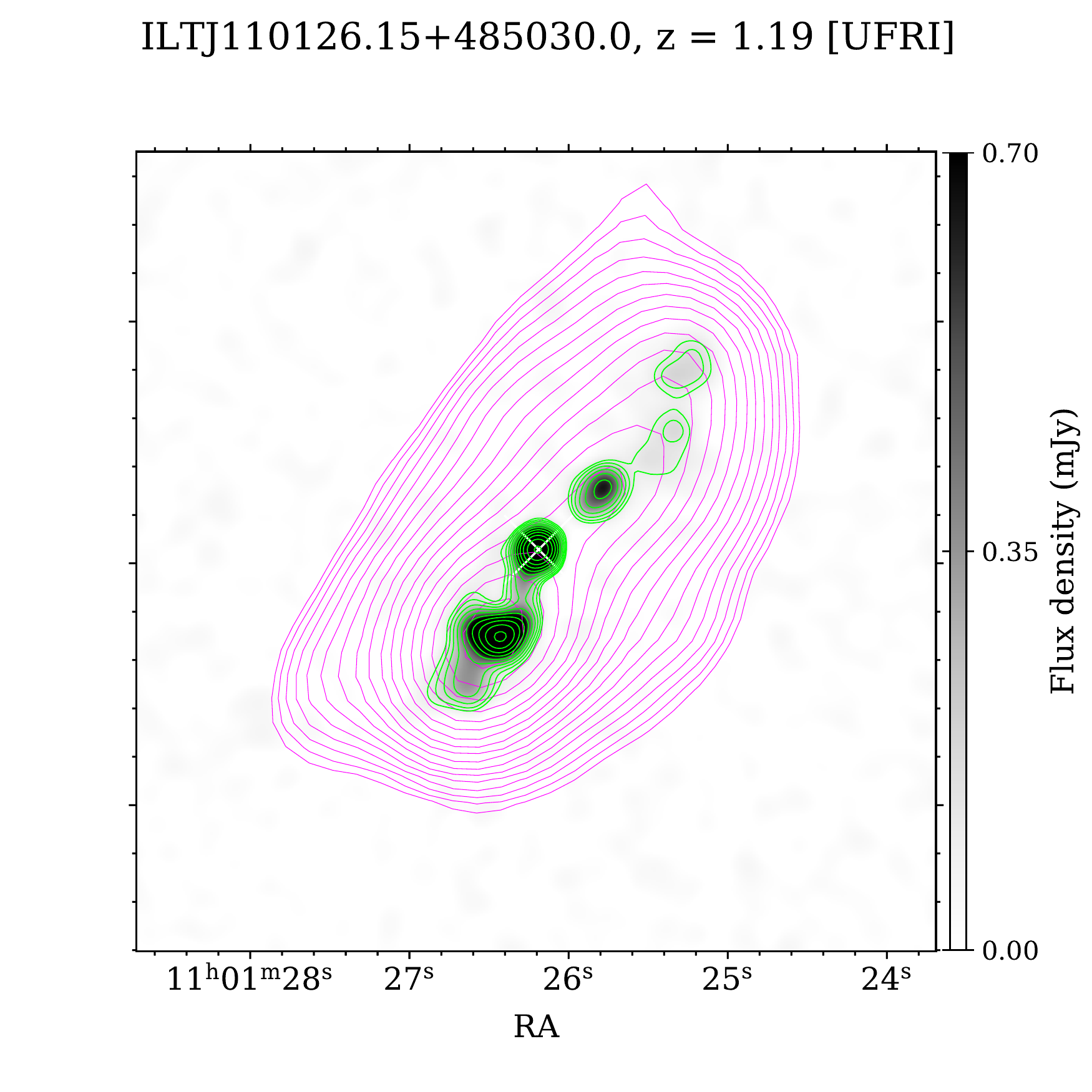}
\hspace{-5.em}\includegraphics[width=6.8cm]{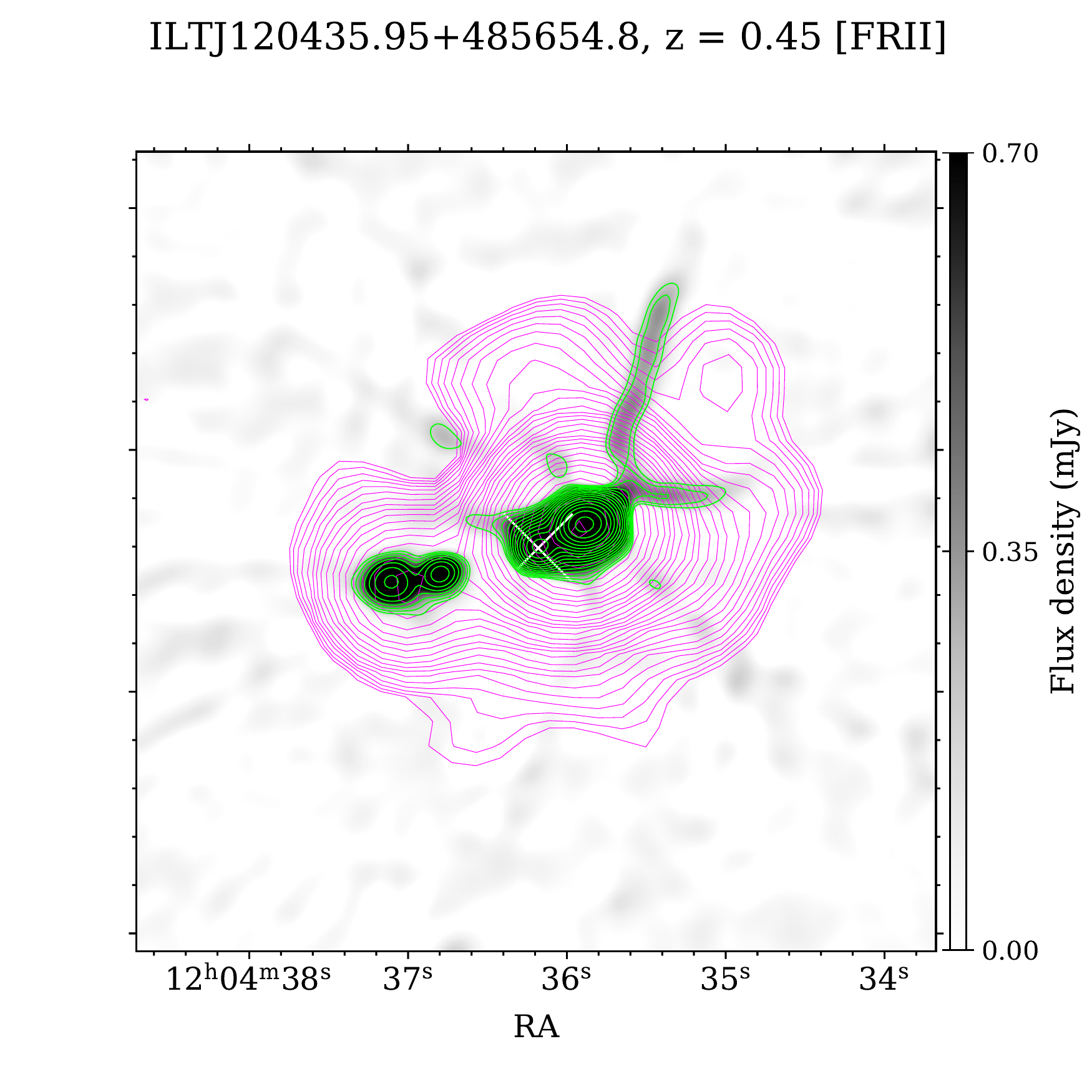}
\hspace{0.em}\includegraphics[width=6.8cm]{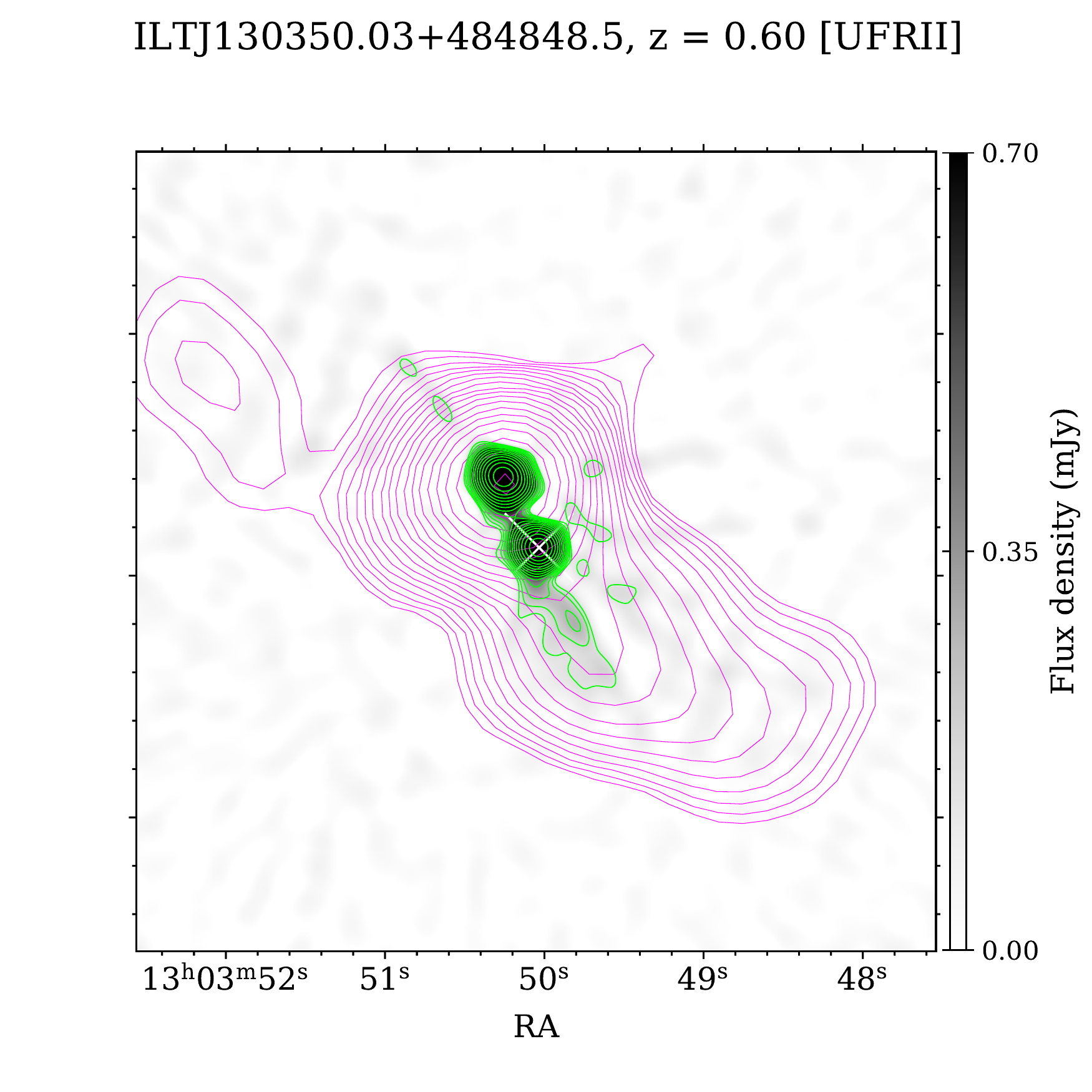}
\hspace{-5.em}\includegraphics[width=6.8cm]{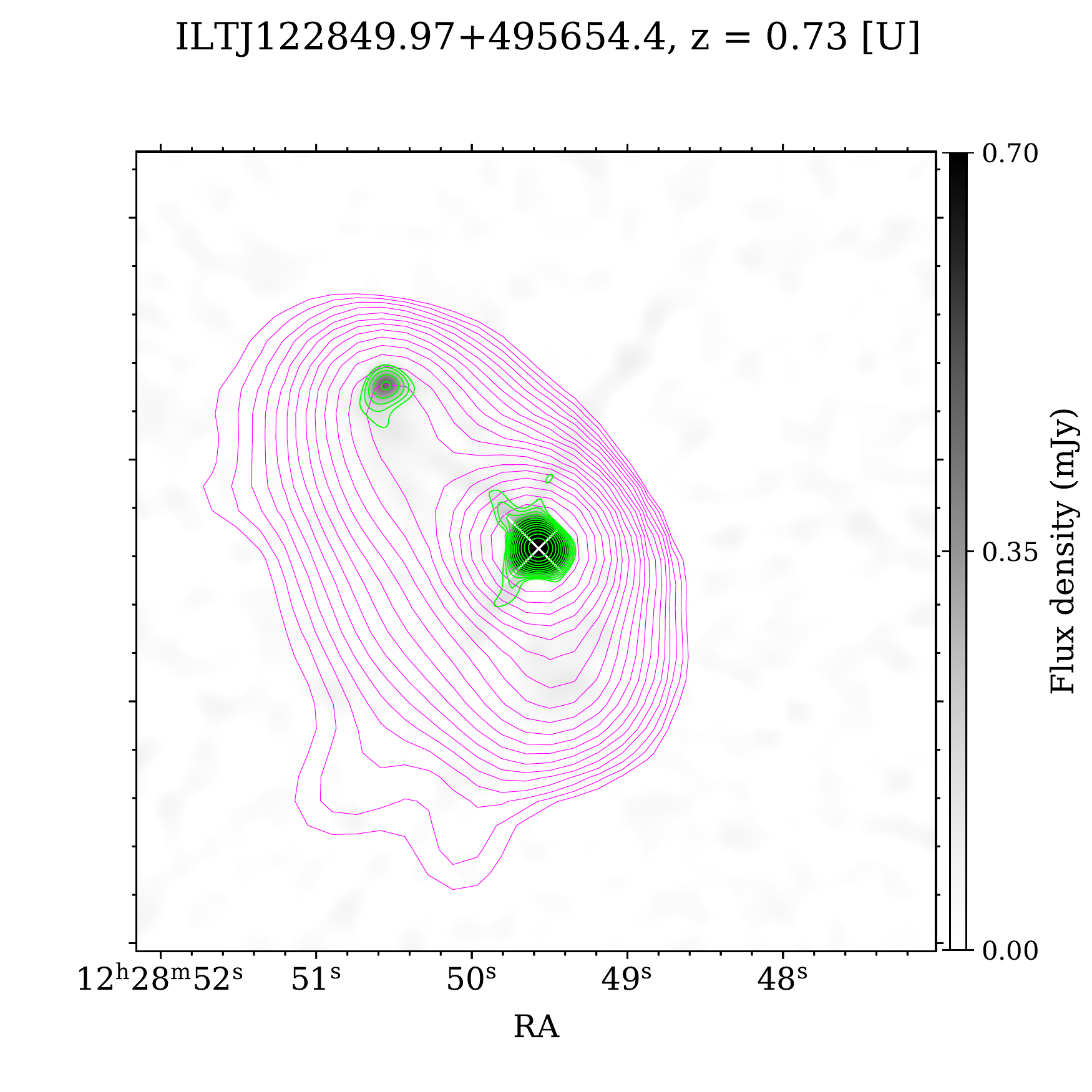}
\hspace{0.em}\includegraphics[width=6.8cm]{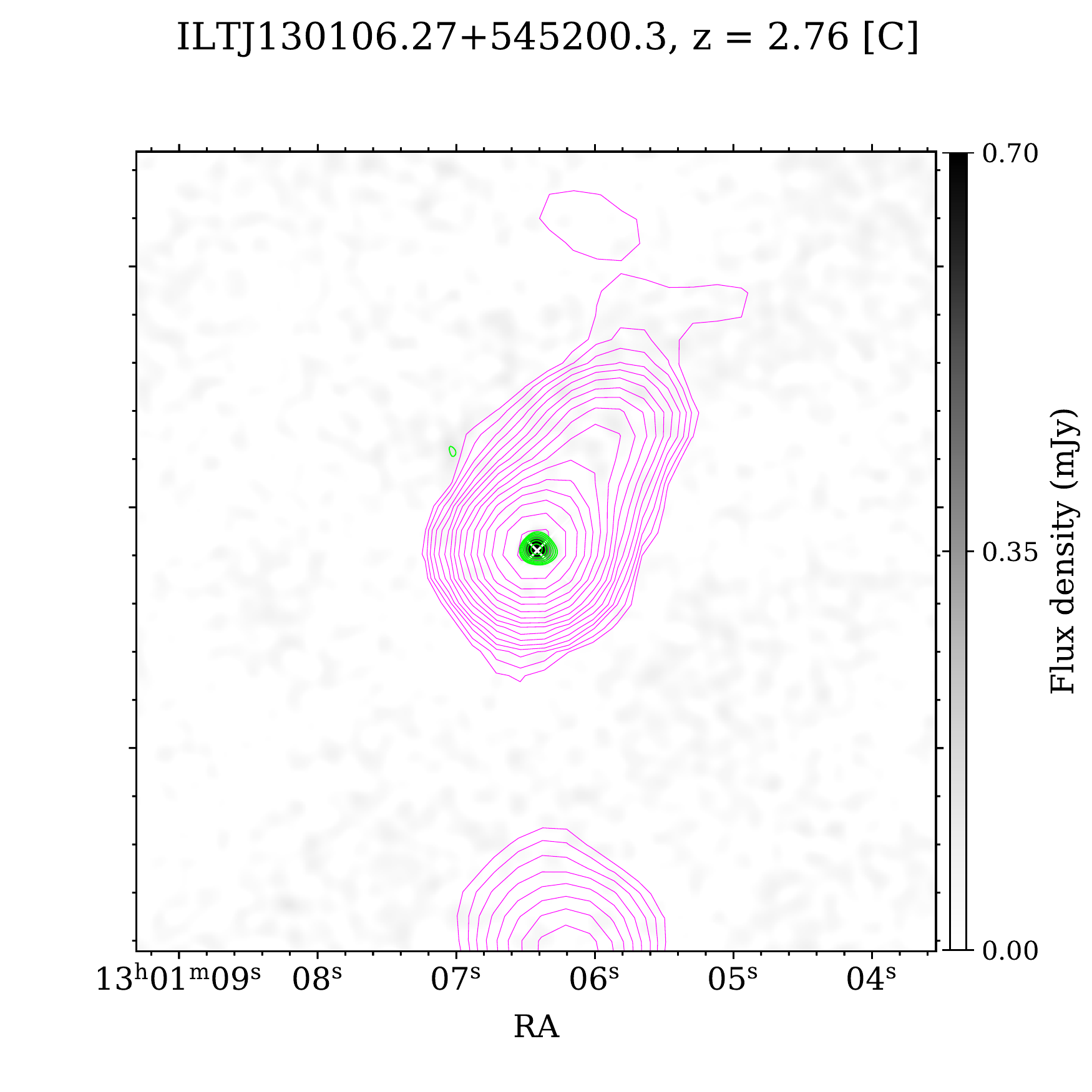}
\caption{Example images for different morphological classes. VLA contours (green), LOFAR contours (magenta) on VLA images (gray).  UFRI: uncertain FRI, UFRII: uncertain FRII, U:  uncertain, C: core-dominated  The LOFAR and VLA contours denote the surface brightness levels starting at 3$\sigma$and increasing at various powers of 3$\sigma$, where $\sigma$ denotes the local root-mean-square (RMS) noise. \label{fig:exam}} 
\end{figure}

We initially used the final VLA images of the target sources to have better constraints on their radio morphology. The VLA images of our target sources along with other radio data (LOFAR, FIRST, and VLASS) are shown in Figure \ref{fig:app} in the Appendix \ref{appendix}. We visually inspected these images to classify their radio morphology and these results can be found in Table \ref{tbl1}. Due to the limitation of the snapshot observations along with the effect of the source orientation to the line of sight the inner structures of some target quasars were not clearly revealed. The decision on a classification was made based on collective agreement from individual classifications by six authors. If a source was classified in the same category by more than three classifiers it is considered to be a genuine FRI or FRII. If three classifiers agreed upon a category then this source was labelled as uncertain FRI/FRII class. There are sources with no resolved components in the VLA images which were classified as core dominated sources and the remaining form the uncertain class. Example images for each morphological class can be seen in Figure \ref{fig:exam}. A summary of our target sources along with VLA measurements and their classifications are given in Table \ref{tbl:summary} in the Appendix \ref{appendix}.

\begin{table}[H]
	\centering
	\caption{Classification of the target sample using the follow up VLA observations.}
	\label{tbl1}
\begin{tabularx}{\textwidth}{c C l r m{6cm} p{2cm} b{2cm}}

		\midrule
		\textbf{Morphology Type} & \textbf{Counts}\\
		\midrule
		FRI & 15\\
		UFRI (Uncertain FRI) & 13\\
		FRII & 3\\
		UFRII (Uncertain FRII) & 6\\
		C (Unresolved/core dominated) & 9\\
		U (Uncertain) & 14\\
		\midrule
	\end{tabularx}
\end{table}

\subsection{Nuclear Properties of Quasars Associated with Different Radio Morphologies}
Although FRIs make up almost 5 percent of the LoTSS clean AGN sample \citep{mingo19}, among the extended quasars the fraction of FRIs is much lower: we found 32 FRIs (30 per cent) and 77 FRIIs (70 per cent). In this study out of the 60 quasars (we selected using SDSS DR14 and LoTSS DR1) the follow-up VLA snapshot observations allowed us to confirm 15 of these to be genuine FRIs. There are 13 quasars that show extended kpc structures similar to FRIs but there are bright locations in these jets which might be a hot spot or a jet knot so these were classified as uncertain FRIs. There are also three clear FRIIs and six uncertain FRIIs. Quasars which were not resolved by the VLA (A-config) beam were classified as core dominated. There are 14 objects which do not fit in any of these classifications so these are labelled as uncertain. The sample presented here form the largest sample of FRI quasars (and that of the possible candidates) to date. Moreover, our sample is not limited to powerful, bright objects in the sky as they are selected from the most sensitive low-frequency radio survey (going down to L$_{150\,MHz} > $10$^{25}$ W Hz$^{-1}$). Taking into account the size of our sample and its selection these statistics suggest that quasars with FRI radio morphology are not common in the universe. These results reinforces the idea that it is possible for a radiatively efficient system to form FRI jets but this probably requires certain conditions.

\begin{figure}[H]

\begin{adjustwidth}{-\extralength}{0cm}
\centering 
\hspace{18.mm}\includegraphics[width=8.cm]{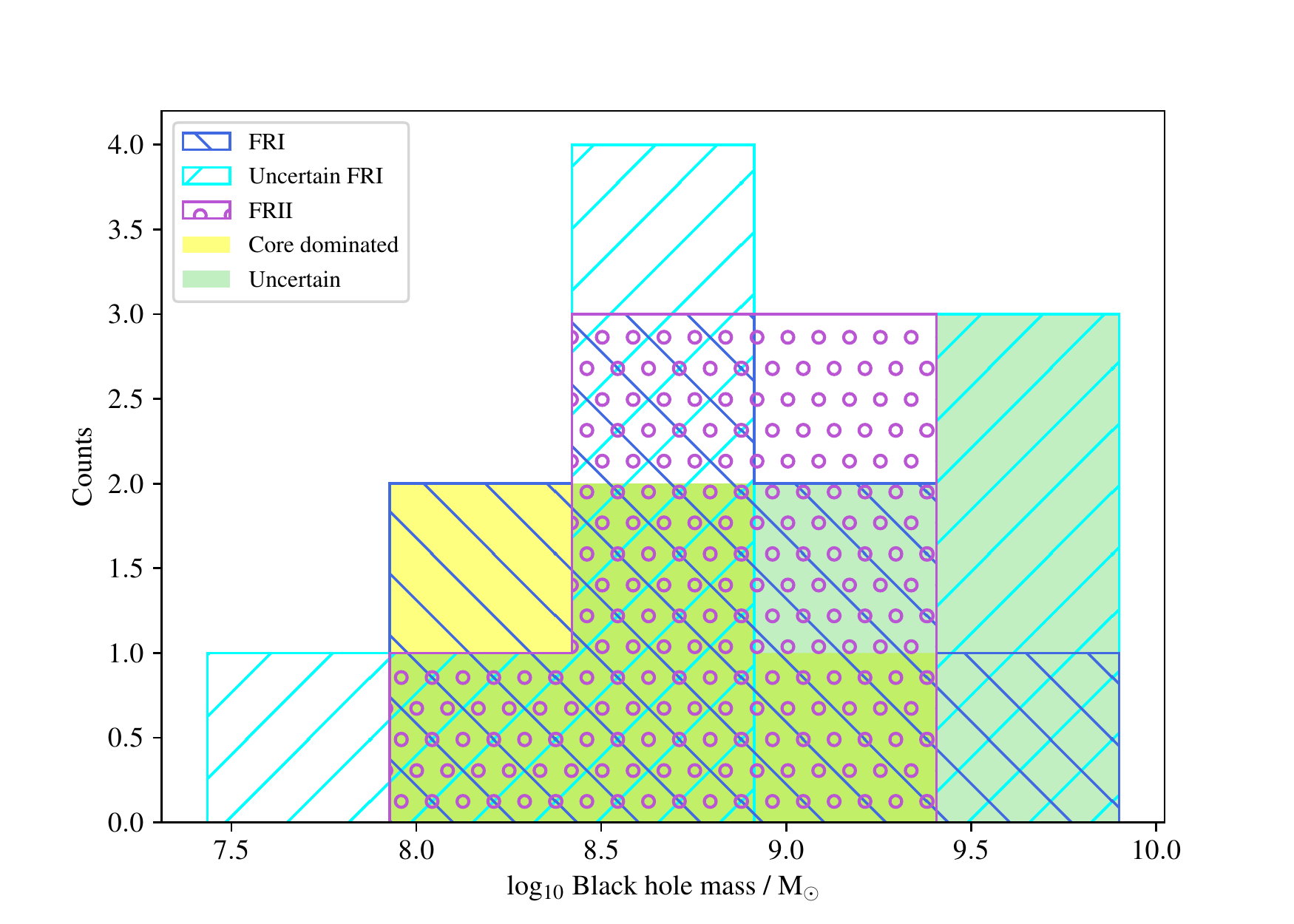}
\hspace{1.5em}\includegraphics[width=8.cm]{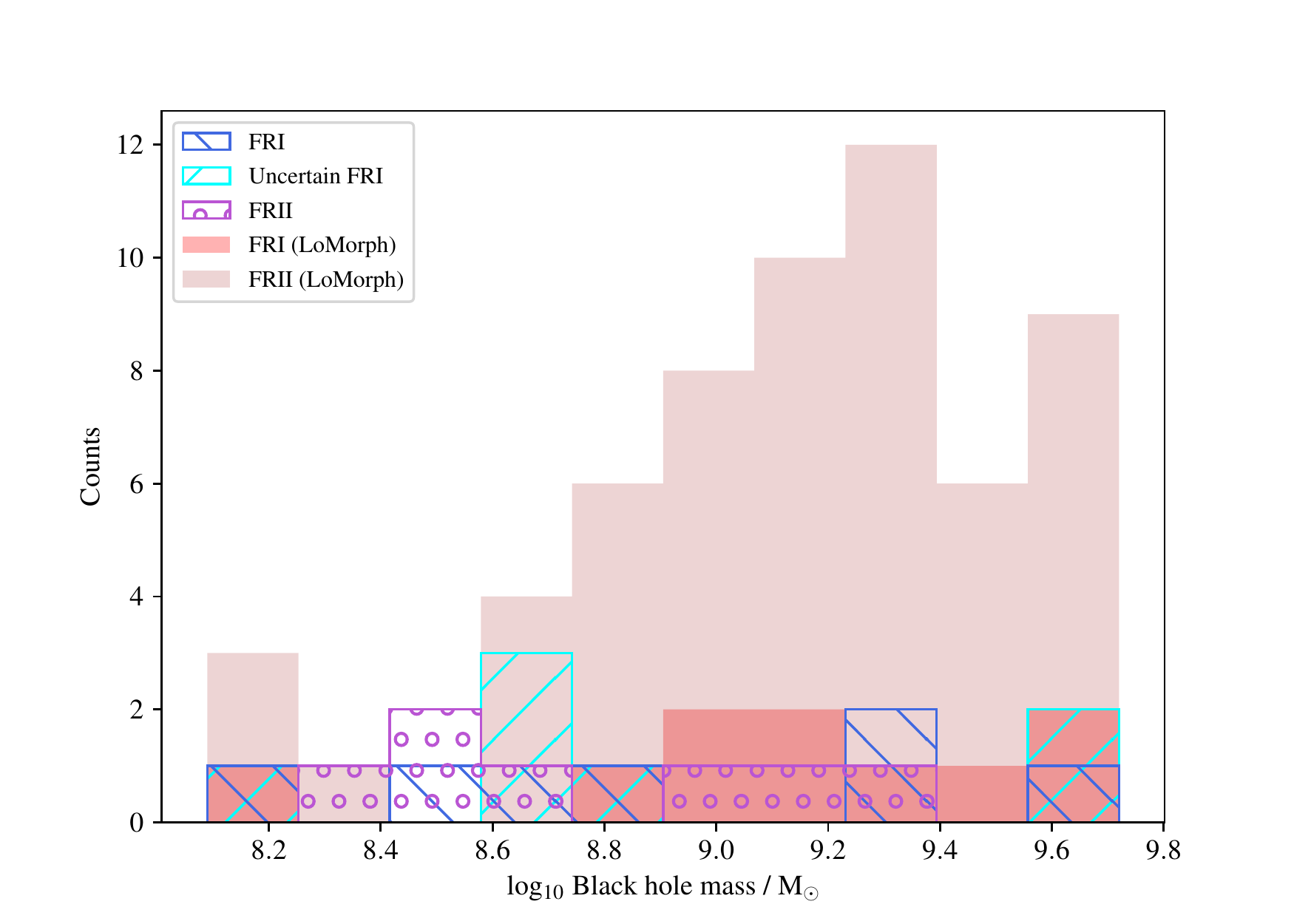}
\end{adjustwidth}
\caption{\textbf{Left-panel}: The black hole mass histogram of quasars that were included in the pilot survey with the VLA. \textbf{Right-panel}: The black hole mass histogram of quasars that were classified as FRI, FRII and uncertain FRI as part of our pilot study along with FRI and FRII quasars classified by  
\citep{mingo19} using LoMorph. For a clean representation, we show FRII quasars and uncertain FRIIs using the same symbol. \label{fig2}} 
\end{figure}

It is important to compare the properties of quasars associated with different radio morphology (FRI and FRII). We know that by definition quasars have radiatively efficient accretion so the accretion mode is not a parameter that plays a role here. The properties of the central engine in AGN might be important for understanding the rarity of FRI quasars. In Figure \ref{fig2} we show the histogram of quasar black hole masses, where available. In the left panel we show the black hole mass distribution of our sample observed as part of the VLA snapshot survey. From this panel only, we do not see any apparent difference in the range of black hole masses of quasars associated with different radio morphology, though the sample size is small. In the right panel of Figure \ref{fig2} we only show FRI, FRII and uncertain FRI quasars as well as FRI and FRII quasars from LoTSS sample classified by LoMorph. Here we see a similar result to the left panel: there is no clear difference between the black hole masses of FRI and FRII quasars. 

Although our sample size is small we used (k-sample) Anderson-Darling (AD) test to assess whether underlying probability distributions of black hole masses of FRI and FRII quasars differ from each other (Table \ref{tbl:adbhmass}). As can be seen from the results we cannot reject the null hypothesis (that two samples are drawn from the same population) at 5\% level for the black hole masses of quasars classified as FRI or FRII using our VLA observations. In terms of comparing FRI quasars (including uncertain FRIs) with FRII quasars similarly we fail to reject the null hypothesis at 5\% level. The underlying probability distribution of the black hole masses of all FRI quasars (FRIs, uncertain FRIs and LoMorph FRIs) and that of all FRII quasars (FRIIs and LoMorph FRIIs) do not differ from each other at 5\% level.

\begin{table}[H]
	\caption{Results of the Anderson-Darling test that has been used to assess the black hole masses of FRI and FRII quasars.}
		
	\label{tbl:adbhmass}
\begin{tabularx}{\textwidth}{C C C C m{3cm} p{2cm} b{2cm}}
\toprule
         \multicolumn{2}{c}{\textbf{Compared Populations and the Sample Size}} & \textbf{AD Statistic }& \textbf{Critical Value}\\
     \midrule
     FRI (9) & FRII (7) & $-$0.002 &  2.401 (0.05)\\ 
    \midrule
    FRI + UFRI (18) & FRII  (7) & $-$0.20 & 1.961 (0.05)\\ 
    \midrule
    All FRIs  (19)  & All FRIIs  (67) & $-$0.22 & 1.961 (0.05)\\ 
    \midrule
\end{tabularx}
\end{table}

\begin{figure}[H]

\begin{adjustwidth}{-\extralength}{0cm}
\centering 
\hspace{7.5em}\includegraphics[width=8cm]{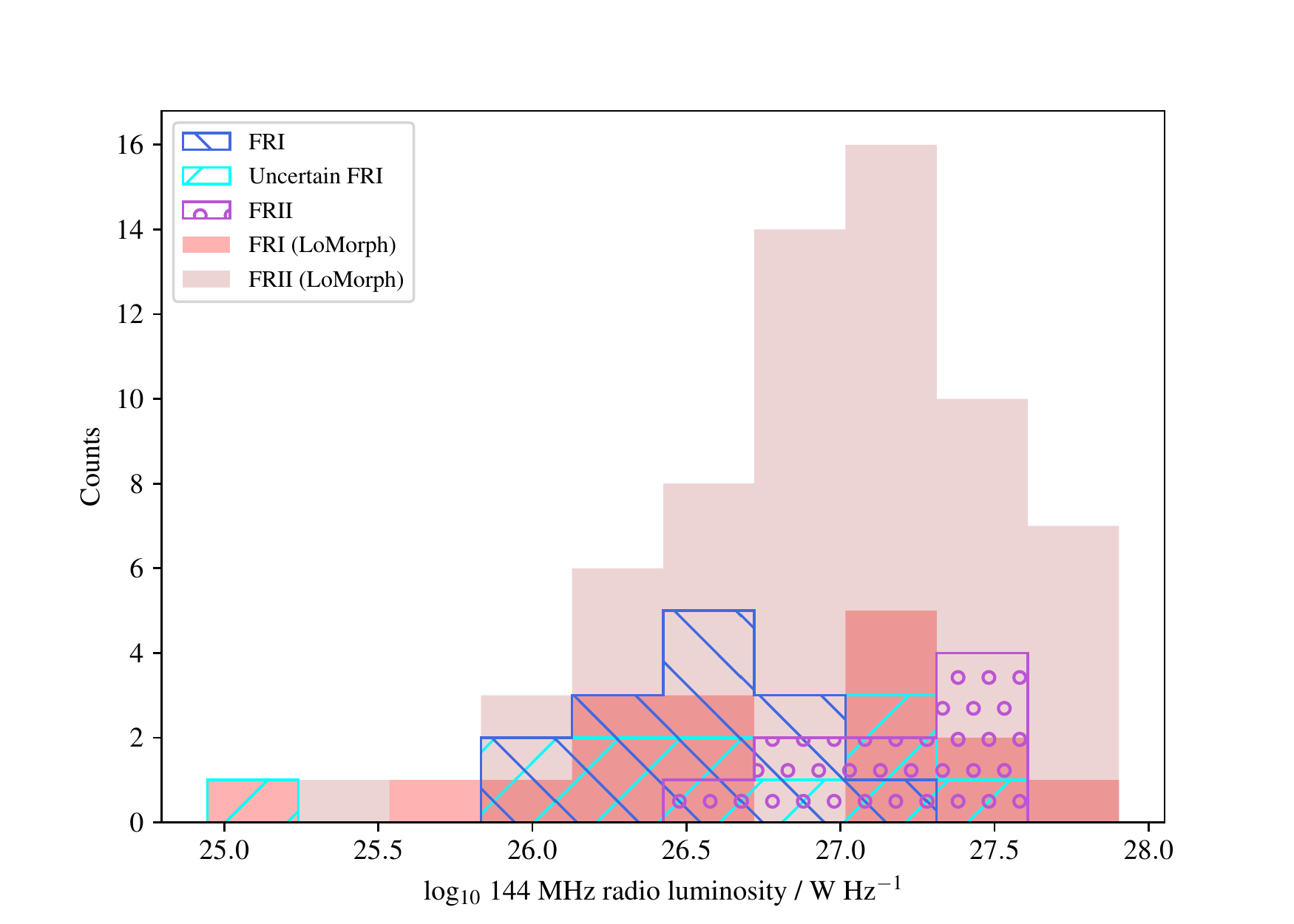}
\hspace{-1.em}\includegraphics[width=8cm]{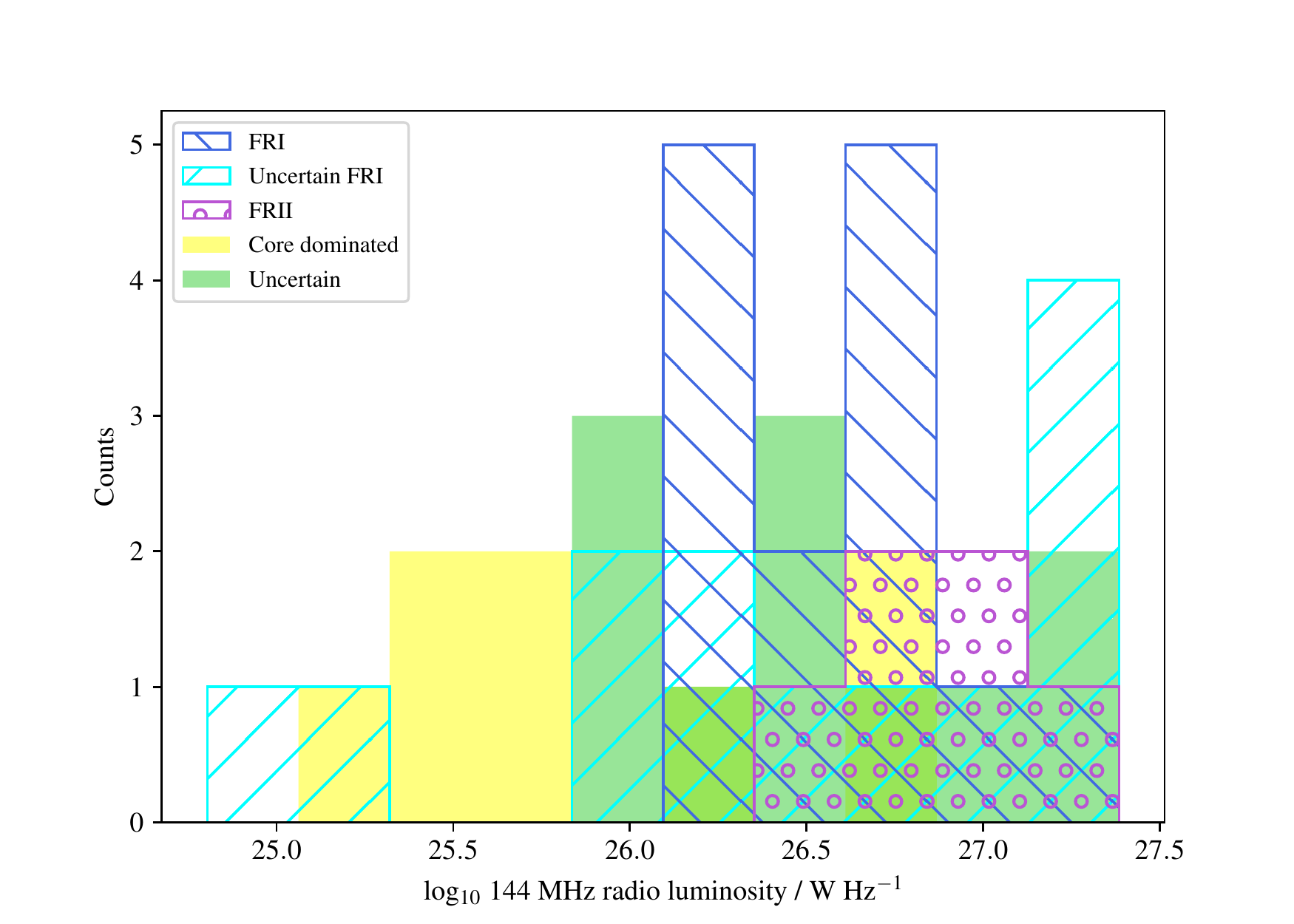}
\end{adjustwidth}
\caption{\textbf{Left-panel}: The low-frequency radio luminosity (an indicator for the jet power) histogram of quasars that were included in the pilot survey with the VLA. \textbf{Right-panel}: The low-frequency radio luminosity histogram of quasars that were classified as FRI, FRII and uncertain FRI as part of our pilot study along with FRI and FRII quasars classified by LoMorph. For a clean representation, we show FRII quasars and uncertain FRIIs using the same symbol. \label{fig3}} 
\end{figure}

\textls[-25]{AGN jet power is another key parameter (particularly the jet interaction with AGN environment) that defines the structural difference between FRI and FRII sources. The low-frequency radio luminosity is not only related to the lifetime-averaged jet kinetic power but also to the age of the source and to the properties of the external environment, e.g., \cite{hk13}. {As it is not straightforward to estimate the lifetime-averaged jet power we initially evaluate the low-frequency radio luminosity (which can be used to approximate the instantaneous jet power with the caveat of uncertainties) of our sources.} The low-frequency radio luminosity distribution of our sample sources is seen in the left panel of Figure \ref{fig3}. As can been seen in this panel, core-dominated quasars have a range of radio luminosities (10$^{25} < $L$_{150}$ / W Hz$^{-1}<$10$^{27}$) whereas} FRI quasars tend to have lower radio luminosities and FRII quasars have relatively higher radio luminosities. This is more apparent in the right panel of Figure \ref{fig3} where we have better statistics in terms of classification, albeit not all quasars in this panel have follow-up observations with the VLA. The results of the Anderson-Darling hypothesis testing also support this picture (Table \ref{tbl:adjet}). We can reject the null hypothesis at 5 per cent level that the radio luminosities of FRI and FRII quasars are drawn from the same distribution. We obtain similar results when we assess the radio luminosities of FRIs (including uncertain FRIs) and FRIIs (at 5 per cent significance level). Taking into account all FRIs (FRIs, uncertain FRIs and LoMorph FRI quasars) and FRIIs (FRIIs and LoMorph FRIIs) do not indicate any different result: the low-frequency radio luminosities (an approximation for AGN jet power) of FRI and FRII quasars are not drawn from the same distribution (at 5\% significance level). These results do not change when we use the jet powers of our sources estimated using a relation given by \citep{willott99} instead of their low-frequency radio luminosity (also see below).
\vspace{-12pt}

\begin{table}[H]
	\centering
	\caption{Results of the Anderson-Darling test that has been used to assess the low-frequency radio luminosity (as an indicator for AGN jet power) of FRI and FRII quasars.}
	\label{tbl:adjet}
\begin{tabularx}{\textwidth}{C C C C m{3cm} p{2cm} b{2cm}}
\toprule
         \multicolumn{2}{c}{\textbf{Compared Populations}} & \textbf{AD Statistic} & \textbf{Critical Value}\\
     \midrule
     FRI (15) & FRII (9) & 3.02 & 1.961 (0.05)\\ 
    \midrule
    FRI + UFRI (28) & FRII (9) & 3.50 & 1.961 (0.05) \\ 
    \midrule
    All FRIs (45) & All FRIIs (86) & 9.51 & 1.961 (0.05)\\ 
    \midrule
\end{tabularx}
\end{table}

\begin{figure}[H]
\includegraphics[width=15cm]{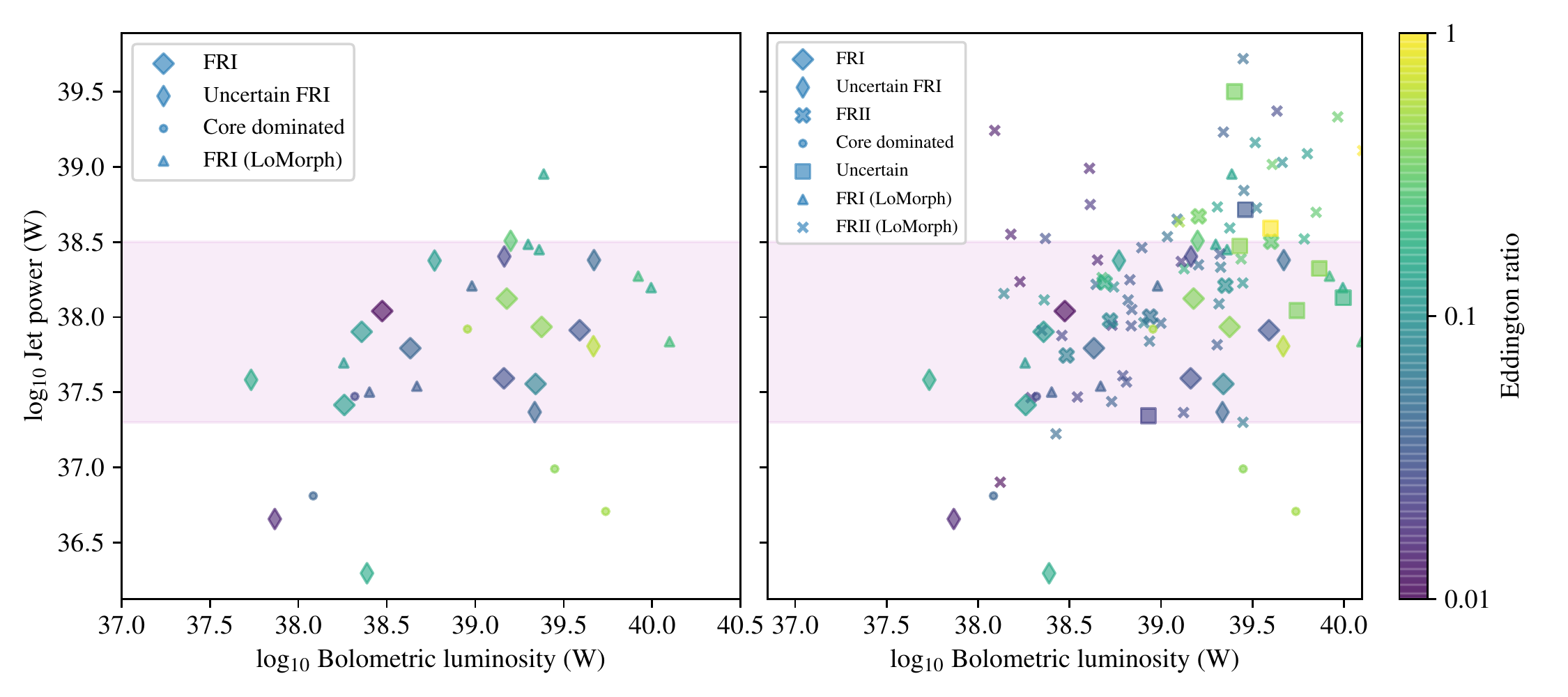}\\
\caption{The distribution of total bulk kinetic power as function of radiative power of our target sources: FRI, uncertain FRI, core-dominated quasars, respectively. We also show FRI quasars classified by LoMorph 
\textbf{left})  \citep{mingo19}. The right panel shows the same distribution as the right-hand side but we further include quasars with uncertain morphological classification as well as LoMorph-FRII quasars. For a clean representation, we show FRII quasars and uncertain FRIIs using the same symbol. Pink shaded regions just highlight the narrow range of jet powers of FRIs across a wider range of their radiative powers. Colour coding shows the Eddington ratios (i.e.,  Bolometric luminosity/Eddington luminosity).
\label{fig4}} 
\end{figure}

Radio-loud quasars, as radiatively efficient systems, produce both radiative and kinetic output as the matter is channelled onto the supermassive black hole via the accretion disk. In order to compare the radiative power of FRI and FRII quasars with their jet powers we estimate the total bulk kinetic power of quasars using the low-frequency radio luminosity following the relation given by \cite{willott99}. It is worthwhile to remind the reader that jets are capable of producing radio emission at a very wide range of luminosities, with the radio luminosity depending on the jet power, the environment and the source age \citep{Hardcastle18} and the jet power$-$radio luminosity relation might be expected to be different for radio populations classified as FRI and FRII \cite{Croston+18}. In Figure \ref{fig4} we compare the radiative (i.e.,  the bolometric luminosity) and the kinetic (i.e.,  jet power) products of quasars, colour coded by the Eddington ratios ({{Eddington ratio here is the ratio of the AGN bolometric luminosity to the Eddington luminosity.}}). Qualitatively the Eddington ratios of quasars with different radio morphology do not show any significant difference. The results of AD test (Table \ref{tbl:adedd}) support this qualitative result: Eddington ratios of FRI and FRII quasars are not different from each other.

In the left panel we show only FRI, uncertain FRI and core-dominated quasars from our sample with VLA follow-up observations along with FRIs quasars from LoTSS. All cleanly classified FRI quasars, the majority of uncertain FRIs and almost all of FRIs from the LoTSS exhibit a narrow range of jet powers whereas their accretion powers have a wider range of values (37$<$ log$_{10}$L$_{\mathrm{bol}}<$40.5). This picture emerges more clearly in the right panel of Figure \ref{fig4} where we include FRII quasars from both our sample and the LoTSS sample. 

\begin{table}[H]
	\centering
	\caption{Results of the Anderson-Darling test that has been used to assess the Eddington ratios of FRI and FRII quasars.}
	\label{tbl:adedd}
\begin{tabularx}{\textwidth}{C C C C m{3cm} p{2cm} b{2cm}}
\toprule
         \multicolumn{2}{c}{\textbf{Compared Populations}} &\textbf{ AD Statistic} & \textbf{Critical Value}\\
     \midrule
     FRI (9) & FRII (7) & 0.1 &  2.401 (0.05)\\ 
    \midrule
    FRI + UFRI (18) & FRII (7) & 0.09 & 1.961 (0.05) \\ 
    \midrule
    All FRIs (19) & All FRIIs (67) & 0.34 & 1.961 (0.05)\\ 
    \midrule
\end{tabularx}
\end{table}

\vspace{-12pt}

\subsection{Host Galaxy Colours}
\begin{figure}[H]
\includegraphics[width=13cm]{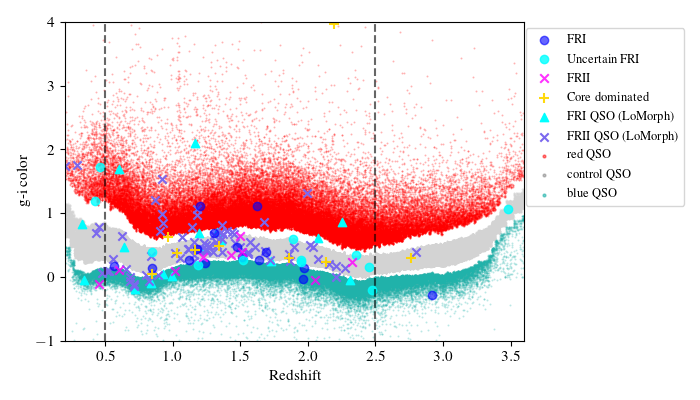}\\
\caption{The optical colour (SDSS $g-i$) distribution of quasars from the DR16Q catalogue, our target sample and the quasars sample with clean classification by LoMorph. Red quasar are quasars with red optical colours, control quasar indicate the control sample (the 10\% of quasars around the median of the $g-i$ distribution) and pure blue colour quasars are given as blue quasar. The optical colour classification is based on
\citep{klindt19,rosario20}. The dashed lines show the range of redshift within which the optical colours are reliable as described by \citep{klindt19,rosario20}. Symbols of our quasars sample are as above. For a clean representation, we show FRII quasars and uncertain FRIIs using the same symbol. \label{fig5}} 
\end{figure}

Recent studies focusing on SDSS quasars by \cite{klindt19} using FIRST data and \citep{rosario20} using LoTSS data investigated quasars with red and blue optical colours and found that quasars with red optical colours present enhanced radio emission from AGN. In their morphological classification most of the red quasars show compact features and some have FRI-like morphology. It is therefore interesting to evaluate the optical colours of our sample sources to see whether there is any trend in the optical colours of FRI quasars and how these compare to FRII quasars. Figure \ref{fig5} shows the optical colour distribution of quasars from the SDSS DR16 (this is most recent version which was not available at the time a sample for this project was selected), the quasar sample classified by LoMorph and our target sources from the VLA survey. Here the optical colours have been defined in the same way as done by~\citep{rosario20} using SDSS $g$ and $i$ magnitudes (corrected for Galactic extinction using the extinction values provided in the DR16Q catalogue \citep{lyke20}). Here in Figure \ref{fig5} red quasars are shown as red points and the control sample as defined by \citep{rosario20} are shown as blue and the remaining quasars are shown as gray points (both quasars that are shown as gray and blue points form the blue quasar sample by the aforementioned work). It is clear from this figure that: (i) The majority of FRI quasars do not have red optical colours and (ii) there is no clear difference between the optical colours of FRI and FRII quasars. 

\section{Discussion}
Based on the available data, we have evaluated various parameters with respect to the nuclear engine in the interest of testing whether there are differences between FRI and FRII quasars. As shown above, FRI and FRII quasars have a similar range of black hole masses (Figure \ref{fig2} and Table \ref{tbl:adbhmass}) and Eddington ratios (Figure \ref{fig4} and Table \ref{tbl:adedd}). There is a tendency for FRII quasars to have relatively higher jet powers than FRI quasars (Figure \ref{fig3}). Both FRI and FRII quasars can produce a variety of accretion power. For a given range of accretion powers FRII quasars can produce a broad range of jet powers ($\sim$10$^{36.6}-$10$^{40}$ W). FRI quasars, on the other hand, exhibit a quite narrow range of jet powers ($\sim$10$^{37.5}-$10$^{38.5}$ W). It is also possible that based on our selection criteria we are limited with less luminous objects (with the full survey we will be able to cover a wider range of luminosities). 

The formation of different radio morphologies has been argued to be due to either the influence of external environmental factors on the jet structure, e.g., \cite{bicknell95,ka97} or by parameters associated with the jet production mechanism \citep{meier01}. It is thought that both FRI and FRII sources initially expand supersonically and FRII objects then might evolve to FRIs once their jet disrupted, e.g., \cite{kb07}. There is no reason not to think that both effects can contribute to this dichotomy. In addition this, we know that AGN bolometric luminosity is a function of accretion rate, which is a function of black hole mass and radiative efficiency. For a standard accretion disk the radiative efficiency is expected to depend on black hole spin. Taking into account black holes masses of FRI and FRII quasars are not different for our sample and they are capable of producing a similar range of accretion powers quasars showing jet structures in the sample (independent of their radio morphology) should have spin values above a certain threshold. On the other hand, the radio luminosity of AGN is a complex function of AGN jet power (along with radiative losses, time, cosmic epoch and finally AGN environment) which depends on black hole spin and magnetic flux. As indicated above, quasars both with FRI and FRII morphology should have a black hole spin above a certain threshold in order to produce the relativistic jets but the intrinsic scatter on black hole spins and the strength of magnetic fields might be the key factors in determination of the low jet powers in FRI quasars, e.g., \cite{bz77}. Quasars with jets have been found be clustered more strongly than their radio-quiet counterparts across a wide redshift range, e.g., \cite{retanar17}. It has not been possible to compare the clustering/environments of FRI and FRII quasars to date but this is likely to change with the current and upcoming surveys. Nevertheless, previous studies on the radio galaxy environments found that FRI radio galaxies reside in denser (host scale) environments than FRII radio galaxies \citep{bicknell95,laing02,tchekbromber16},~but see \citep{lietzen11}. Therefore, we think that a combination of the scatter on black hole spin and magnetic flux, and the influence of external environmental factors define the radio morphology. Additionally, it is worth noting that based on the unification model quasar jets are seen at small angles to the line of sight [$\Theta<45$ deg;] \cite{barthel89} so it is possible that the projection effects coupled with the sensitivity limited observations restrict the number of sources we can classify.

We do not see any particular difference in the host galaxy colours probed using $g-i$ (extinction corrected) magnitudes from the SDSS data (Figure \ref{fig5}): the majority of FRI and FRII quasars have mainly blue optical colours. All these results indicate that another key parameter is important for quasars to produce FRI type of jets. A potential parameter for this is the galaxy environment. The current paradigm suggests that \citep{laing02,laing03,laing14} jets decelerate by entrainment of matter from relativistic speeds on pc scales to sub-relativistic speeds on kpc scales to form FRI jets. It is still debated whether entrainment is defined by the power of the jets and the galaxy external environment or it is due to a more fundamental difference. With the current data available to us we are unable to comment on the environment properties of quasars as our target sources are mainly at moderate and high redshifts. 

\section{Conclusions}
We homogeneously selected 60 FRI quasar candidates using the LoTSS DR1 data and further observed these target sources at higher angular resolution using the VLA with the aim of revealing the inner structures of the AGN jets. Our sensitivity-limited snapshot survey allowed us to validate the FRI classification of these quasars and increase the sample of known FRI quasars to date. We utilised the available SDSS, LoTSS and value-added data for understanding this population further and our results are as follows: 

\begin{itemize}

\item {25 per cent of the selected target quasars have been confirmed to be FRIs and around 25~per~cent show extended structures similar to FRI jets but with the current observations it is not clear whether the bright jet parts are jet knots of an FRI jet or hot spots in FRII jets,} 

\item{Although our sample size is small we do not see any difference in the range of black hole masses and Eddington ratios of FRI and FRII quasars. On the other hand, FRI quasars have relatively low jet powers compared to FRII quasars in the samples explored in this study, consistent with the expectation from low-z radio galaxies,} 

\item{Investigation of optical galaxy colours of quasars showed that FRI and FRII quasars mainly have blue optical colours. Based on the sample studied here we conclude that quasars, as the most powerful, radiatively efficient systems among the AGN population are more likely to produce FRII type jets. For a quasar to form FRI jets should have low jet power (which is expected to be characterized by black hole spin and magnetic flux) and reside in a significantly dense environment.}
\end{itemize}

Future studies using the VLA polarisation data of our target sources as well as e-MERLIN observations of selected quasars will provide more insights into the nature of these objects. As the ongoing/near future radio surveys progress the sample of FRI quasars will increase and eventually with the SKA (in conjunction with the optical surveys) we will advance our understanding of this rare population.

\acknowledgments{We thank the anonymous referees for their comments which improved the manuscript. MJH acknowledges support from STFC [ST/V000624/1]. JHC and BM acknowledges support from the UK Science and Technology Facilities Council (STFC) under grants ST/R00109X/1, ST/R000794/1, and ST/T000295/1. WLW  acknowledges support from the CAS-NWO programme for radio astronomy with project number 629.001.024, which is financed by the Netherlands Organisation for Scientific Research (NWO). This research has made use of the University of Hertfordshire high-performance computing facility (\url{http://uhhpc.herts.ac.uk/}) and the LOFAR-UK computing facility located at the University of Hertfordshire and supported by STFC [ST/P000096/1]. This research made use of astropy, a community-developed core Python package for astronomy \citep{astropy13} hosted at \url{http://www.astropy.org/} and of TOPCAT \citep{taylor05}. This publication uses data generated via the Zooniverse.org platform, development of which is funded by generous support, including a Global Impact Award from Google, and by a grant from the Alfred P. Sloan Foundation. This research has made use of the CIRADA cutout service at \url{cutouts.cirada.ca}, operated by the Canadian Initiative for Radio Astronomy Data Analysis (CIRADA). CIRADA is funded by a grant from the Canada Foundation for Innovation 2017 Innovation Fund (Project 35999), as well as by the Provinces of Ontario, British Columbia, Alberta, Manitoba and Quebec, in collaboration with the National Research Council of Canada, the US National Radio Astronomy Observatory and Australia’s Commonwealth Scientific and Industrial Research Organisation}

\captionsetup[table]{margin=9pt,justification=raggedright,singlelinecheck=off}

\startlandscape
\appendixstart
\appendix
\section{}
\label{appendix}
\begin{table}[H]
\caption{Properties of our target sources that were observed as part of the VLA snapshot survey. Unique source name (column 1), radio morphological class (column 2), right ascension \\and declination (column 3 $\&$ 4), redshift (column 5), 150 MHz flux density (column 6), 1.5 GHz flux density and its error (column 7 $\&$ 8), black hole mass (column 9), \\SDSS $i-$band magnitude (column 10), physical angular scale (column 11) and 5 GHz flux density (column 12).}
\centering
\setlength{\cellWidtha}{\textwidth/12-2\tabcolsep+1.2in}
\setlength{\cellWidthb}{\textwidth/12-2\tabcolsep+0.0in}
\setlength{\cellWidthc}{\textwidth/12-2\tabcolsep+0.0in}
\setlength{\cellWidthd}{\textwidth/12-2\tabcolsep+0.0in}
\setlength{\cellWidthe}{\textwidth/12-2\tabcolsep+0.0in}
\setlength{\cellWidthf}{\textwidth/12-2\tabcolsep+0.3in}
\setlength{\cellWidthg}{\textwidth/12-2\tabcolsep+0.3in}
\setlength{\cellWidthh}{\textwidth/12-2\tabcolsep+0.4in}
\setlength{\cellWidthi}{\textwidth/12-2\tabcolsep+0.0in}
\setlength{\cellWidthj}{\textwidth/12-2\tabcolsep+0.0in}
\setlength{\cellWidthk}{\textwidth/12-2\tabcolsep+0in}
\setlength{\cellWidthl}{\textwidth/12-2\tabcolsep+0.1in}
\scalebox{0.75}[0.75]{\begin{tabular}{>{\PreserveBackslash\raggedright}m{\cellWidtha}>{\PreserveBackslash\raggedright}m{\cellWidthb}>{\PreserveBackslash\raggedright}m{\cellWidthc}>{\PreserveBackslash\raggedright}m{\cellWidthd}>{\PreserveBackslash\raggedright}m{\cellWidthe}>{\PreserveBackslash\raggedright}m{\cellWidthf}>{\PreserveBackslash\raggedright}m{\cellWidthg}>{\PreserveBackslash\raggedright}m{\cellWidthh}>{\PreserveBackslash\raggedright}m{\cellWidthi}>{\PreserveBackslash\raggedright}m{\cellWidthj}>{\PreserveBackslash\raggedright}m{\cellWidthk}>{\PreserveBackslash\raggedright}m{\cellWidthl}}
\toprule
\textbf{Source Name}&\textbf{Class}&\textbf{RA (deg)}&\textbf{DEC (deg)}&\textbf{z}&\textbf{150 MHz Flux Density (mJy)}&\textbf{1.5 GHz Flux Density (mJy)}&\textbf{1.5 GHz Flux Density Error (mJy)}&\textbf{BH Mass log$_{10}$(M\boldmath$_{\odot}$)}&\textbf{\boldmath$i-$Band (mag)}&\textbf{Angular Scale (Mpc)}&\textbf{5 GHz Density (mJy)}\\
\midrule
ILTJ104920.91 + 521010.1&UFRI&162.337&52.169&1.52&8.21&12.35&3.71&$-$&-24.71&0.06&$-$\\
ILTJ105127.19 + 474716.8&C&162.863&47.788&1.03&3.98&0.52&0.06&8.94&-27.05&0.17&0.0\\
ILTJ110126.15 + 485030.0&UFRI&165.359&48.842&1.19&254.27&32.8&9.84&8.58&-24.44&0.32&$-$\\
ILTJ110201.90 + 533911.6&U&165.508&53.653&4.31&25.96&6.18&0.09&8.36&-27.34&0.08&7.025\\
ILTJ110309.20 + 475418.9&FRI&165.788&47.905&1.48&32.24&7.14&0.38&8.9&-24.22&0.07&$-$\\
ILTJ110548.29 + 511201.9&U&166.451&51.201&1.52&8.97&13.34&4.0&9.44&-24.92&0.09&$-$\\
ILTJ111014.00 + 464413.5&FRII&167.558&46.737&1.51&92.97&19.41&5.82&9.27&-25.96&0.1&17.05\\
ILTJ111042.51 + 455744.0&FRI&167.677&45.962&1.97&727.38&776.08&232.82&$-$&-24.66&0.07&$-$\\
ILTJ111312.71 + 483417.7&C&168.303&48.572&1.34&4.3&1.2&0.08&$-$&-23.25&0.08&$-$\\
ILTJ111738.07 + 483012.5&FRI&169.409&48.503&2.35&3.59&1.33&0.08&9.6&-26.05&0.09&$-$\\
ILTJ112003.81 + 463758.3&UFRII&170.016&46.633&1.22&43.67&6.65&2.0&8.53&-23.74&0.07&$-$\\
ILTJ112124.28 + 540227.8&UFRI&170.351&54.041&0.84&58.17&3.66&1.1&$-$&-22.73&0.6&$-$\\
ILTJ112621.48 + 555259.7&FRII&171.589&55.883&2.05&163.87&23.7&7.11&$-$&-26.05&0.08&$-$\\
ILTJ112913.38 + 501146.9&FRI&172.306&50.196&1.18&23.98&3.36&1.01&$-$&-24.2&0.11&$-$\\
ILTJ113251.06 + 541031.8&U&173.213&54.176&1.62&309.78&70.0&1.7&9.84&-26.23&0.3&88.56\\
ILTJ113836.90 + 524511.3&FRI&174.654&52.753&1.62&35.73&62.15&18.64&9.97&-26.72&0.26&15.81\\
ILTJ114047.85 + 462200.6&UFRI&175.199&46.367&0.11&182.18&19.9&2.4&8.09&-23.36&0.08&53.2\\
ILTJ114202.51 + 500229.3&C&175.51&50.041&2.19&113.91&21.82&0.36&$-$&-25.47&0.35&$-$\\
ILTJ114312.36 + 505016.3&U&175.801&50.838&0.91&59.41&41.46&0.58&$-$&-24.1&0.98&$-$\\
ILTJ114749.90 + 504605.4&C&176.958&50.768&2.13&18.4&9.66&0.24&$-$&-24.53&0.08&$-$\\
ILTJ114837.17 + 554500.6&UFRI&177.155&55.75&0.42&27.5&30.3&9.09&8.6&-22.56&0.23&$-$\\
ILTJ114845.60 + 554058.5&UFRI&177.19&55.683&0.46&265.06&22.14&6.64&7.43&-22.39&0.43&$-$\\
ILTJ115415.25 + 484454.7&UFRII&178.564&48.749&1.5&235.0&37.76&11.33&$-$&-24.35&0.17&$-$\\
ILTJ120132.45 + 552408.7&C&180.385&55.402&1.86&2.21&0.28&0.07&8.72&-26.19&0.0&$-$\\
ILTJ120435.95 + 485654.8&FRII&181.15&48.949&0.45&1615.12&241.03&72.31&8.34&-24.48&0.65&175.5\\
ILTJ120644.89 + 464336.6&FRI&181.687&46.727&0.84&157.96&28.13&8.44&8.2&-23.57&0.34&$-$\\

\midrule
\end{tabular}}
\label{tbl:summary}
\end{table}

\begin{table}[H]\ContinuedFloat
\caption{{\em Cont.}}
\centering

\setlength{\cellWidtha}{\textwidth/12-2\tabcolsep+1.2in}
\setlength{\cellWidthb}{\textwidth/12-2\tabcolsep+0.0in}
\setlength{\cellWidthc}{\textwidth/12-2\tabcolsep+0.0in}
\setlength{\cellWidthd}{\textwidth/12-2\tabcolsep+0.0in}
\setlength{\cellWidthe}{\textwidth/12-2\tabcolsep+0.0in}
\setlength{\cellWidthf}{\textwidth/12-2\tabcolsep+0.3in}
\setlength{\cellWidthg}{\textwidth/12-2\tabcolsep+0.3in}
\setlength{\cellWidthh}{\textwidth/12-2\tabcolsep+0.4in}
\setlength{\cellWidthi}{\textwidth/12-2\tabcolsep+0.0in}
\setlength{\cellWidthj}{\textwidth/12-2\tabcolsep+0.0in}
\setlength{\cellWidthk}{\textwidth/12-2\tabcolsep+0in}
\setlength{\cellWidthl}{\textwidth/12-2\tabcolsep+0.1in}
\scalebox{0.75}[0.75]{\begin{tabular}{>{\PreserveBackslash\raggedright}m{\cellWidtha}>{\PreserveBackslash\raggedright}m{\cellWidthb}>{\PreserveBackslash\raggedright}m{\cellWidthc}>{\PreserveBackslash\raggedright}m{\cellWidthd}>{\PreserveBackslash\raggedright}m{\cellWidthe}>{\PreserveBackslash\raggedright}m{\cellWidthf}>{\PreserveBackslash\raggedright}m{\cellWidthg}>{\PreserveBackslash\raggedright}m{\cellWidthh}>{\PreserveBackslash\raggedright}m{\cellWidthi}>{\PreserveBackslash\raggedright}m{\cellWidthj}>{\PreserveBackslash\raggedright}m{\cellWidthk}>{\PreserveBackslash\raggedright}m{\cellWidthl}}
\toprule
\textbf{Source Name}&\textbf{Class}&\textbf{RA (deg)}&\textbf{DEC (deg)}&\textbf{z}&\textbf{150 MHz Flux Density (mJy)}&\textbf{1.5 GHz Flux Density (mJy)}&\textbf{1.5 GHz Flux Density Error (mJy)}&\textbf{BH Mass log$_{10}$(M\boldmath$_{\odot}$)}&\textbf{\boldmath$i-$Band (mag)}&\textbf{Angular Scale (Mpc)}&\textbf{5 GHz Density (mJy)}\\
\midrule
ILTJ121426.22 + 545306.5&FRI&183.609&54.885&2.92&17.04&5.2&1.56&8.47&-25.17&0.08&$-$\\
ILTJ121427.14 + 480403.6&UFRI&183.613&48.068&1.95&17.85&6.29&1.89&8.8&-26.89&0.06&7.346\\
ILTJ121508.06 + 562517.5&C&183.784&56.422&0.96&6.15&1.38&0.06&8.39&-23.13&0.05&$-$\\
ILTJ121548.85 + 522447.2&FRI&183.954&52.413&1.52&61.39&12.39&3.72&$-$&-24.77&0.26&$-$\\
ILTJ121615.85 + 520305.7&U&184.066&52.052&1.15&18.94&1.22&0.37&$-$&-23.62&0.23&$-$\\
ILTJ122219.50 + 524729.2&UFRI&185.581&52.791&3.48&24.66&8.34&2.5&9.65&-25.6&0.07&$-$\\
ILTJ122338.96 + 461120.3&UFRII&185.912&46.189&1.01&516.49&258.42&77.53&9.02&-26.52&0.32&257.3\\
ILTJ122434.28 + 470045.3&C&186.143&47.013&1.16&5.91&0.31&0.05&$-$&-23.13&0.38&$-$\\
ILTJ122437.94 + 500806.1&U&186.158&50.135&0.91&32.31&10.55&0.28&$-$&-24.05&0.08&$-$\\
ILTJ122849.97 + 495654.4&U&187.208&49.948&0.73&174.73&35.82&10.75&$-$&-22.57&0.88&$-$\\
ILTJ122909.25 + 552231.9&U&187.289&55.376&1.41&149.63&88.6&1.3&9.15&-27.05&0.19&52.7\\
ILTJ123315.85 + 492836.1&FRI&188.316&49.477&1.69&130.73&15.28&4.58&$-$&-24.93&0.26&$-$\\
ILTJ123530.80 + 522828.9&U&188.878&52.475&1.66&61.23&97.9&2.0&9.59&-27.63&1.16&74.83\\
ILTJ123639.59 + 494633.7&U&189.165&49.776&1.9&17.65&97.15&29.15&$-$&-25.78&0.29&$-$\\
ILTJ124339.80 + 500653.9&UFRII&190.916&50.115&1.43&61.18&3.72&1.12&9.11&-24.91&0.93&$-$\\
ILTJ124746.08 + 502251.6&UFRII&191.942&50.381&2.33&122.06&16.75&5.03&8.56&-25.91&0.12&18.47\\
ILTJ125959.20 + 560735.0&FRI&194.997&56.126&1.12&117.88&6.29&1.89&9.36&-24.1&0.41&10.66\\
ILTJ130106.27 + 545200.3&C&195.276&54.867&2.76&11.15&1.7&0.09&8.13&-25.09&0.21&$-$\\
ILTJ130350.03 + 484848.5&UFRII&195.958&48.813&0.6&418.74&84.17&25.25&8.71&-24.33&0.31&45.18\\
ILTJ131102.97 + 551353.7&U&197.762&55.232&0.93&586.88&242.7&5.0&8.69&-25.22&0.09&226.3\\
ILTJ131753.68 + 462449.5&UFRI&199.474&46.414&1.95&117.03&30.59&9.18&8.65&-25.66&0.25&$-$\\
ILTJ132348.54 + 540323.3&U&200.952&54.056&2.63&17.25&32.47&9.74&9.06&-26.8&0.11&$-$\\
ILTJ133653.87 + 530511.2&U&204.224&53.086&0.68&136.33&52.14&0.56&$-$&-23.63&0.73&$-$\\
ILTJ133707.41 + 545737.2&C&204.281&54.96&0.84&49.75&6.01&0.2&8.8&-23.78&0.26&4.26\\
ILTJ134350.61 + 480358.9&U&205.961&48.066&2.29&1178.27&310.7&2.9&8.74&-26.16&0.06&$-$\\
ILTJ141317.64 + 480826.8&FRI&213.323&48.141&1.24&28.28&6.26&1.88&9.58&-25.3&0.12&$-$\\
ILTJ141416.33 + 533509.4&UFRI&213.568&53.586&2.45&27.74&12.01&0.14&$-$&-25.12&0.09&$-$\\
ILTJ141419.55 + 491820.4&FRI&213.581&49.306&0.57&106.66&4.58&1.37&8.06&-23.13&0.36&$-$\\
ILTJ142741.62 + 520410.1&FRI&216.923&52.069&1.3&15.58&1.75&0.09&$-$&-24.02&0.11&$-$\\
ILTJ143306.31 + 455527.9&UFRI&218.276&45.924&2.47&13.73&1.8&0.54&$-$&-25.02&0.58&$-$\\
ILTJ144057.33 + 510619.6&FRI&220.239&51.105&1.64&37.22&6.68&2.01&8.64&-26.09&0.11&6.385\\
ILTJ144157.46 + 455453.7&FRI&220.489&45.915&1.97&18.26&2.48&0.74&$-$&-24.46&0.29&$-$\\
ILTJ144611.40 + 484615.2&FRI&221.548&48.771&1.2&27.27&9.6&0.12&9.37&-26.09&0.3&10.53\\
ILTJ151532.53 + 491525.9&UFRI&228.886&49.257&1.89&88.75&182.68&54.8&9.9&-26.86&0.08&35.67\\
\midrule
\end{tabular}}
\end{table}
\finishlandscape

\begin{figure}[H]
\begin{adjustwidth}{-\extralength}{-1cm}
\centering 

\begin{subfigure}

\includegraphics[scale=0.38]{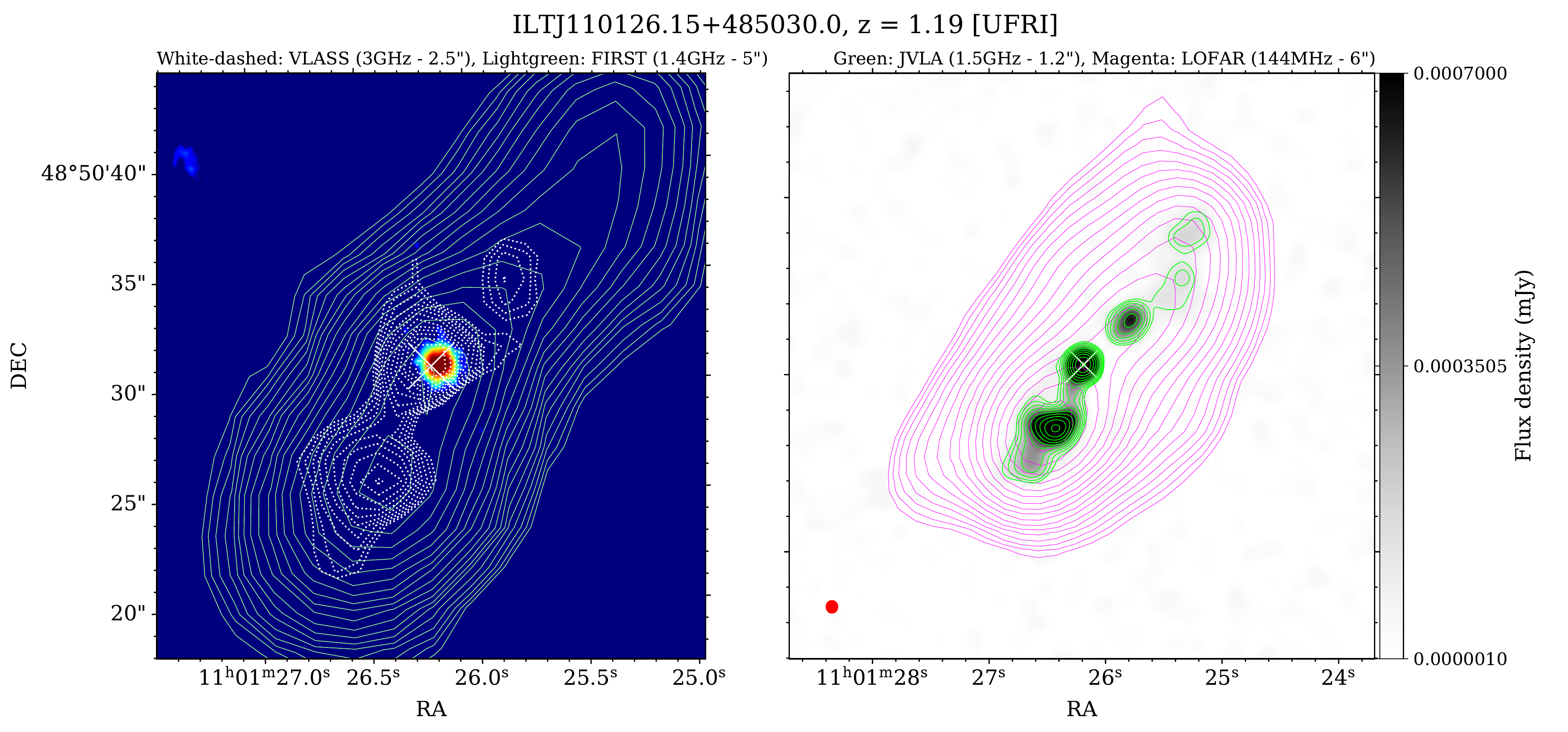}\\
\includegraphics[scale=0.38]{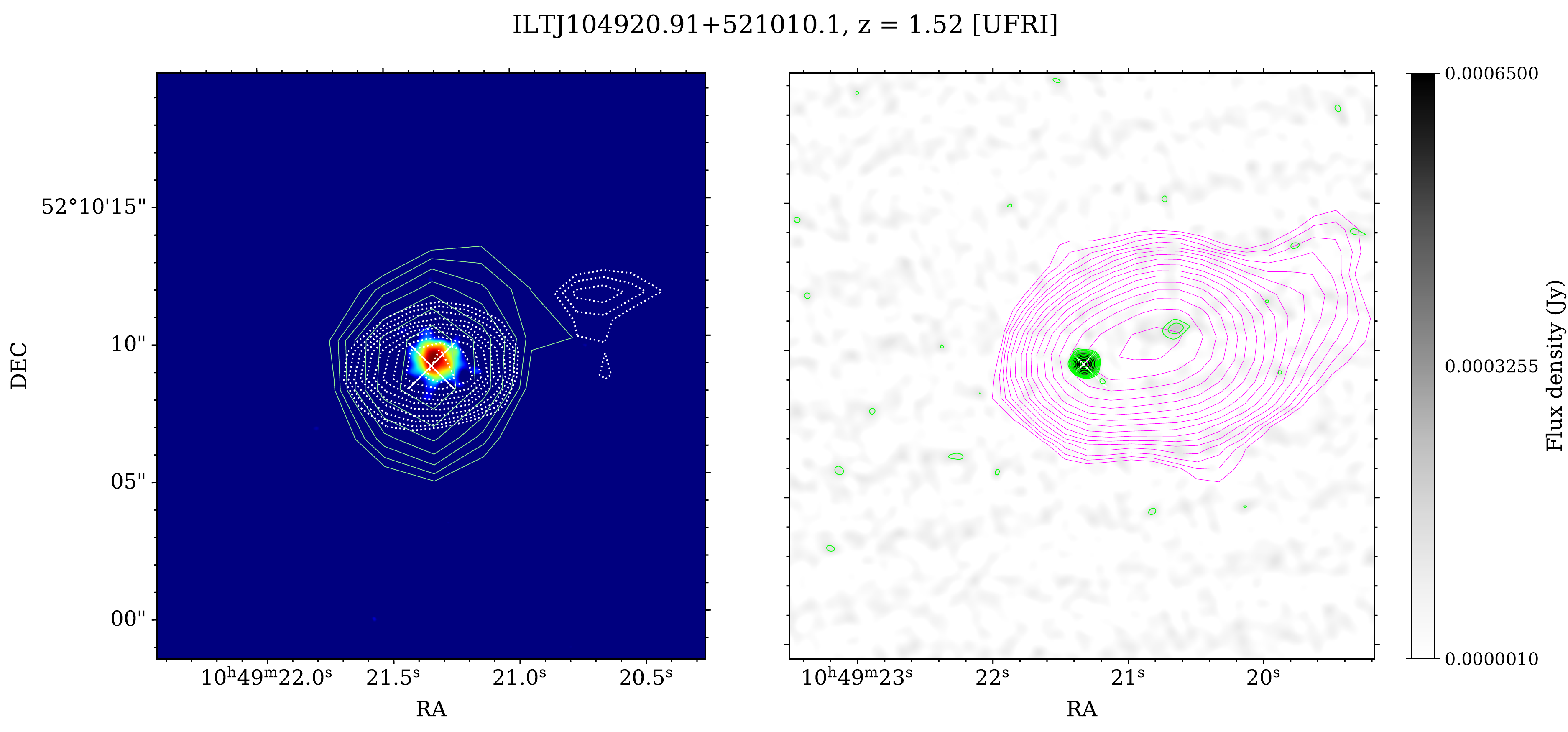}\\
\includegraphics[scale=0.38]{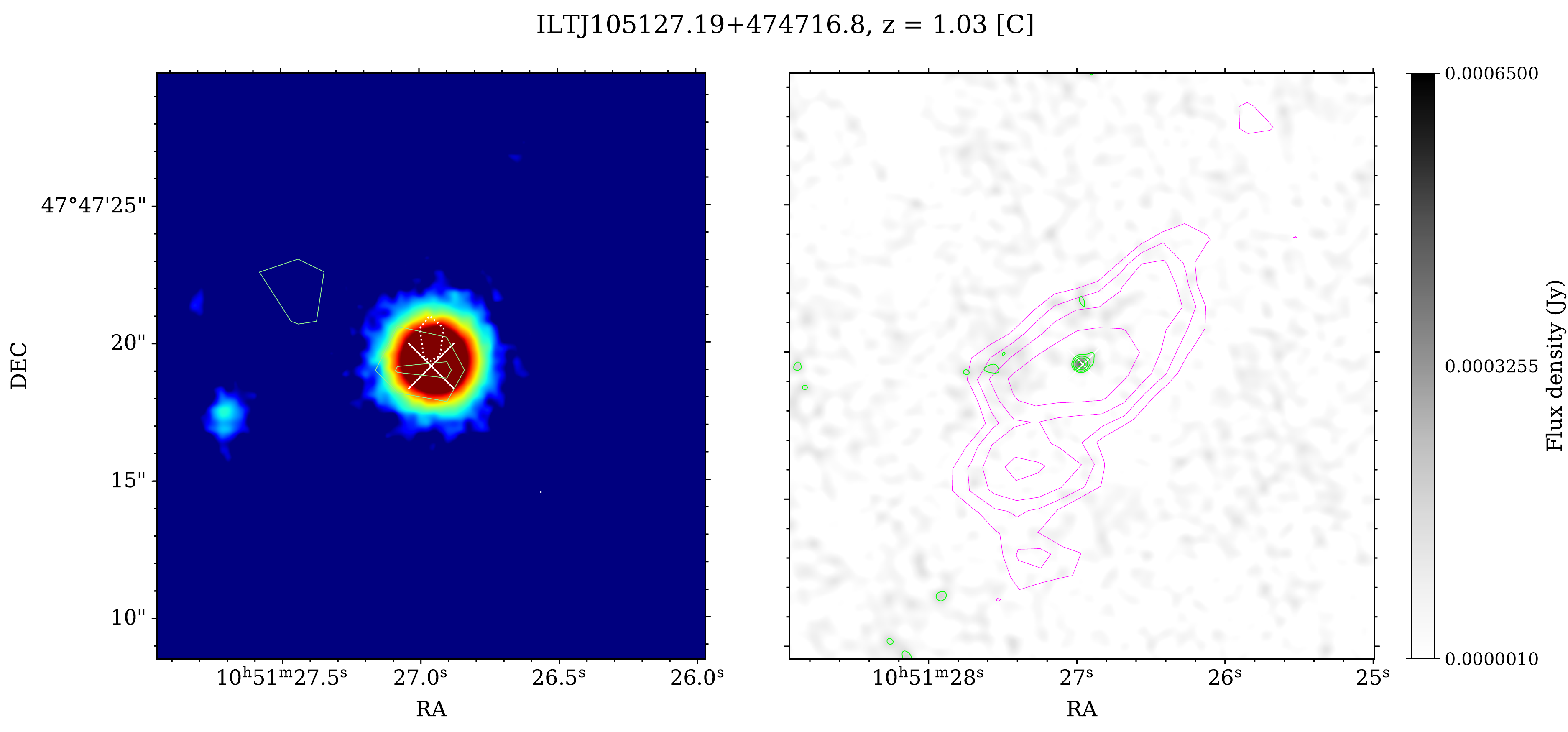}\\
\end{subfigure}
\end{adjustwidth}
\caption{\textls[-25]{Images of our target sources that were observed with the VLA. Left: FIRST contours (lightgreen), the Very Large Array Sky Survey (VLASS) contours (white dotted) on PanSTARRS images (colour). Right: VLA contours (green), LOFAR contours (magenta) on VLA images (gray). The red ellipse in the right panel of the first raw shows the VLA beam. FIRST, VLASS, LOFAR and VLA contours denote the surface brightness levels starting at 3$\sigma$and increasing at various powers of 3$\sigma$, where $\sigma$ denotes the local root-mean-square (RMS) noise in each map. Crosses show the position of the optical counterparts.}}\label{fig:app}
\end{figure}
\begin{figure}[H]\ContinuedFloat

\begin{adjustwidth}{-\extralength}{0cm}
\centering 
\begin{subfigure}
\centering
\includegraphics[scale=0.4]{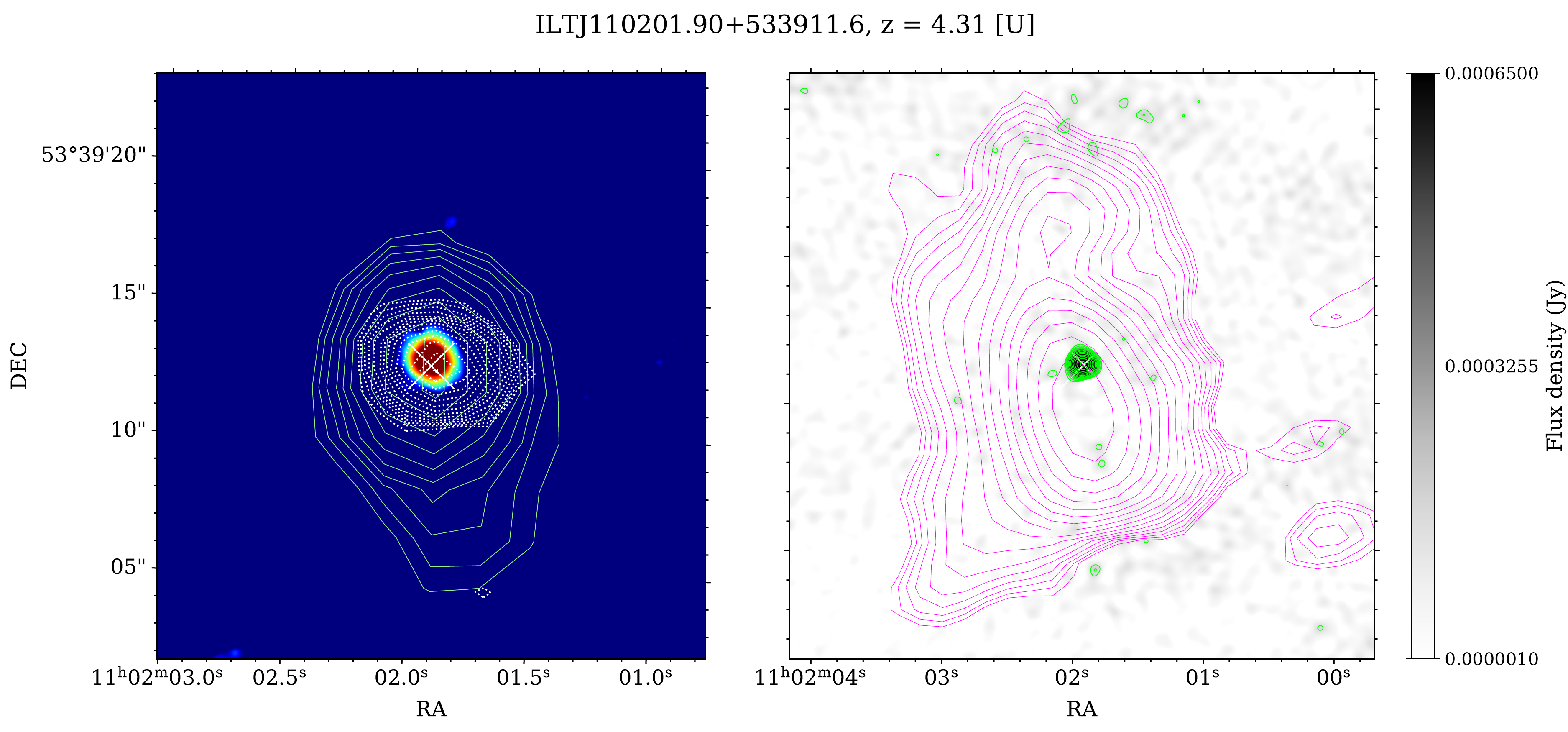}\\
\includegraphics[scale=0.4]{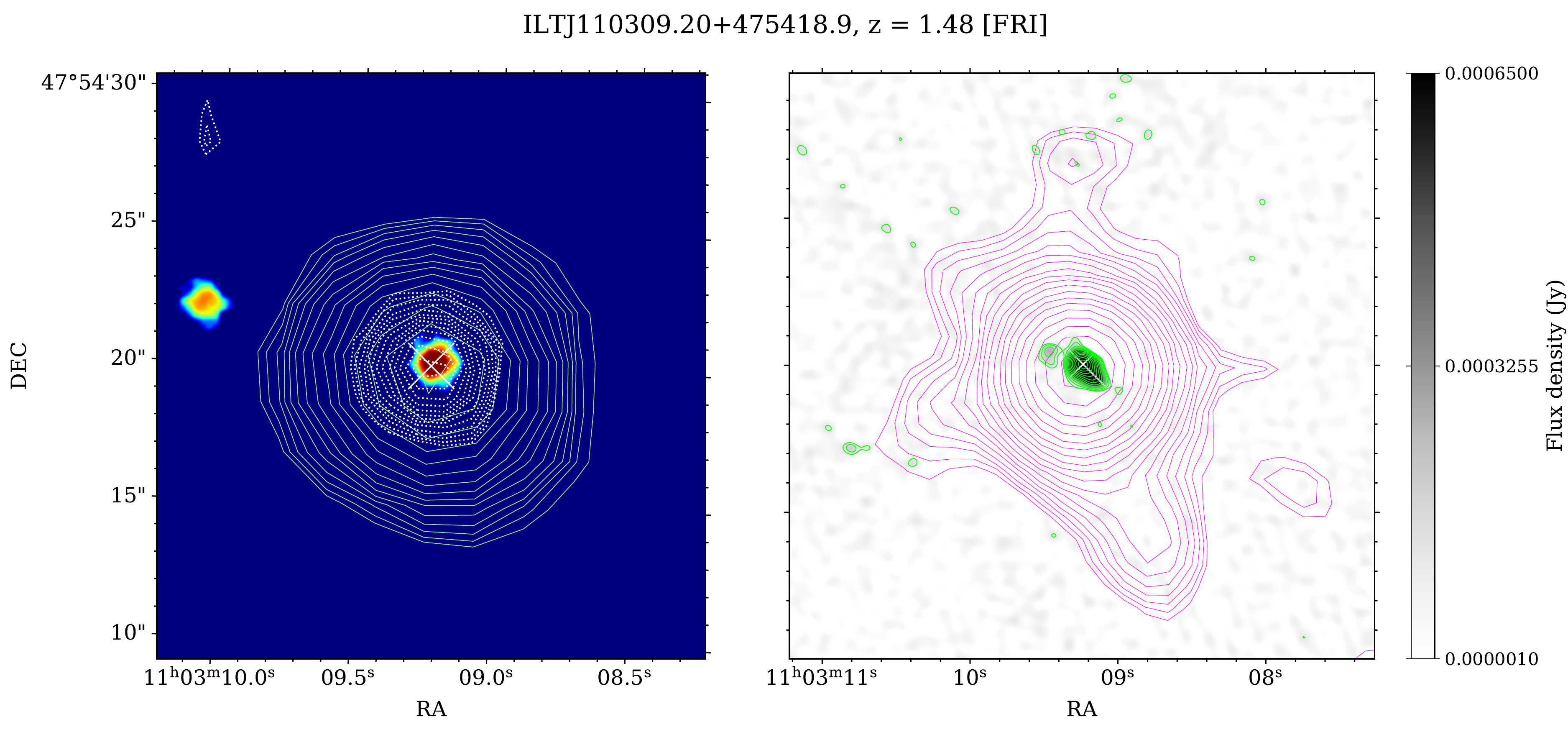}\\
\includegraphics[scale=0.4]{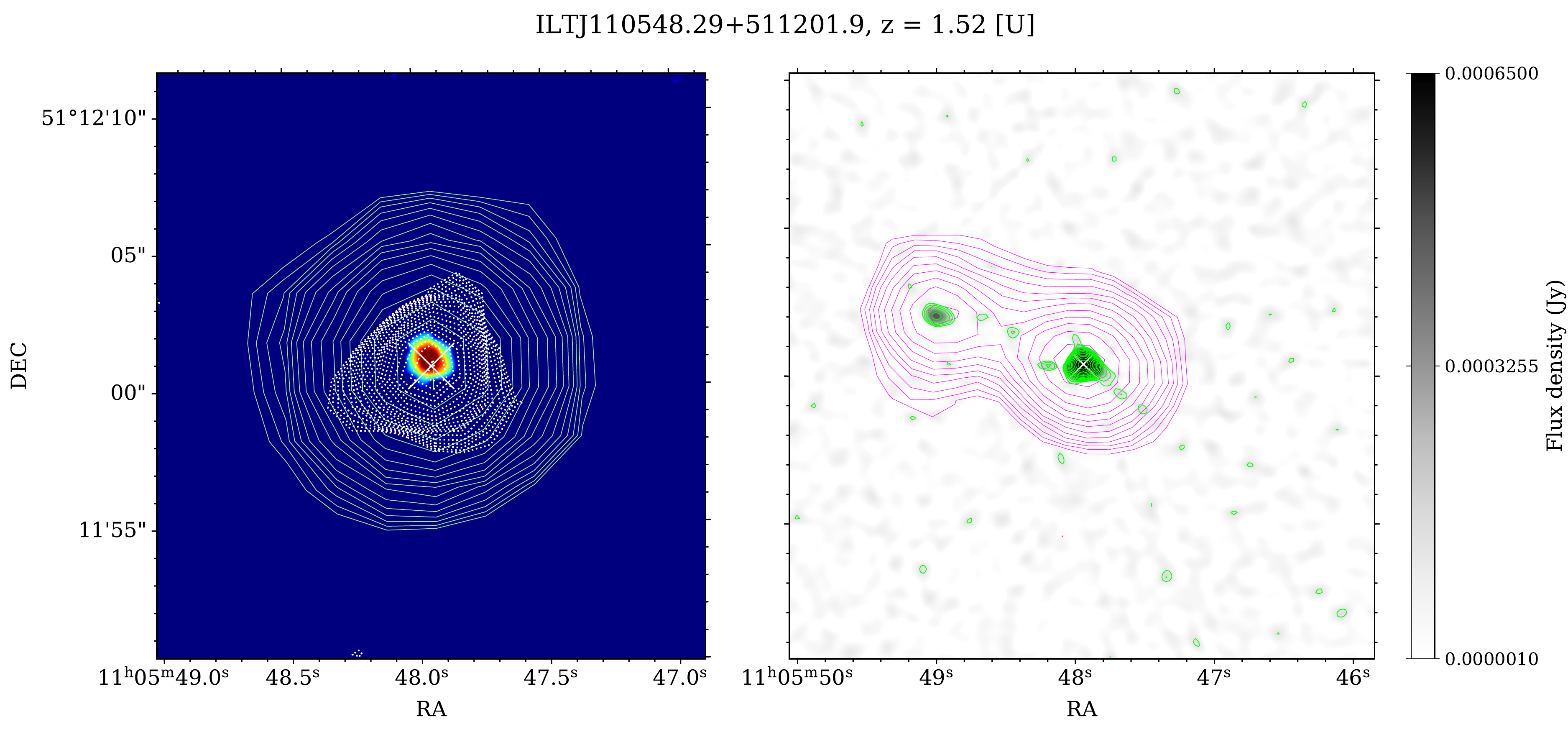}\\
\end{subfigure}
\end{adjustwidth}
\caption{\textit{Cont}.}
\end{figure}

\begin{figure}[H]\ContinuedFloat
\centering

\begin{adjustwidth}{-\extralength}{0cm}
\centering 
\begin{subfigure}
\centering
\includegraphics[scale=0.4]{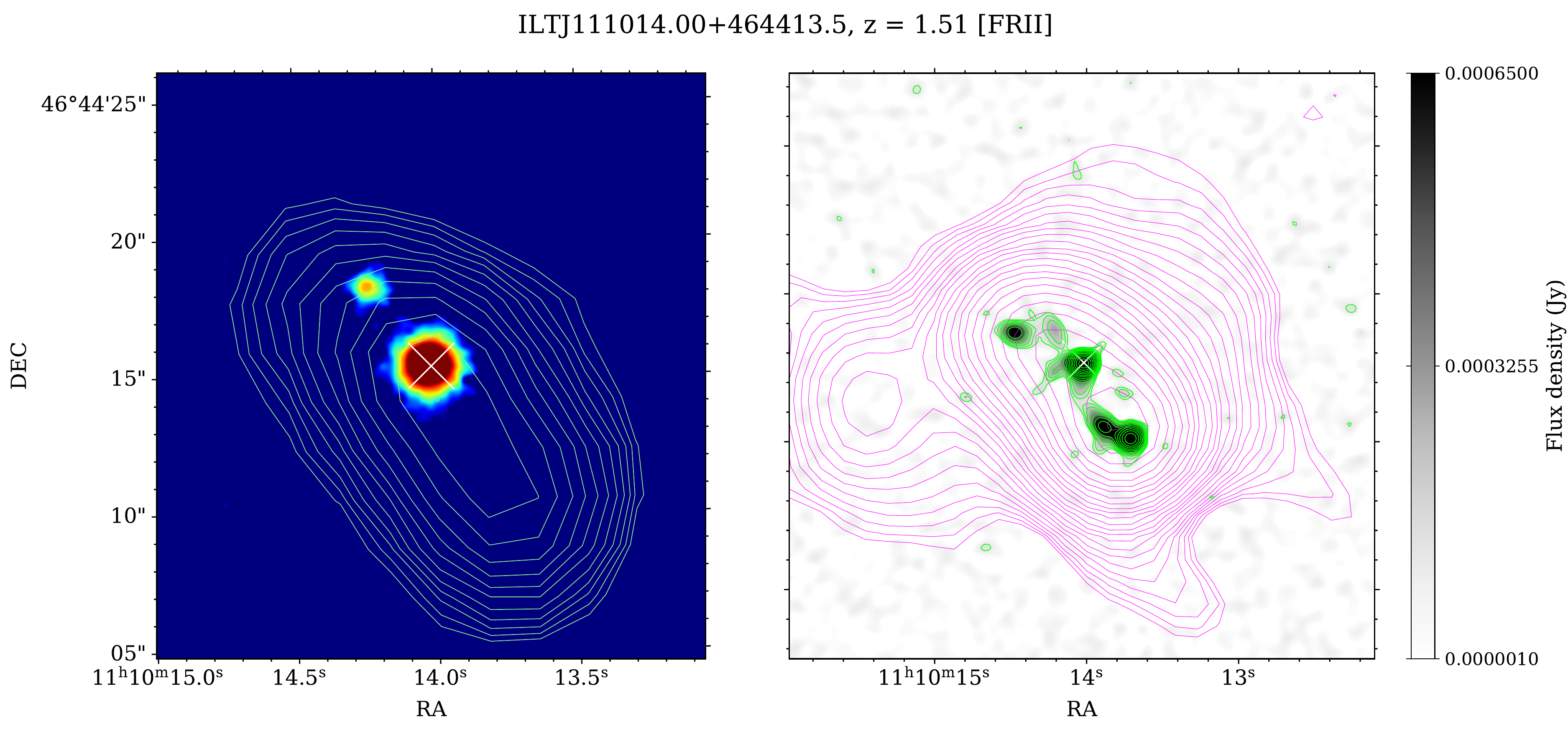}\\
\includegraphics[scale=0.4]{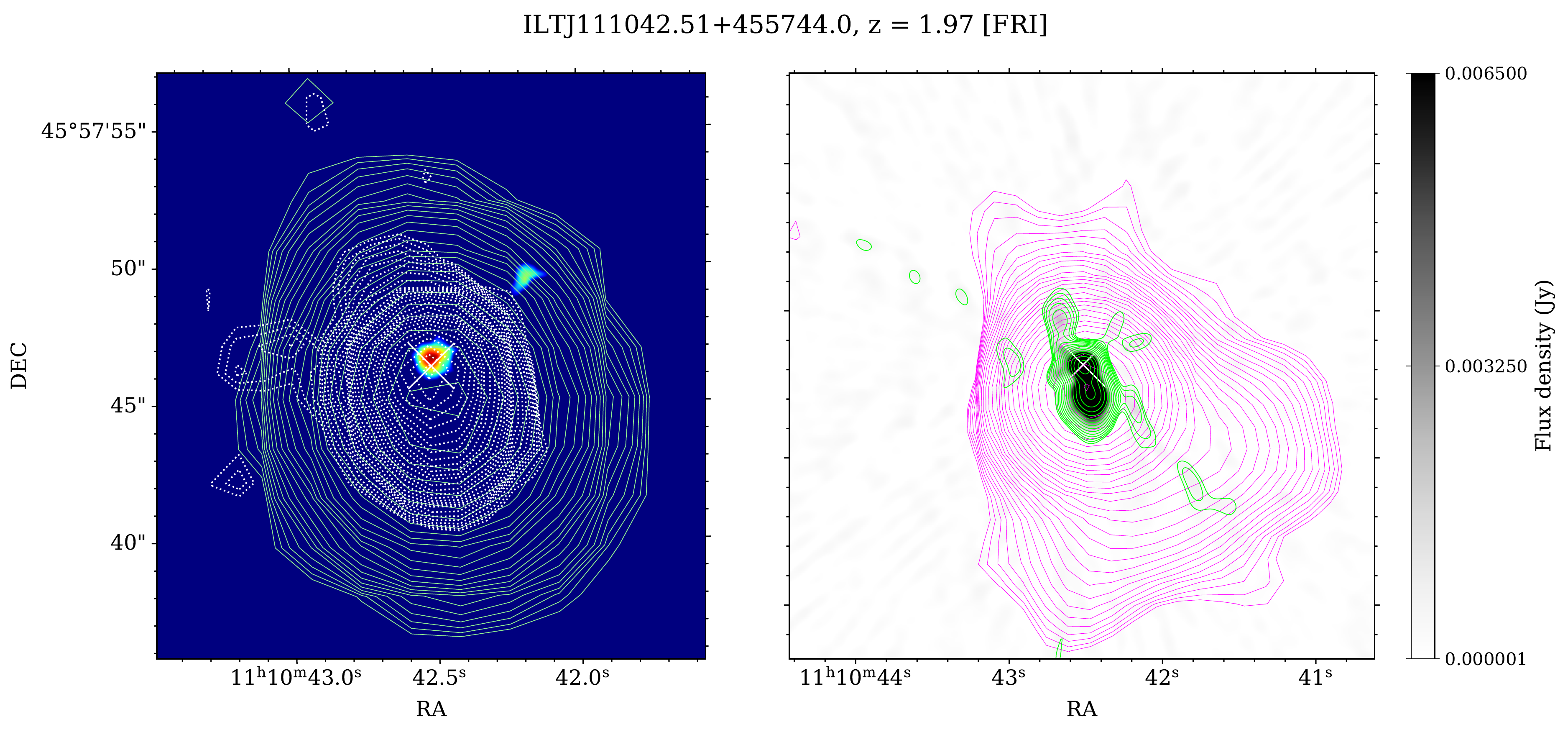}\\
\includegraphics[scale=0.4]{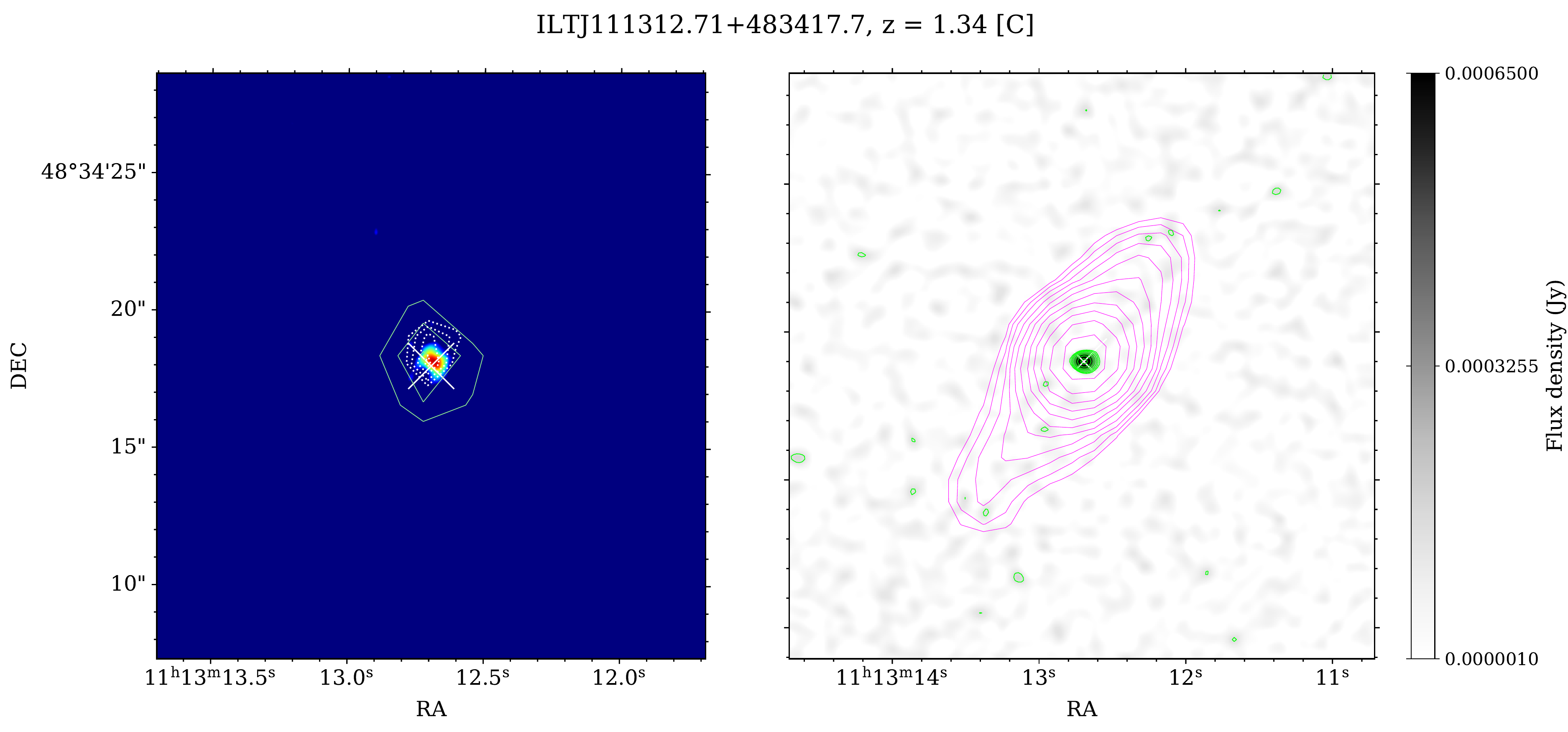}\\
\end{subfigure}
\end{adjustwidth}
\caption{\textit{Cont}.}
\end{figure}

\begin{figure}[H]\ContinuedFloat
\centering
\begin{adjustwidth}{-\extralength}{0cm}
\centering 
\begin{subfigure}

\includegraphics[scale=0.4]{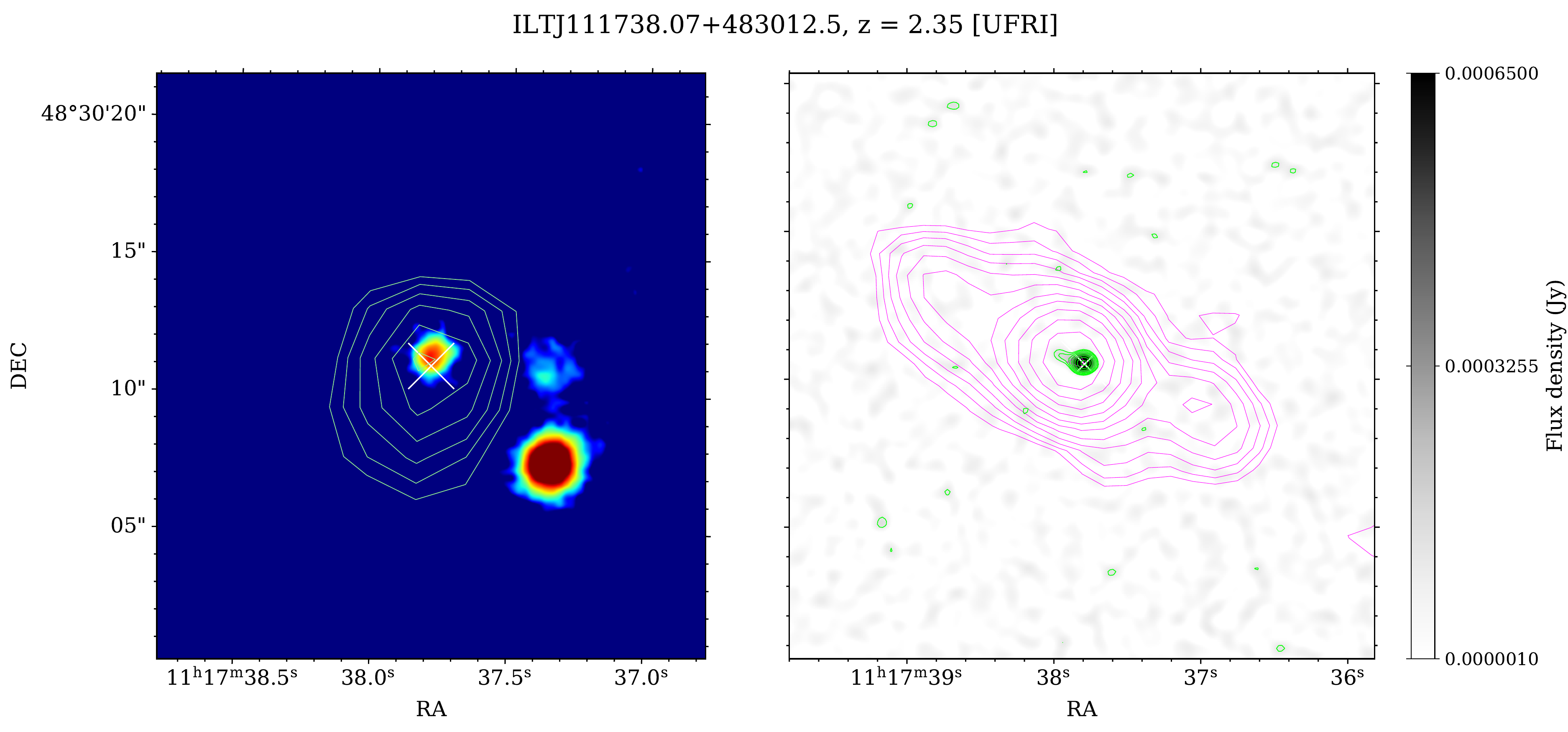}\\
\includegraphics[scale=0.4]{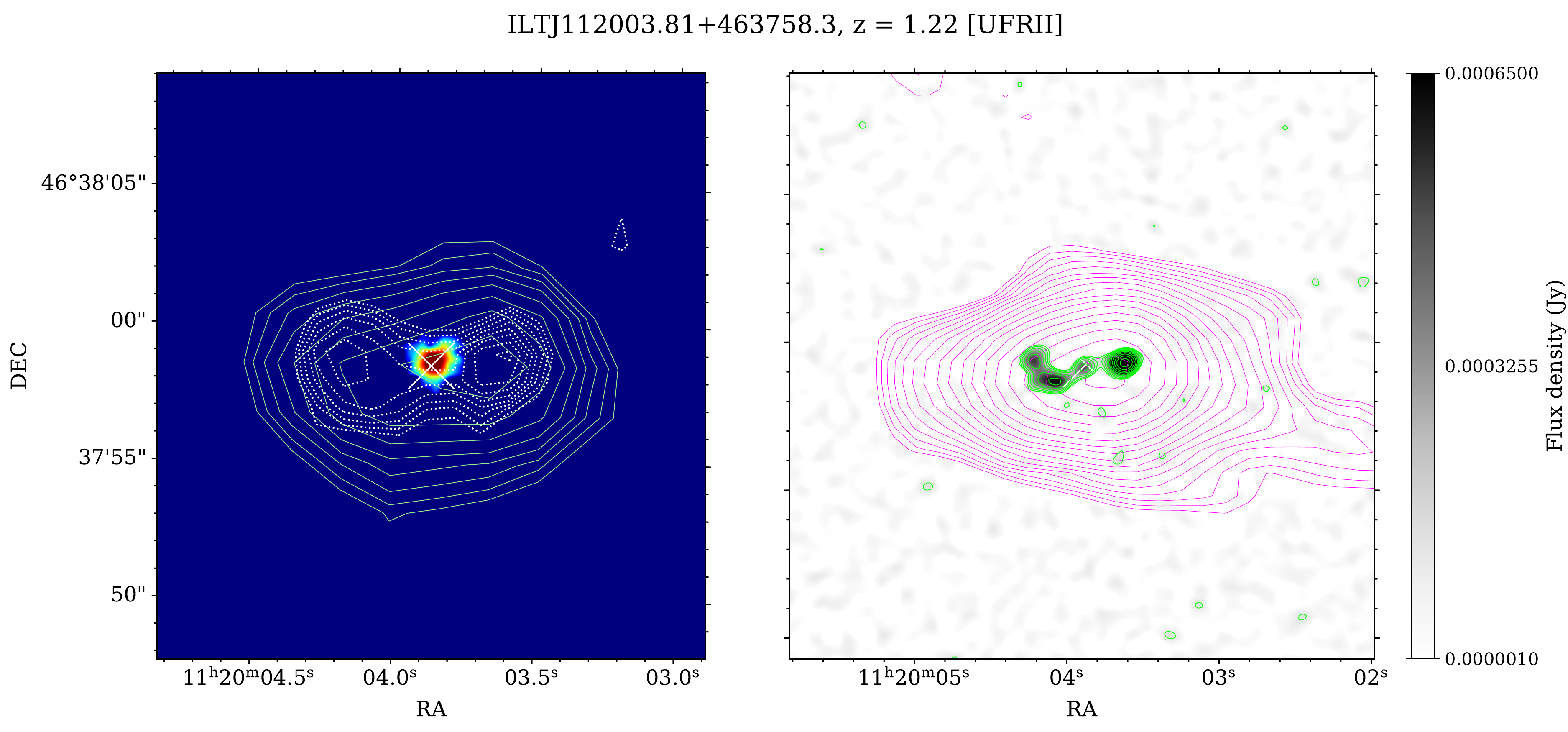}\\
\includegraphics[scale=0.4]{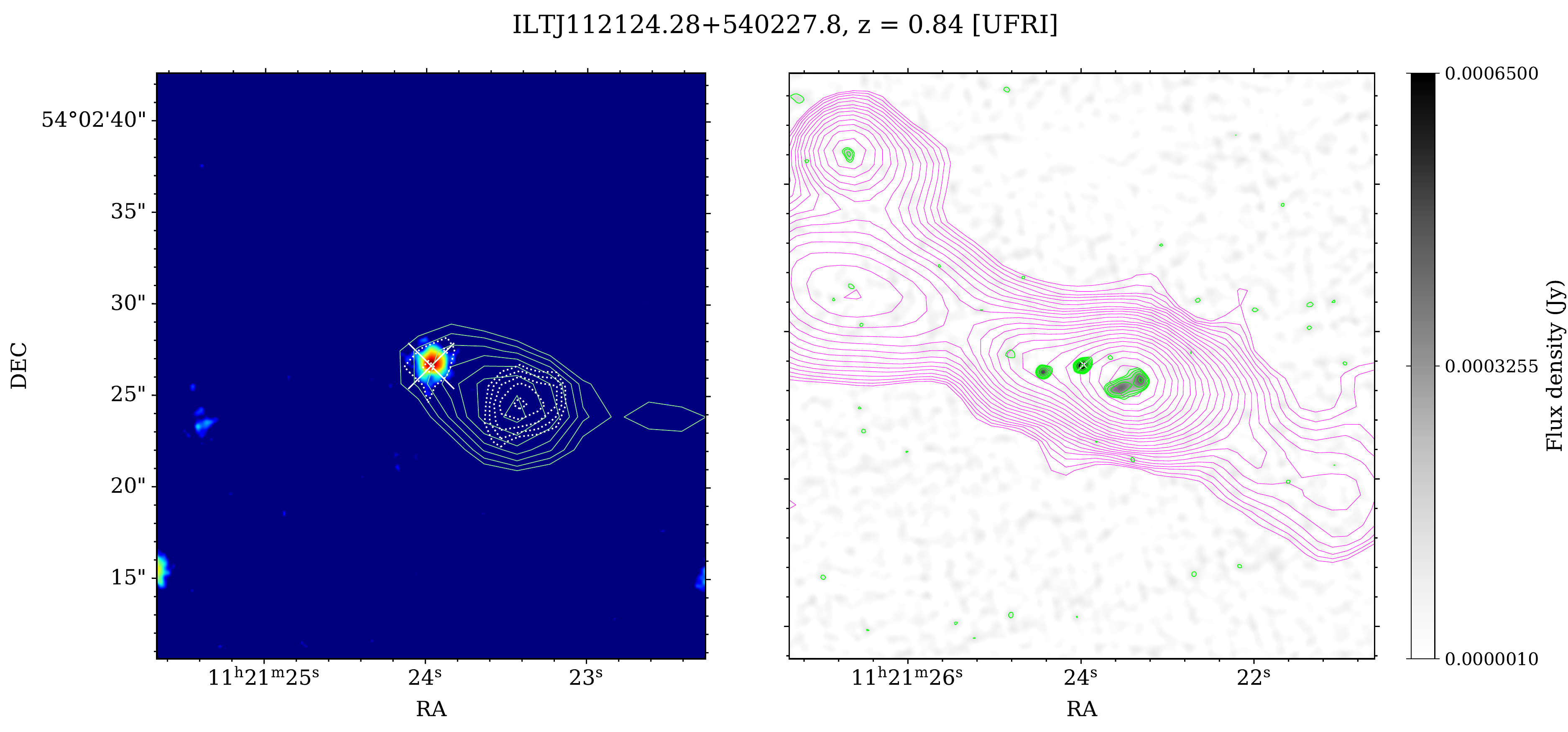}\\
\end{subfigure}
\end{adjustwidth}
\caption{\textit{Cont}.}
\end{figure}

\begin{figure}[H]\ContinuedFloat

\begin{adjustwidth}{-\extralength}{0cm}
\centering 
\begin{subfigure}
\centering
\includegraphics[scale=0.4]{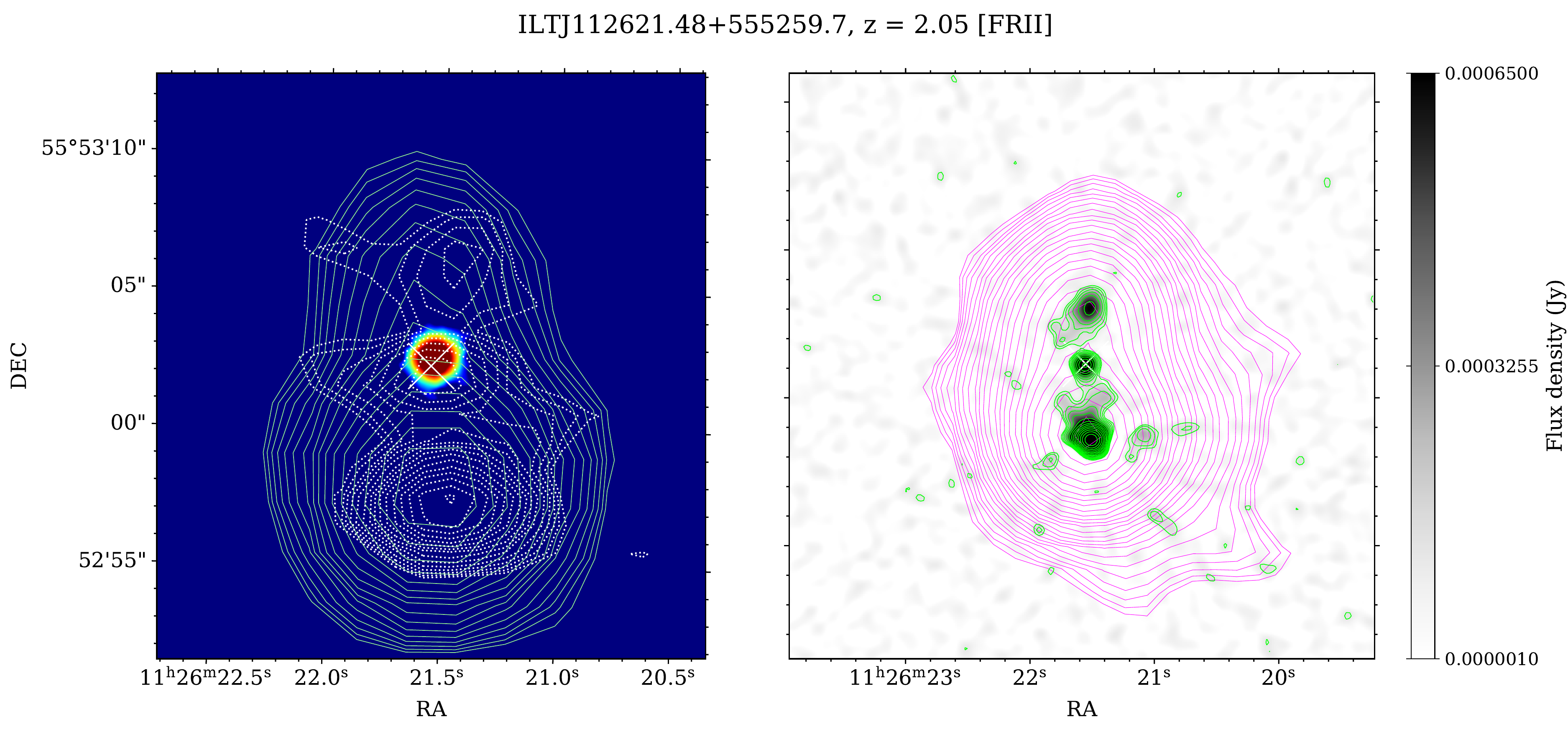}\\
\includegraphics[scale=0.4]{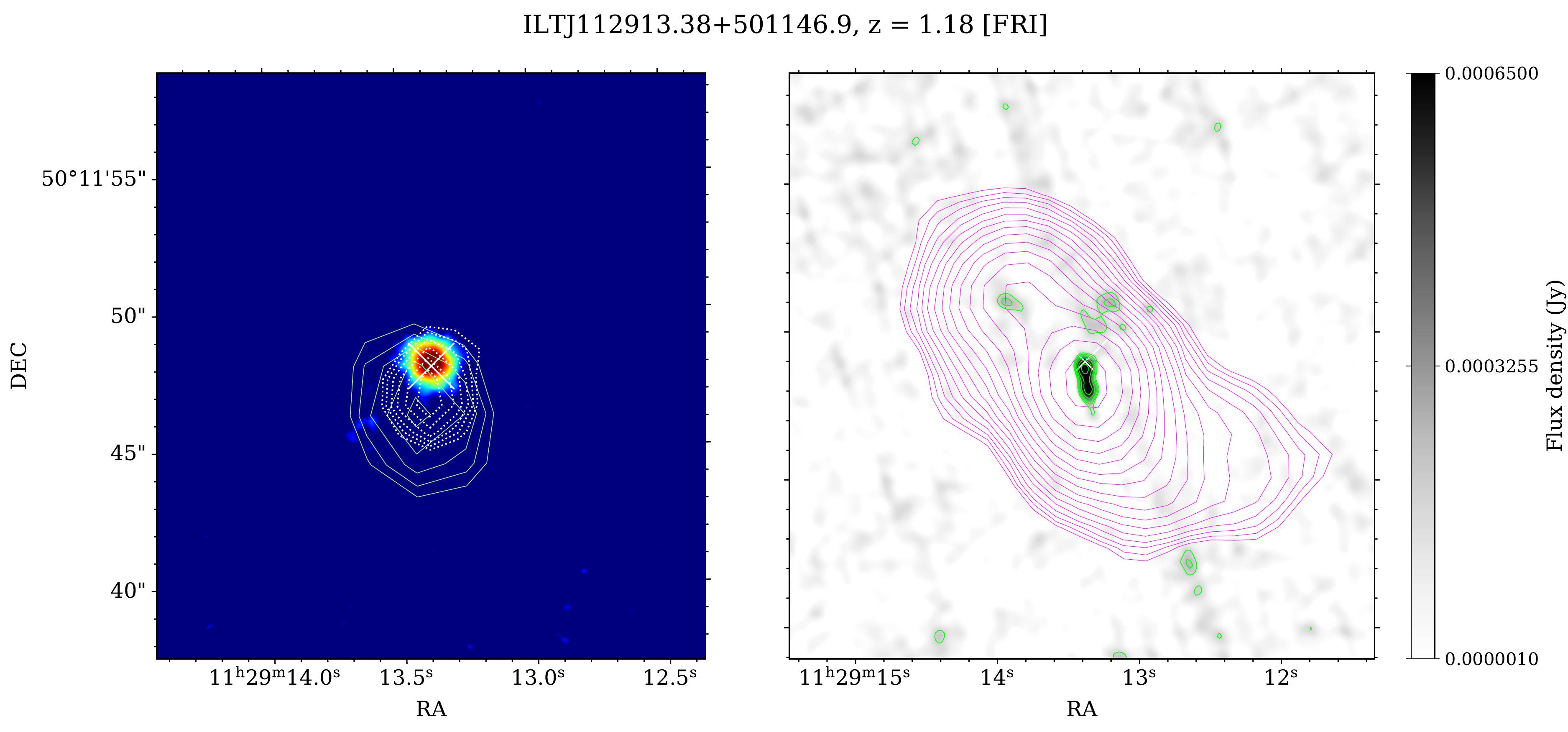}\\
\includegraphics[scale=0.4]{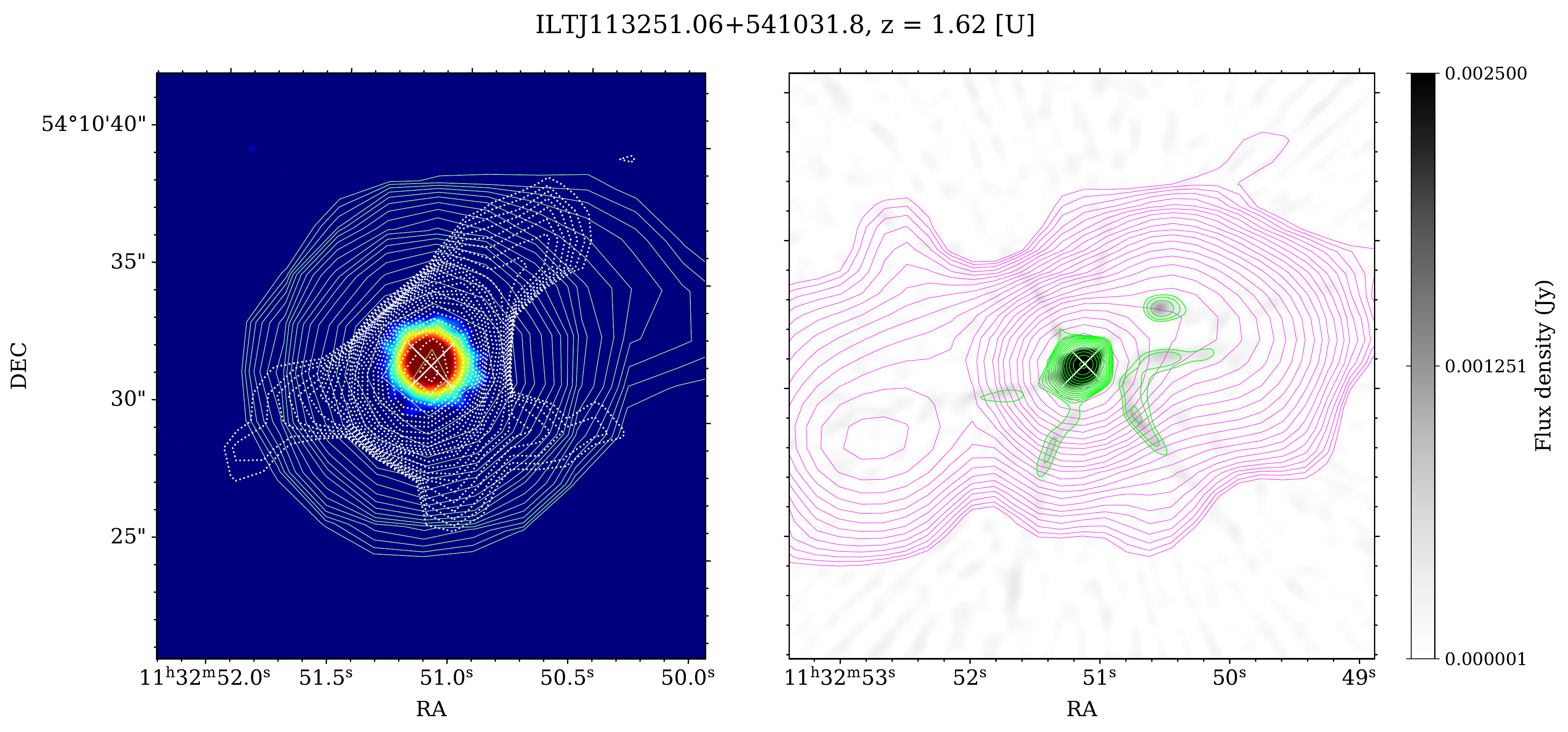}\\
\end{subfigure}
\end{adjustwidth}
\caption{\textit{Cont}.}
\end{figure}

\begin{figure}[H]\ContinuedFloat

\begin{adjustwidth}{-\extralength}{0cm}
\centering 
\begin{subfigure}
\centering
\includegraphics[scale=0.4]{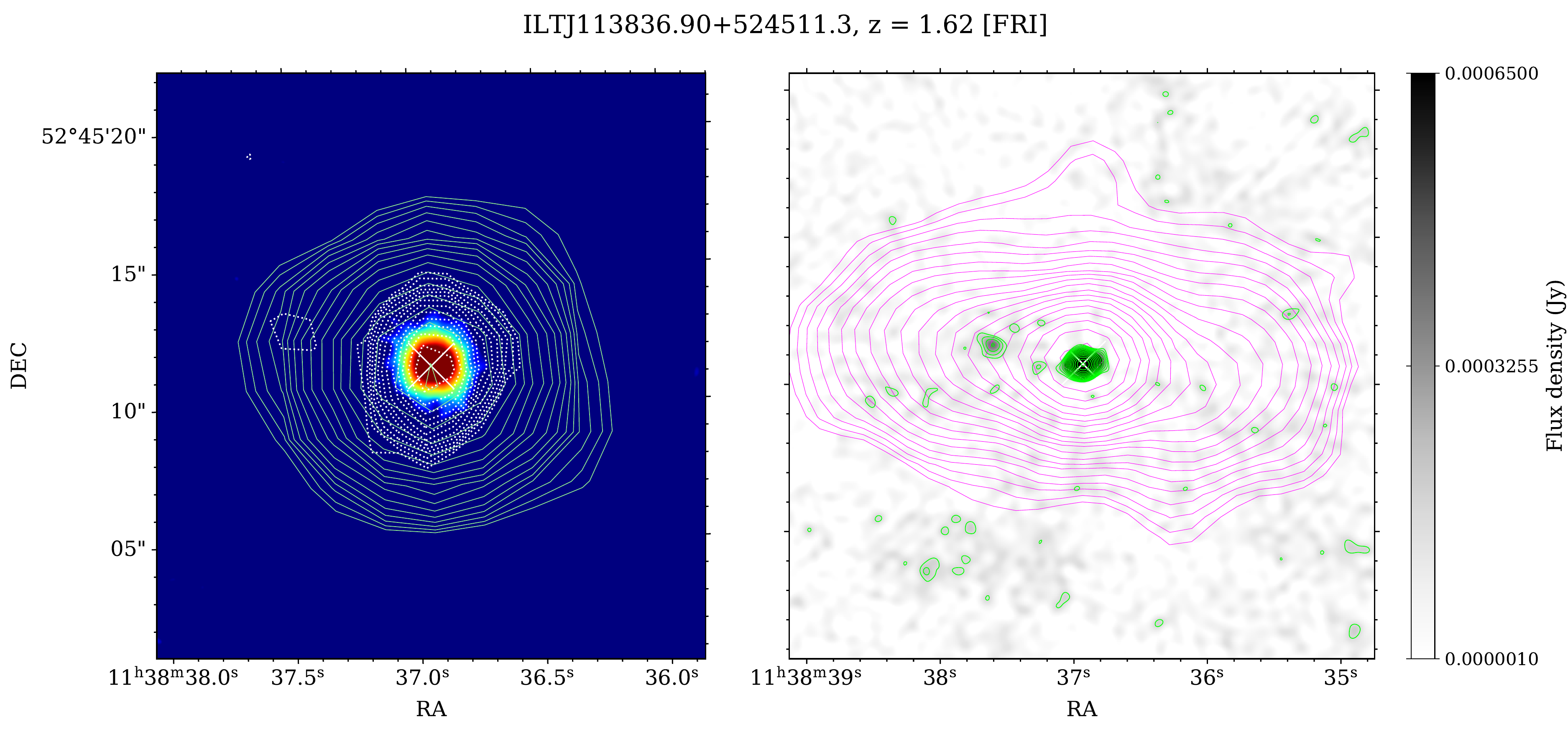}\\
\includegraphics[scale=0.4]{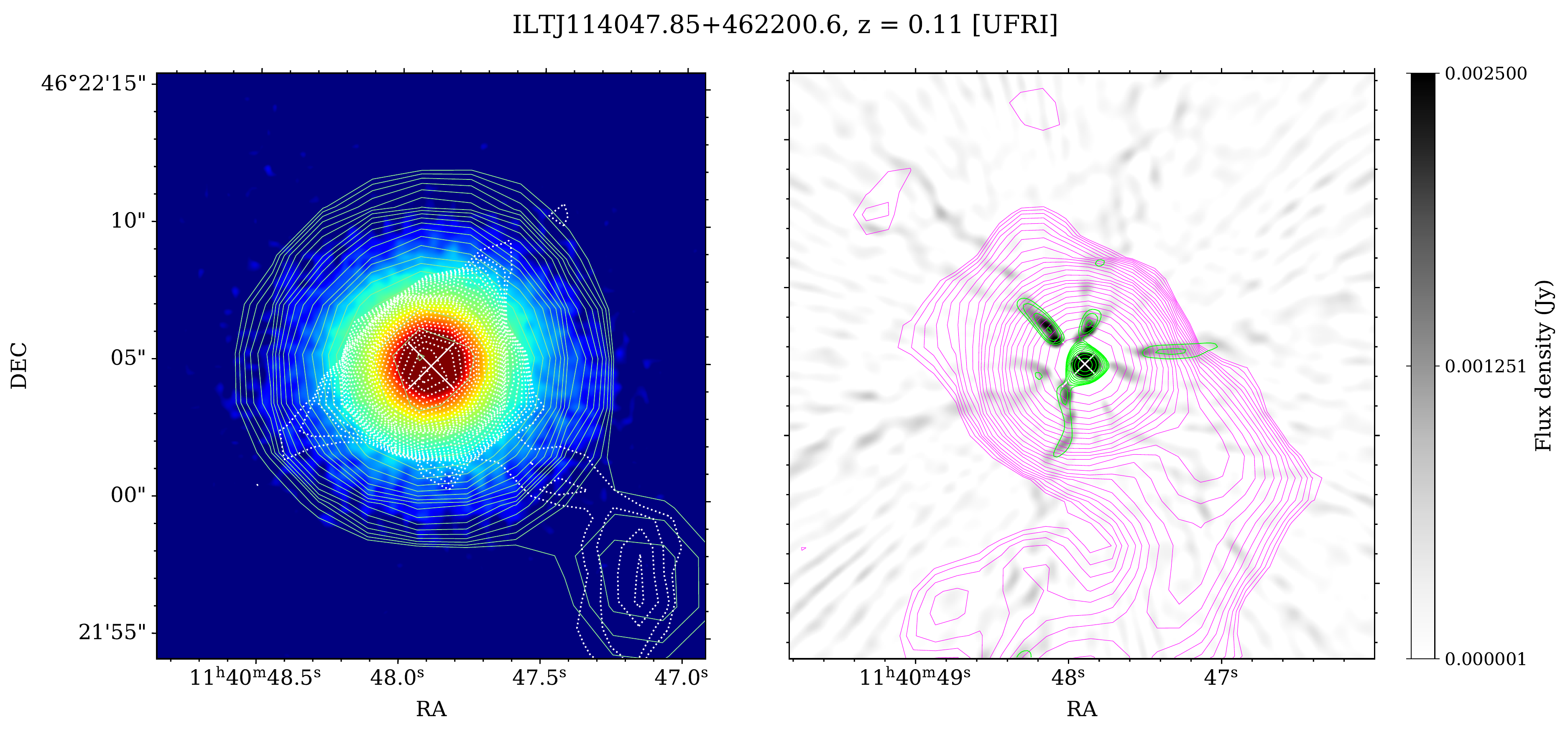}\\
\includegraphics[scale=0.4]{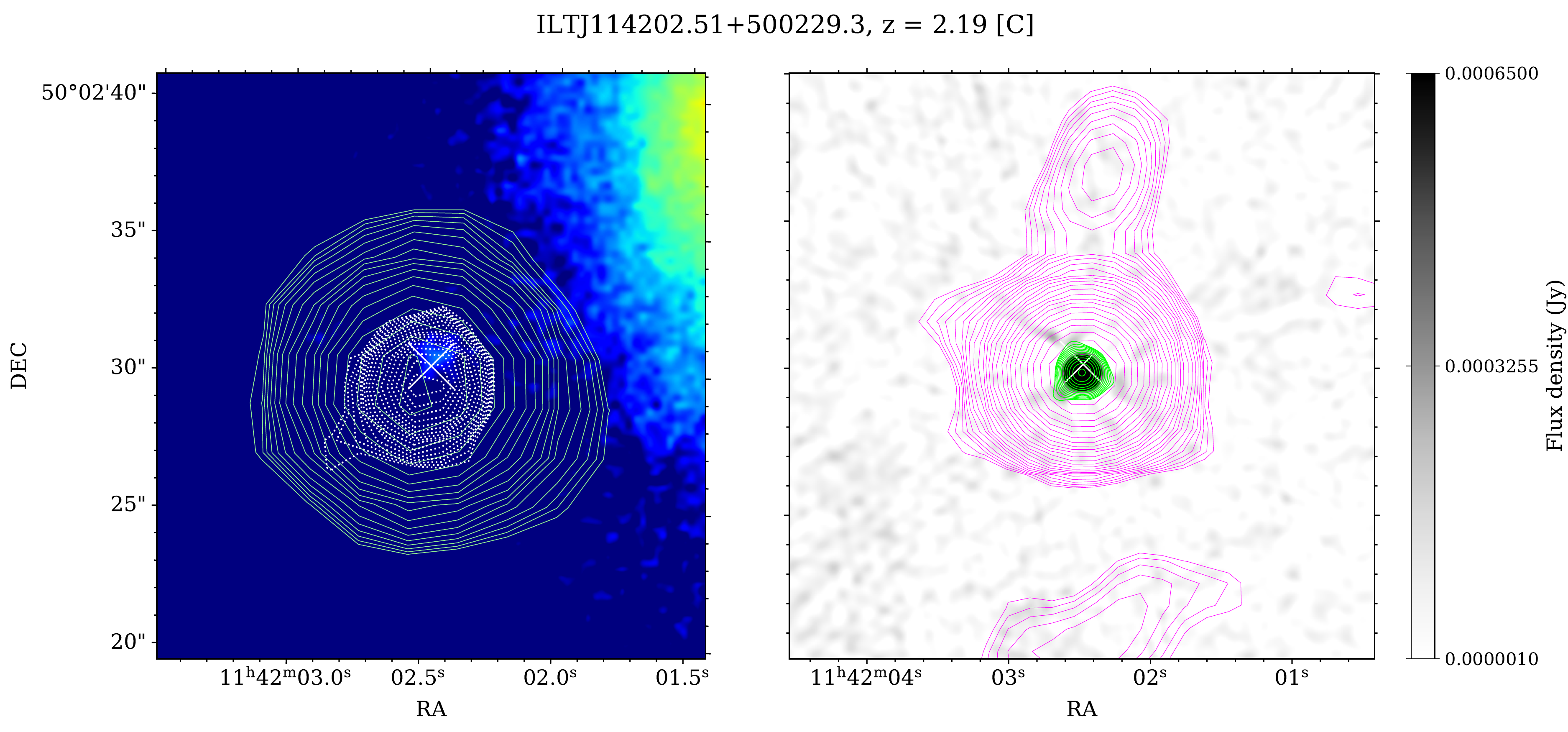}\\
\end{subfigure}
\end{adjustwidth}
\caption{\textit{Cont}.}

\end{figure}

\begin{figure}[H]\ContinuedFloat
\centering

\begin{adjustwidth}{-\extralength}{0cm}
\centering 
\begin{subfigure}
\centering
\includegraphics[scale=0.4]{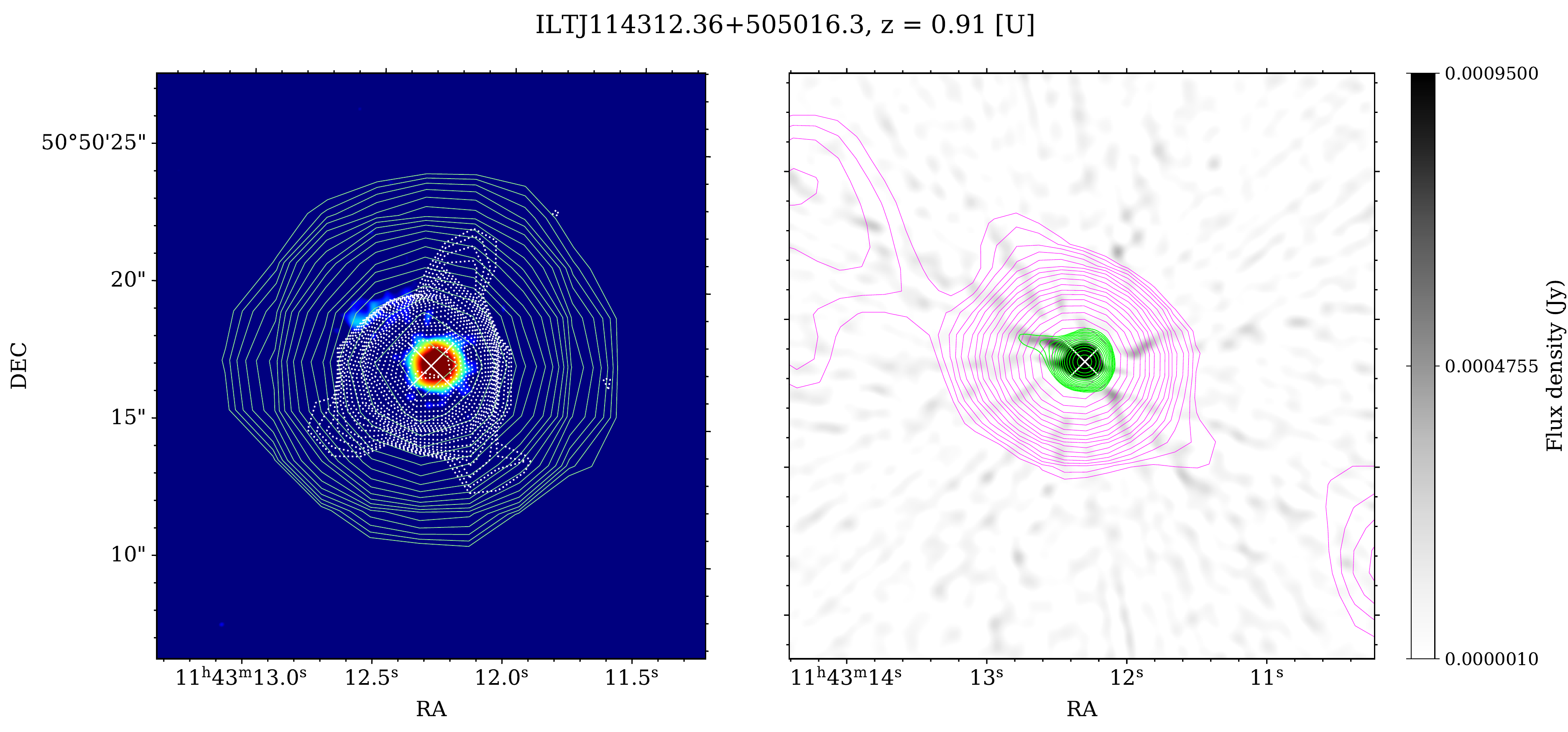}\\
\includegraphics[scale=0.4]{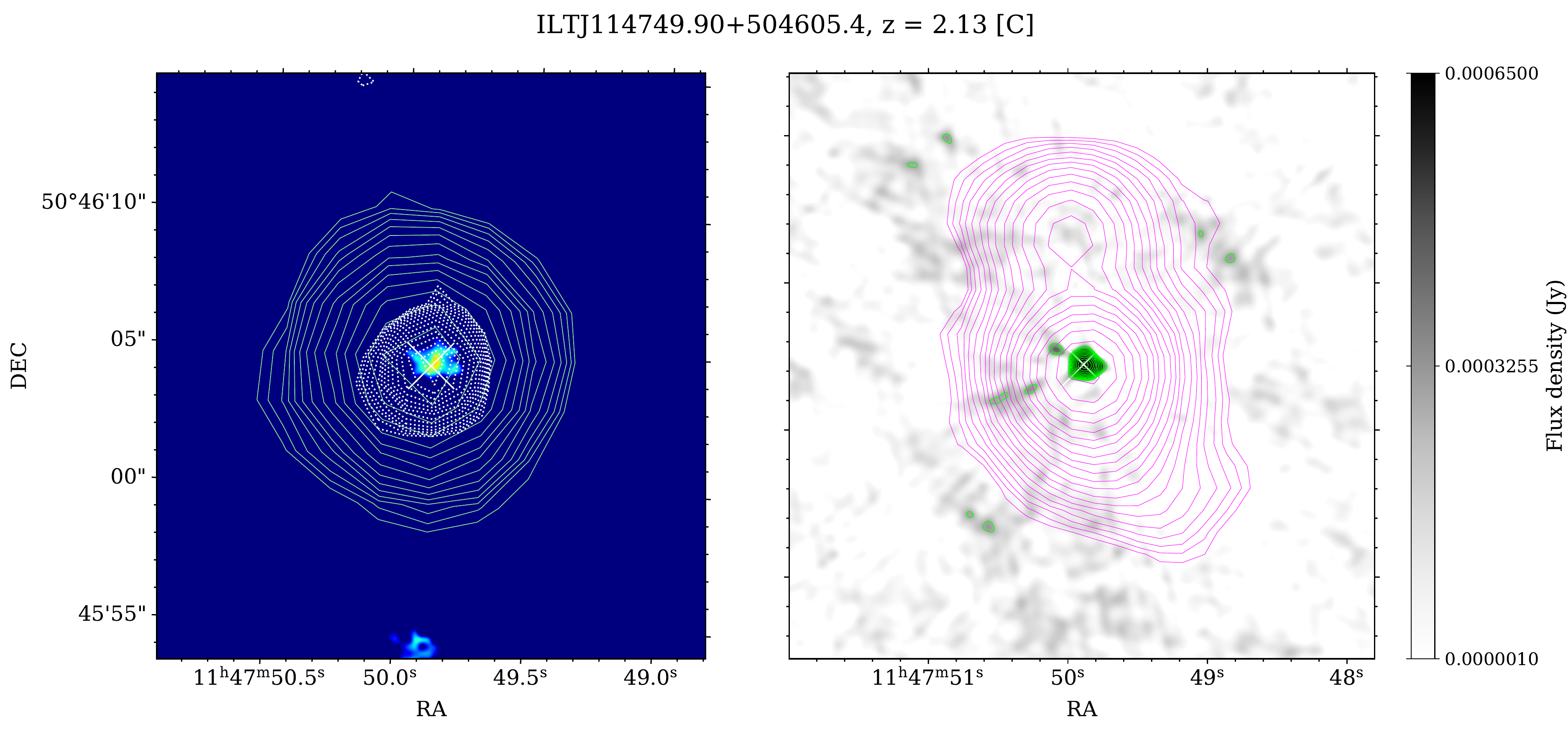}\\
\includegraphics[scale=0.4]{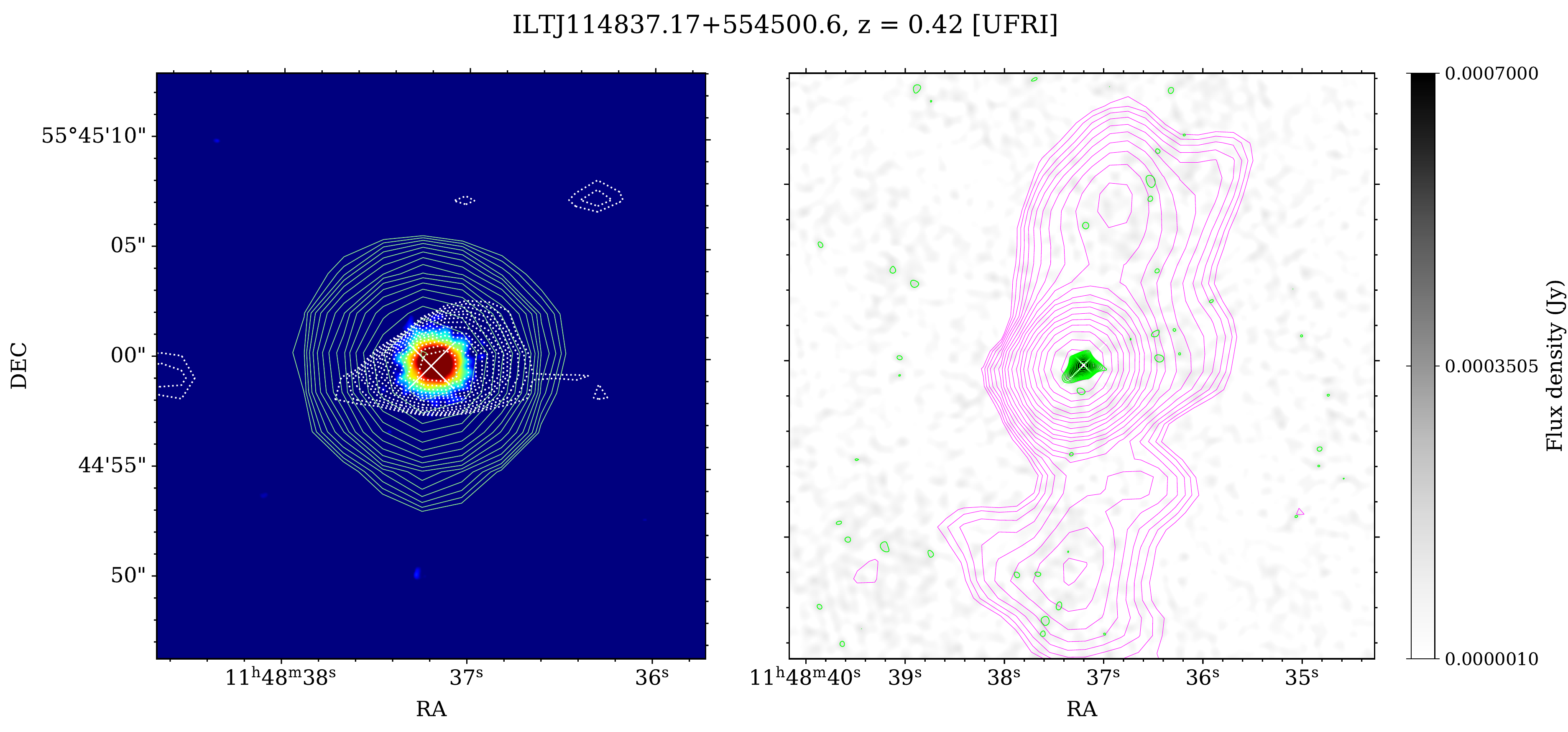}\\
\end{subfigure}
\end{adjustwidth}
\caption{\textit{Cont}.}
\end{figure}

\begin{figure}[H]\ContinuedFloat
\centering

\begin{adjustwidth}{-\extralength}{0cm}
\centering 
\begin{subfigure}

\includegraphics[scale=0.4]{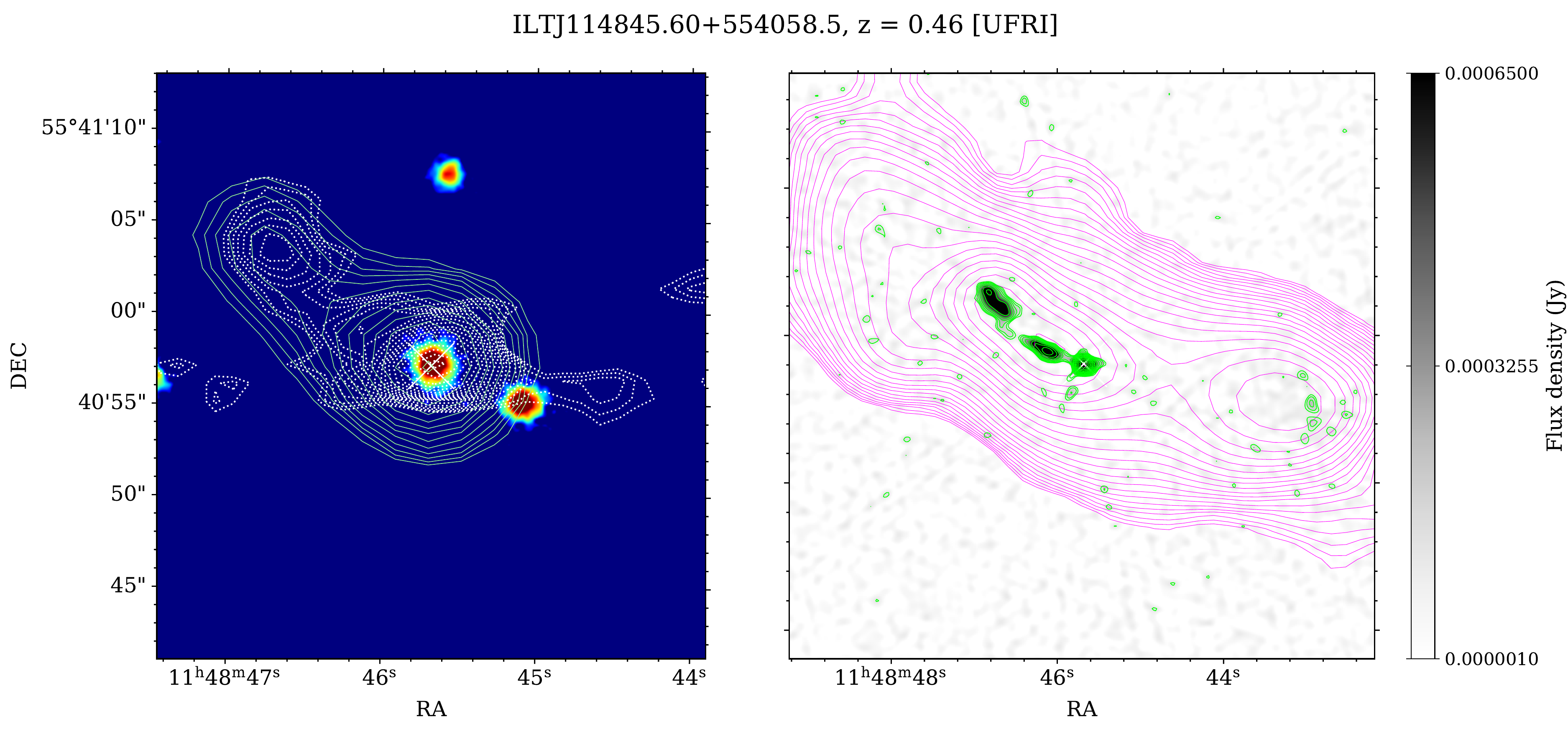}\\
\includegraphics[scale=0.4]{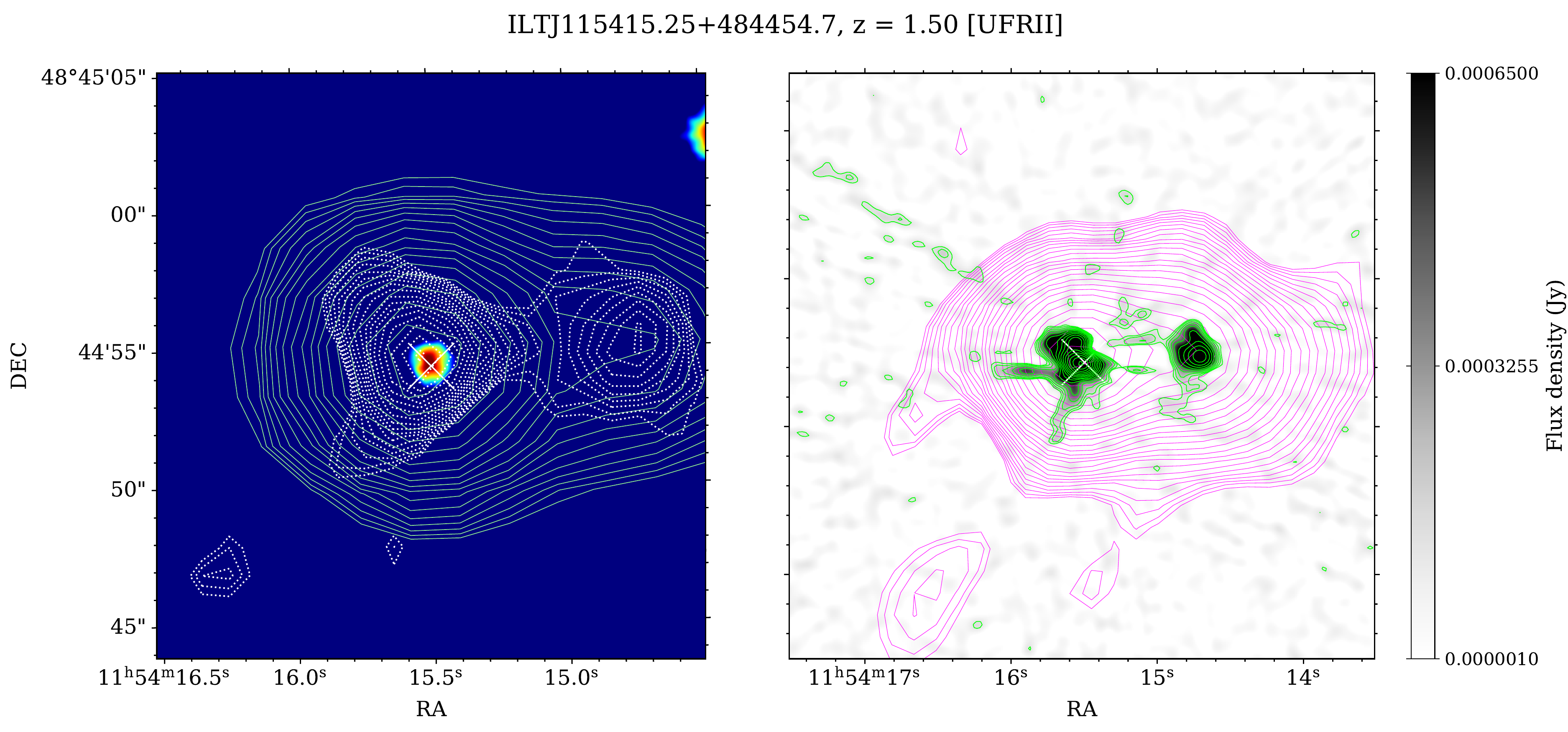}\\
\includegraphics[scale=0.4]{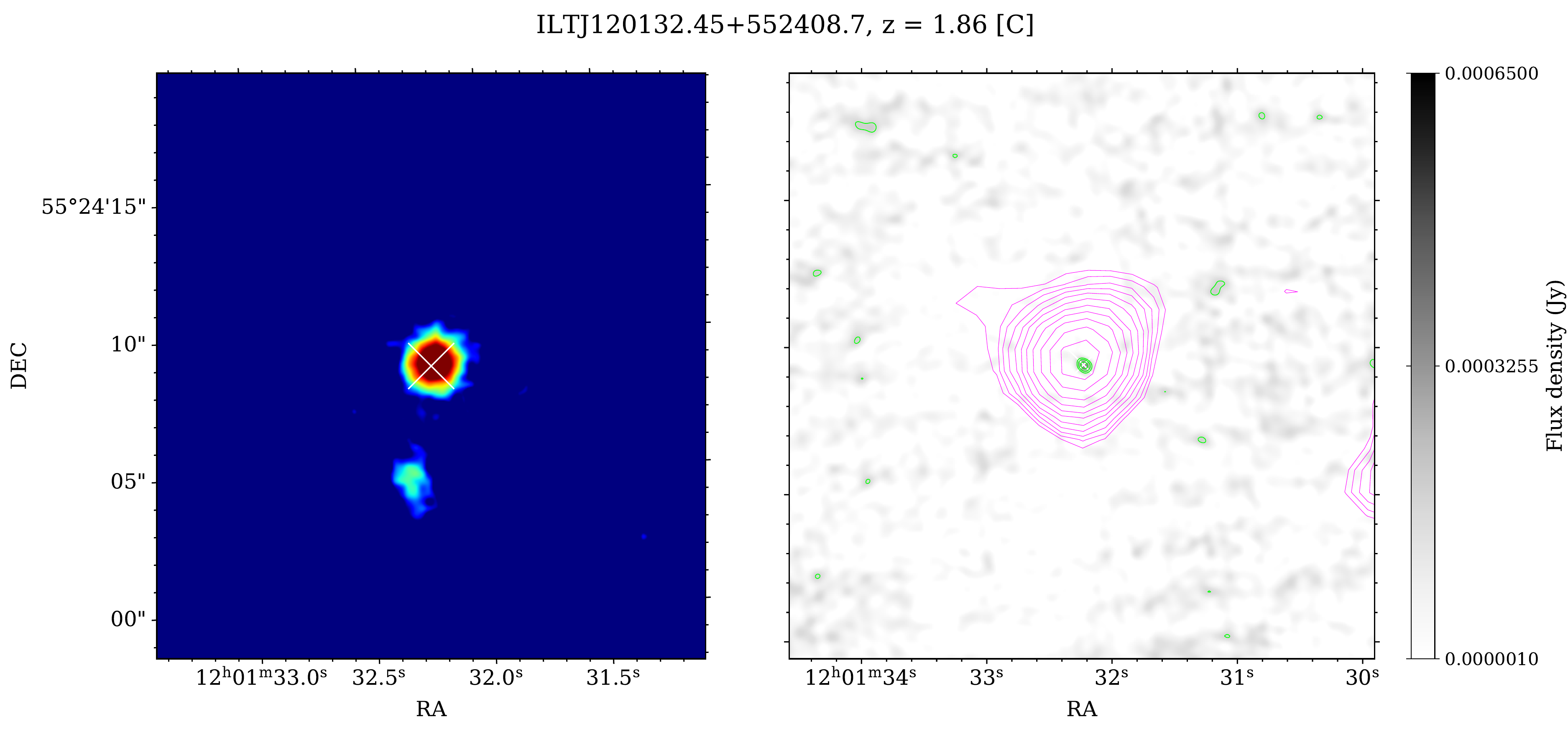}\\
\end{subfigure}
\end{adjustwidth}
\caption{\textit{Cont}.}

\end{figure}

\begin{figure}[H]\ContinuedFloat

\begin{adjustwidth}{-\extralength}{0cm}
\centering 
\begin{subfigure}
\centering
\includegraphics[scale=0.4]{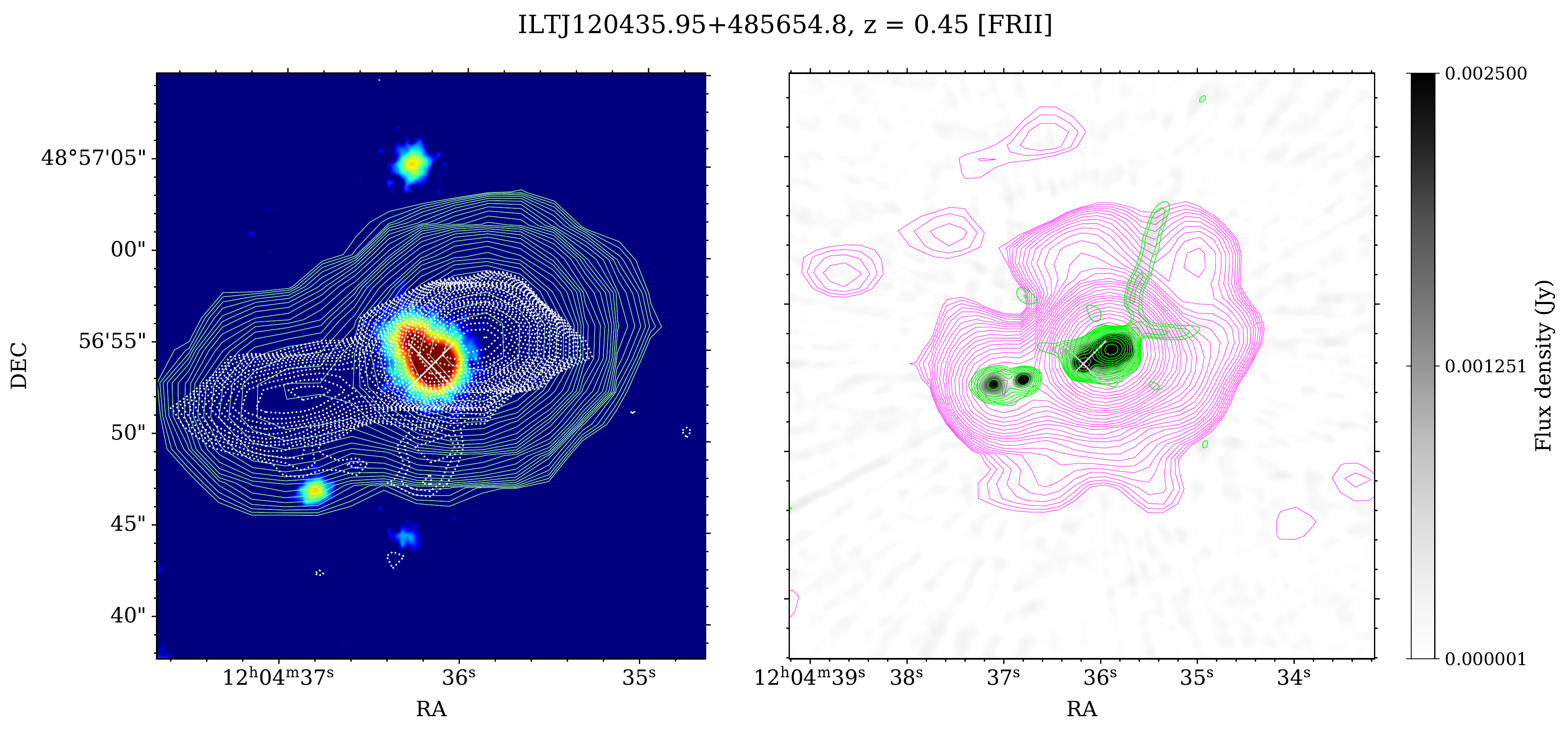}\\
\includegraphics[scale=0.4]{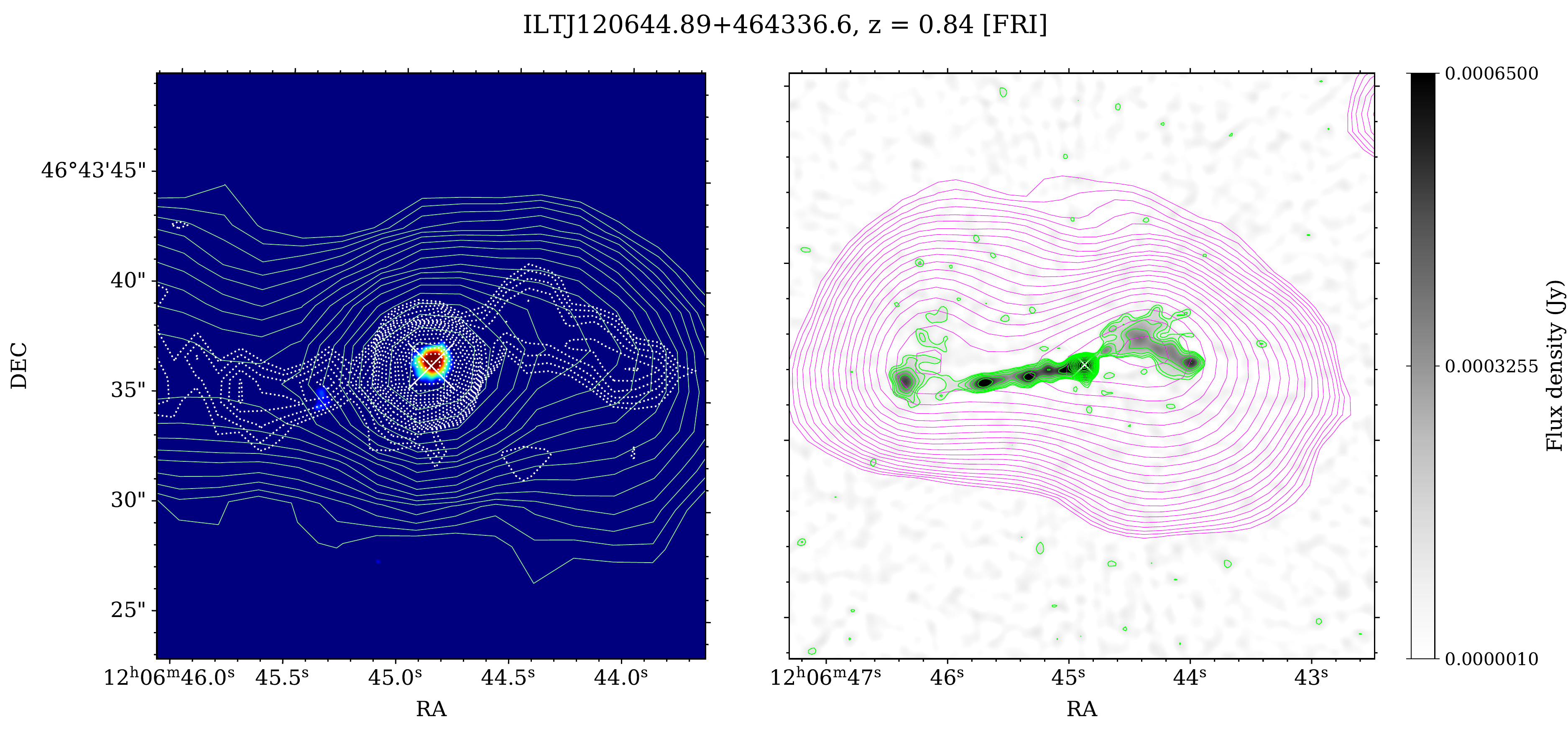}\\
\includegraphics[scale=0.4]{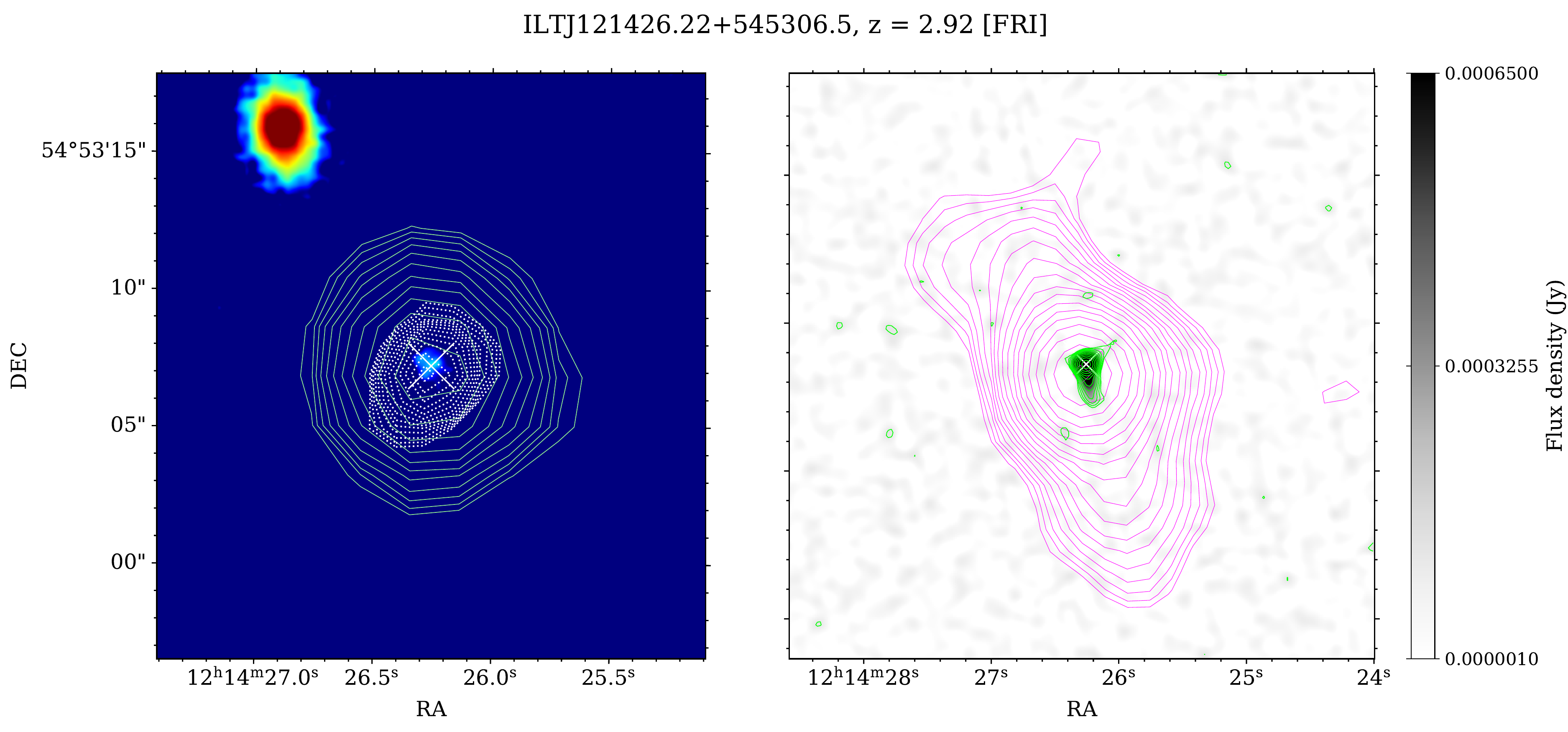}\\
\end{subfigure}
\end{adjustwidth}
\caption{\textit{Cont}.}
\end{figure}
\begin{figure}[H]\ContinuedFloat

\begin{adjustwidth}{-\extralength}{0cm}
\centering 
\begin{subfigure}

\includegraphics[scale=0.4]{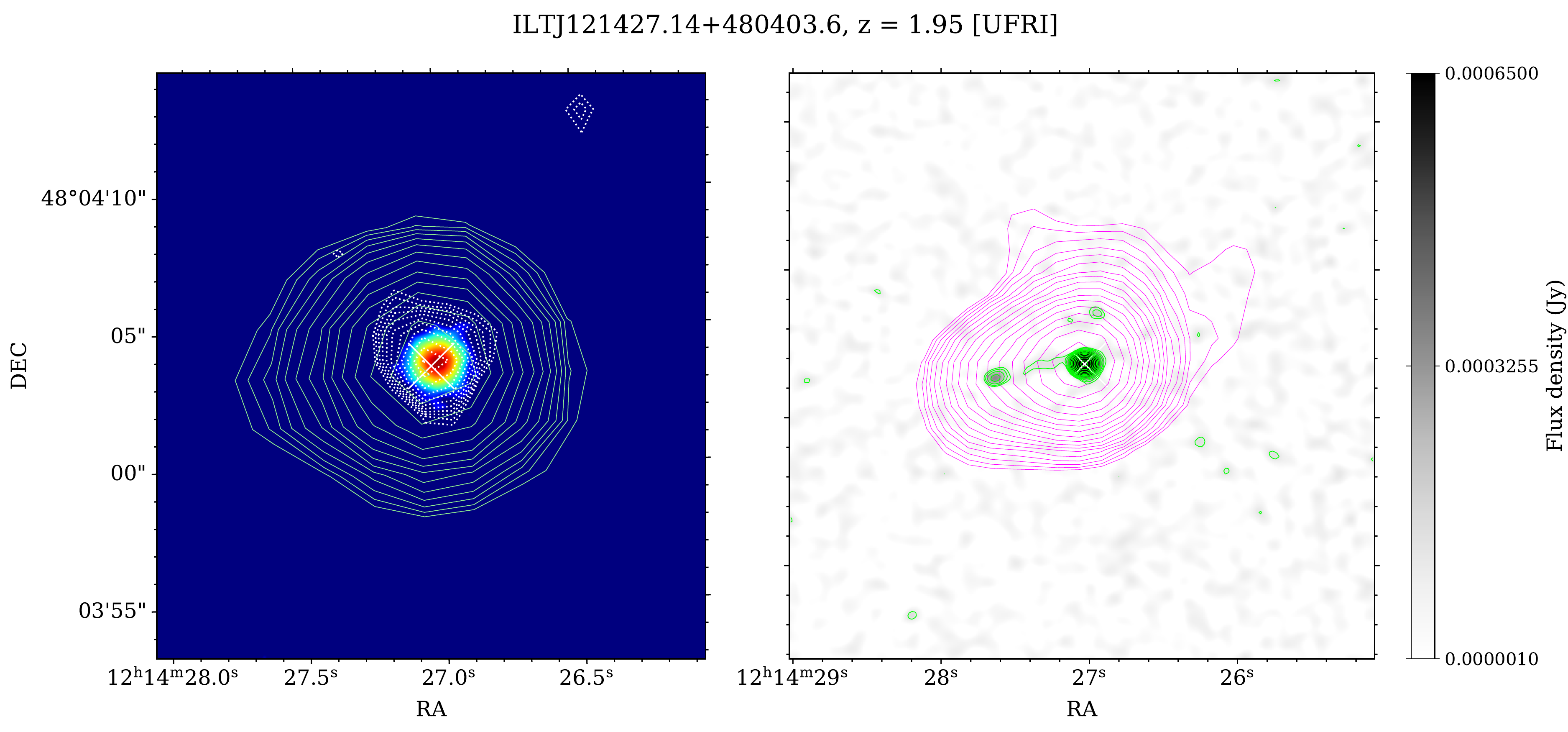}\\
\includegraphics[scale=0.4]{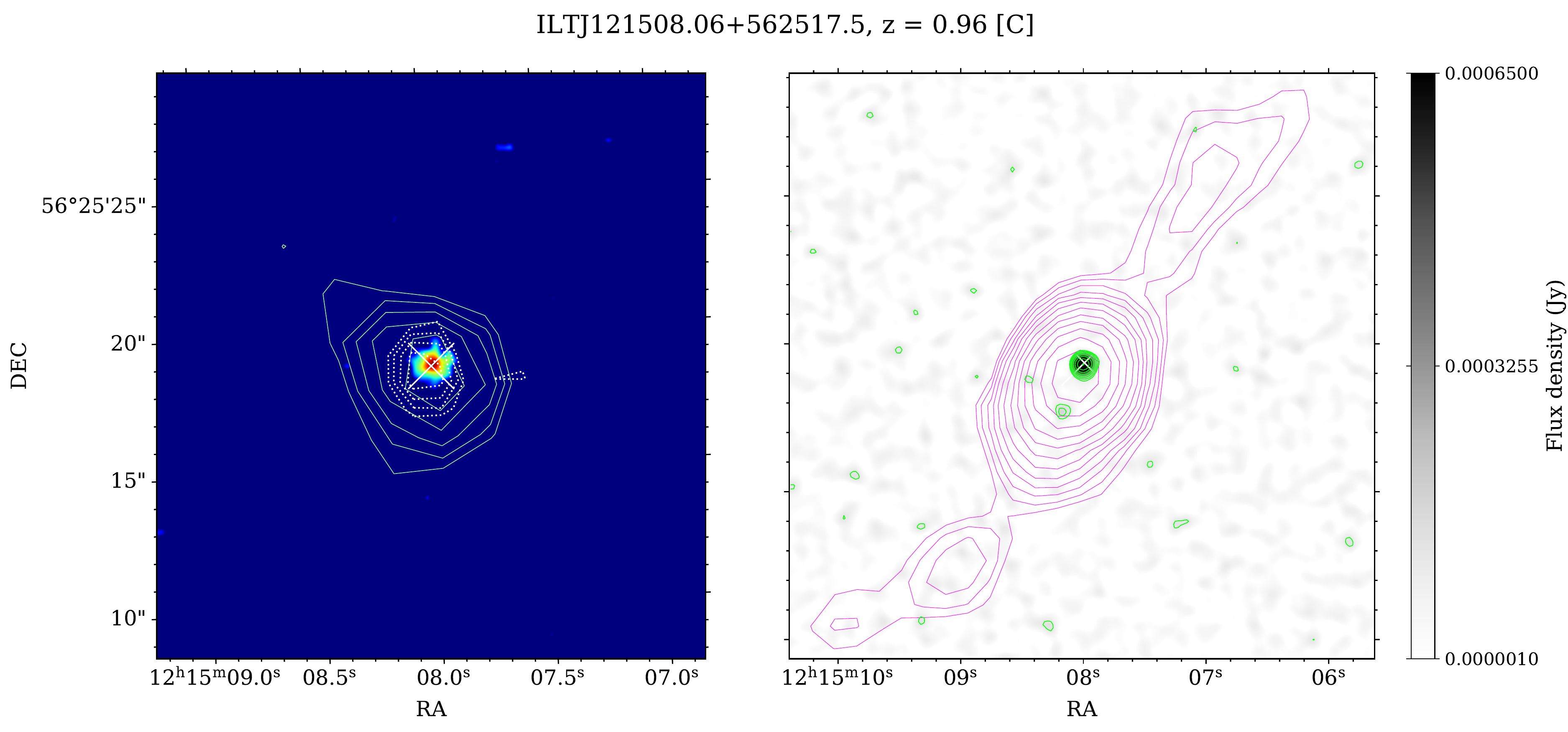}\\
\includegraphics[scale=0.4]{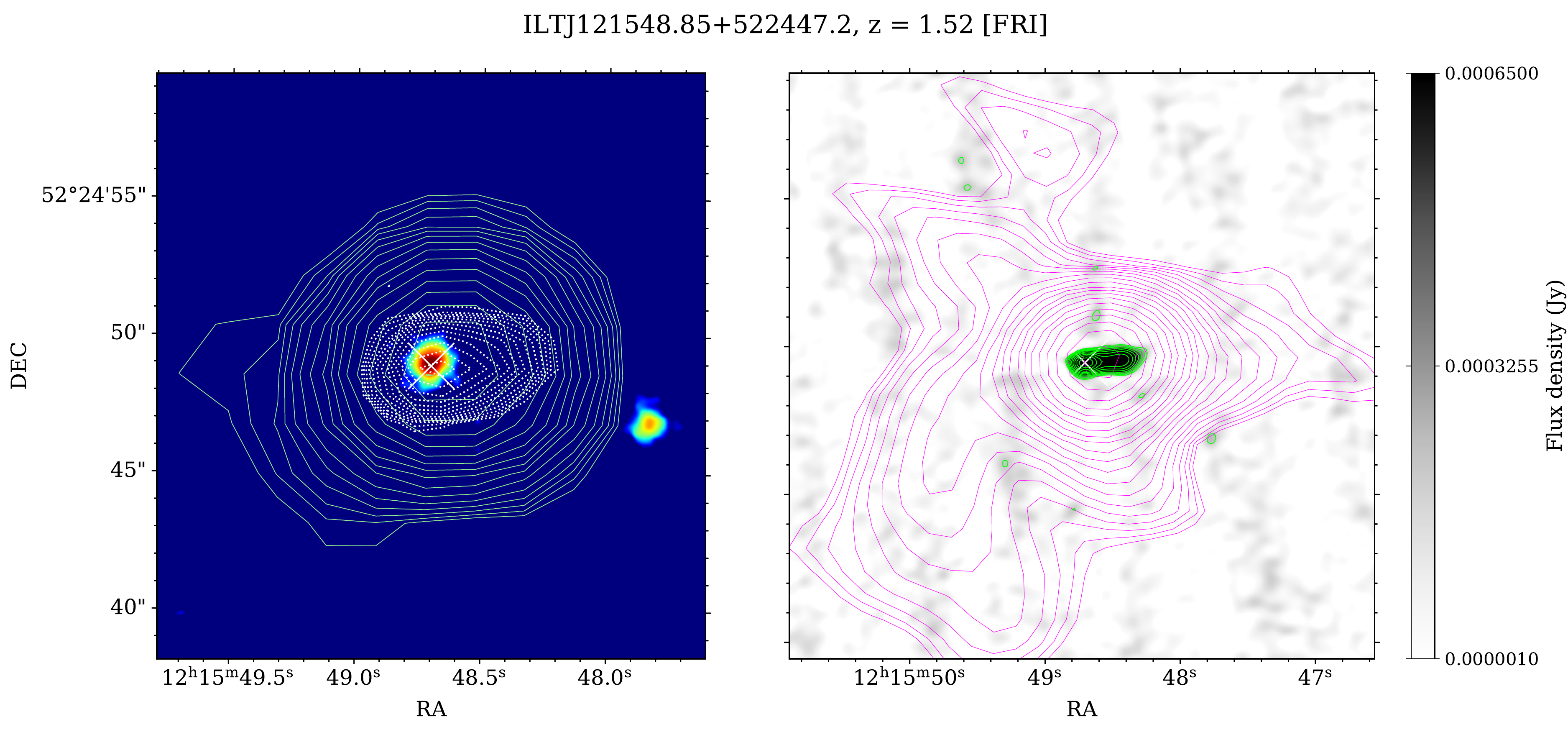}\\
\end{subfigure}
\end{adjustwidth}
\caption{\textit{Cont}.}
\end{figure}
\begin{figure}[H]\ContinuedFloat

\begin{adjustwidth}{-\extralength}{0cm}
\centering
\begin{subfigure}
\centering
\includegraphics[scale=0.4]{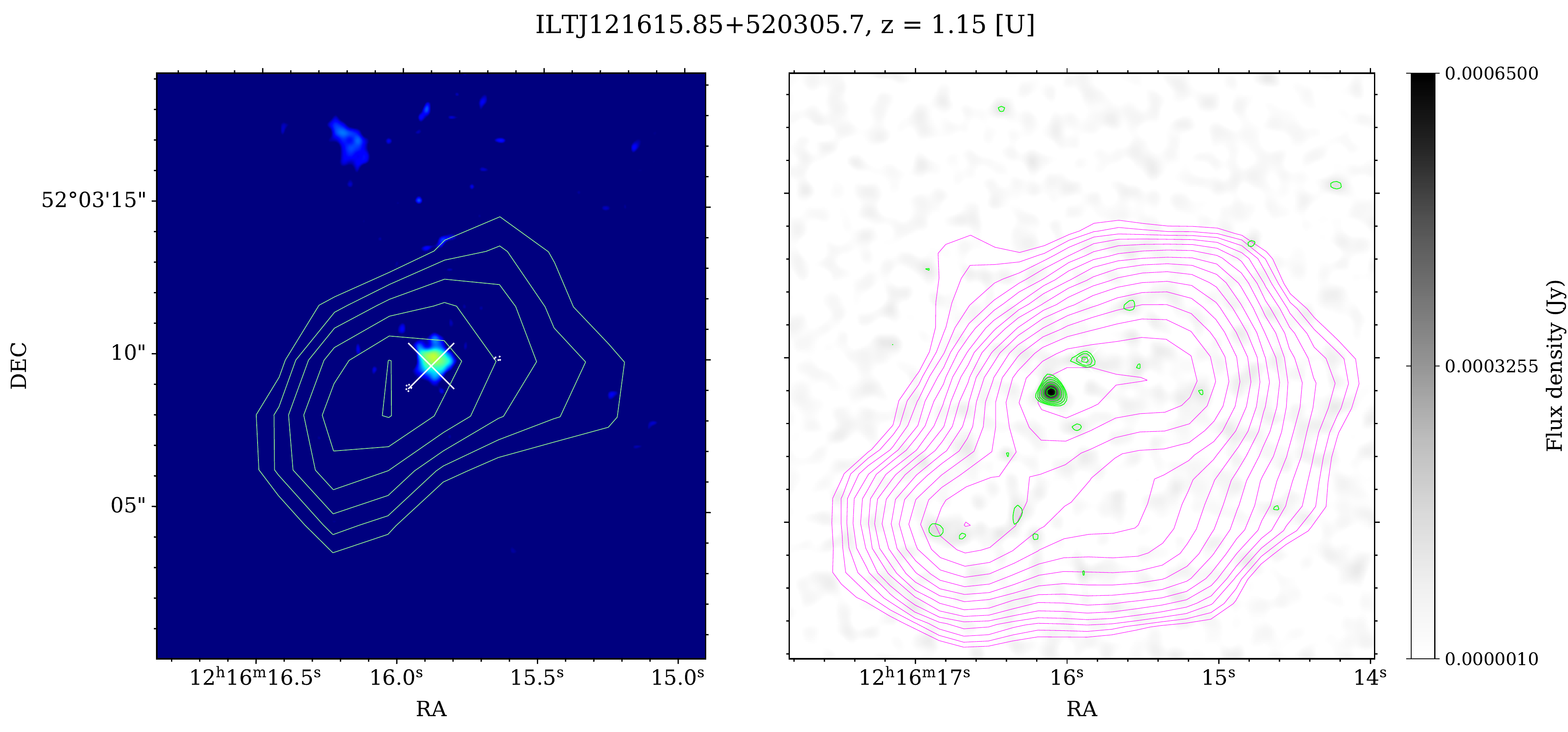}\\
\includegraphics[scale=0.4]{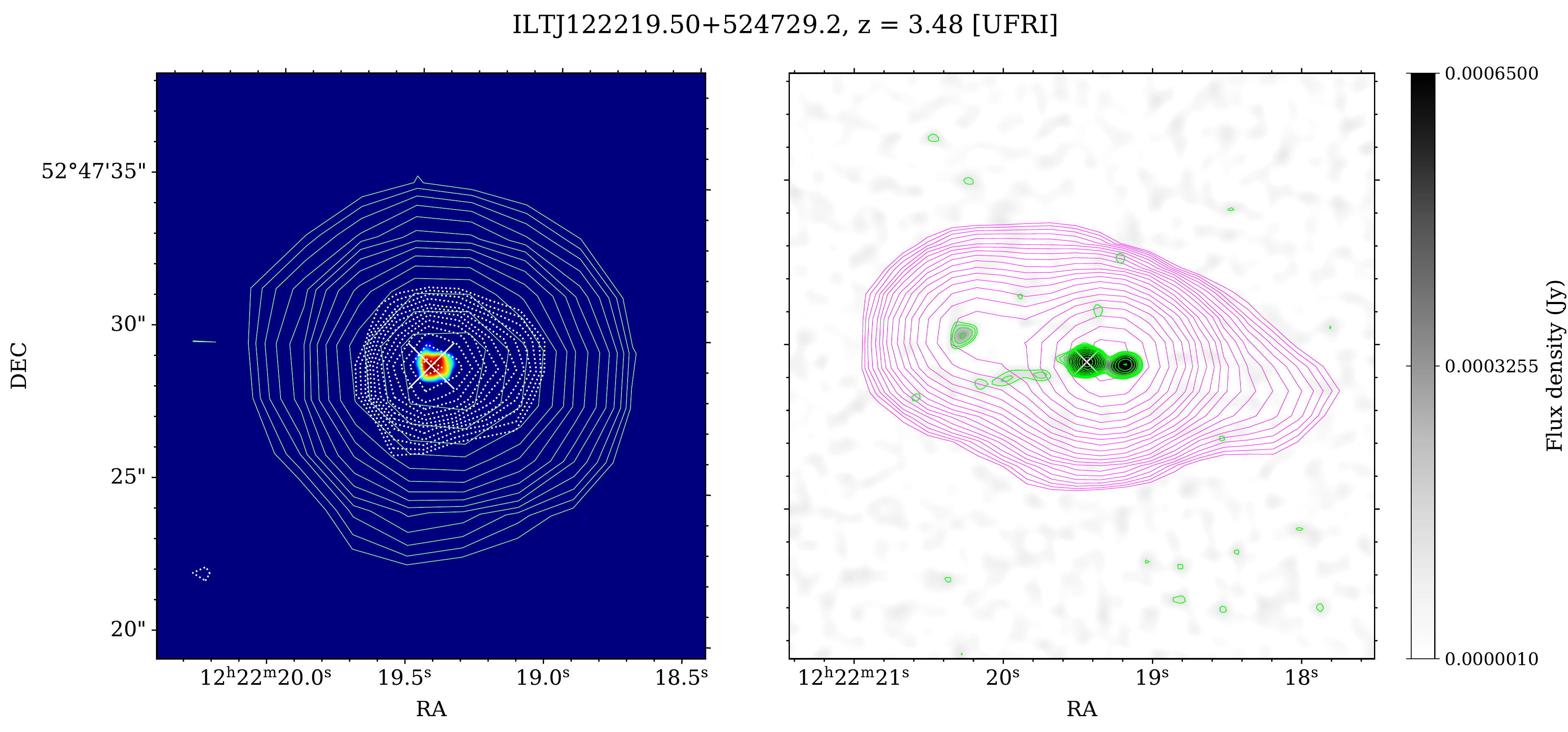}\\
\includegraphics[scale=0.4]{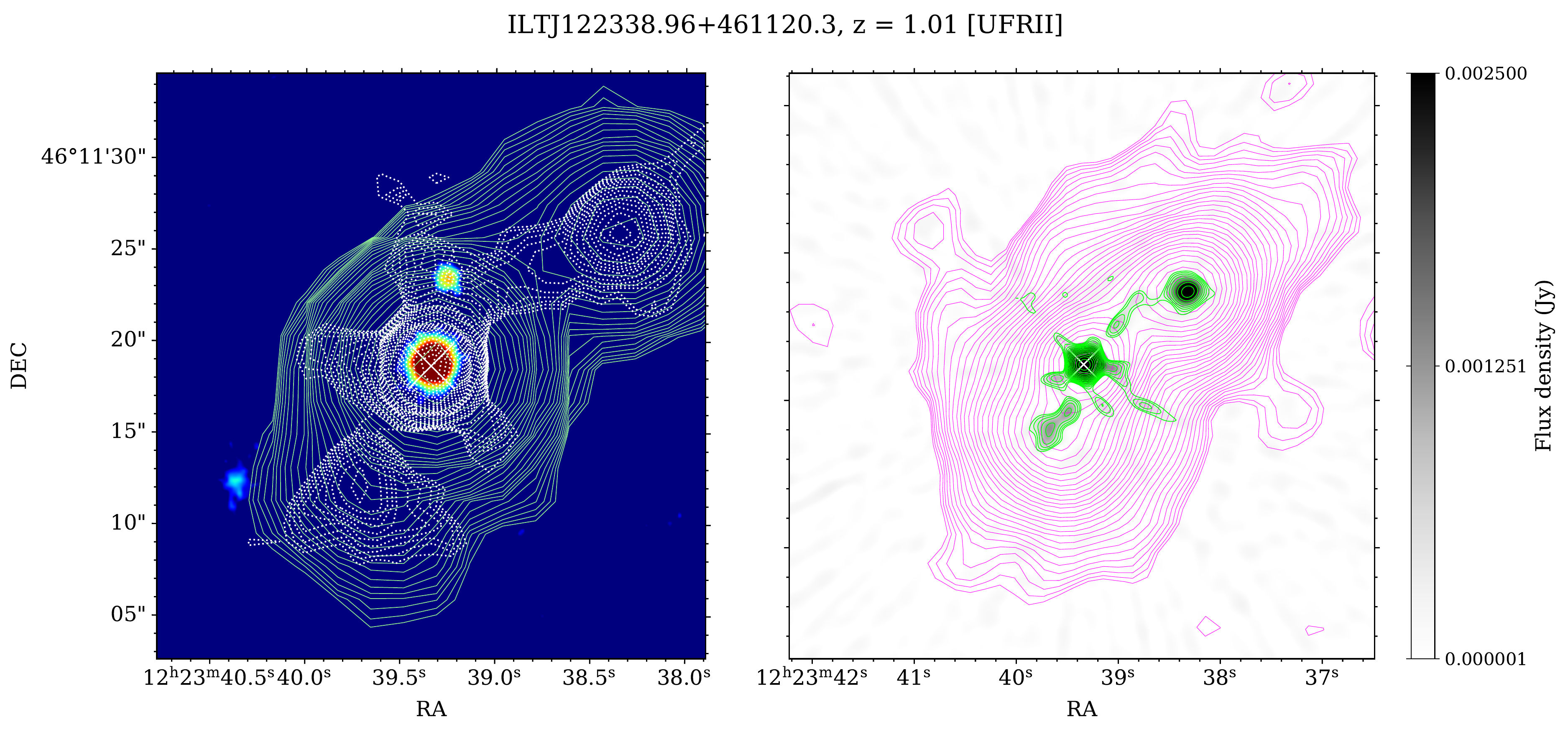}\\
\end{subfigure}
\end{adjustwidth}\caption{\textit{Cont}.}
\end{figure}

\begin{figure}[H]\ContinuedFloat

\begin{adjustwidth}{-\extralength}{0cm}
\centering 
\begin{subfigure}
\centering
\includegraphics[scale=0.4]{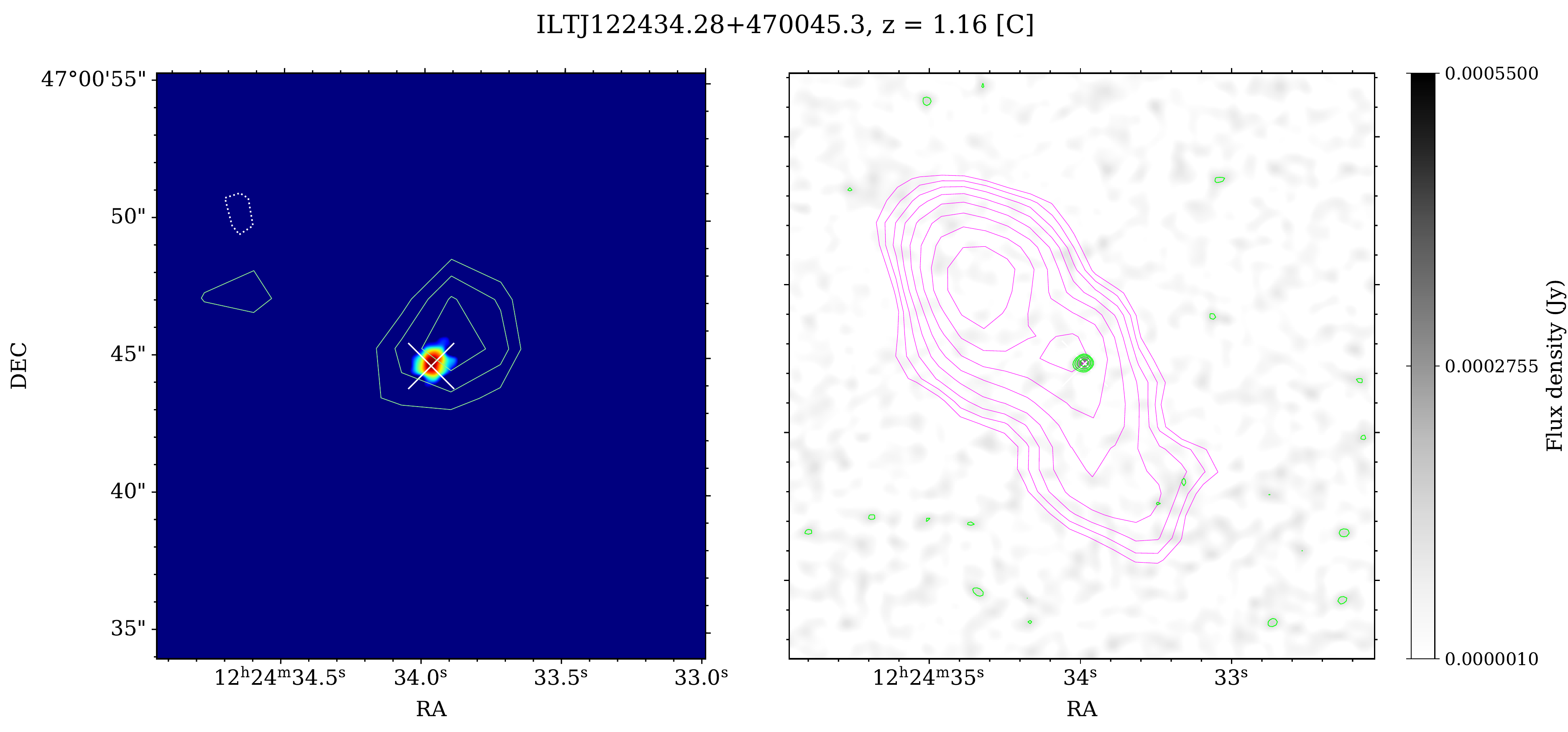}\\
\includegraphics[scale=0.4]{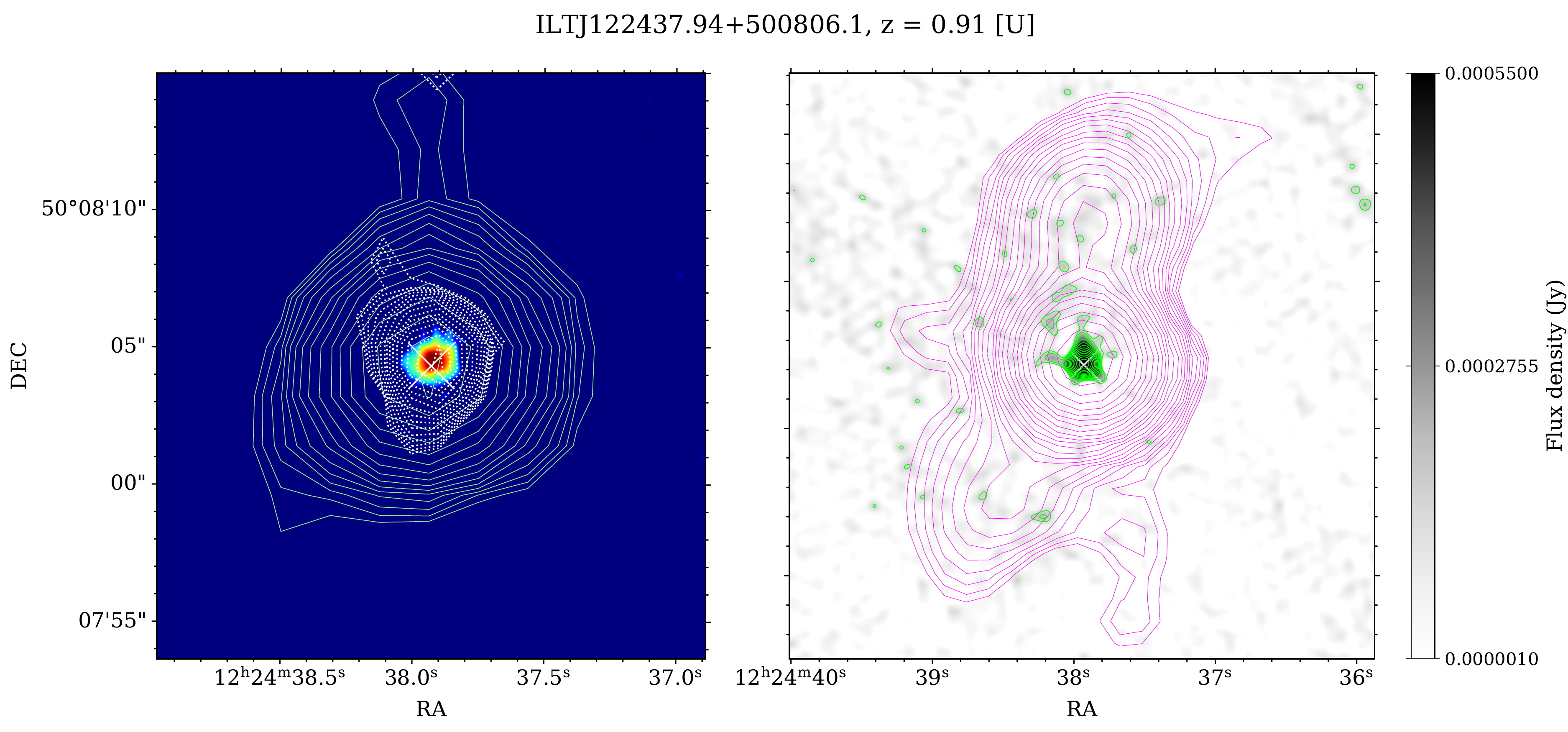}\\
\includegraphics[scale=0.4]{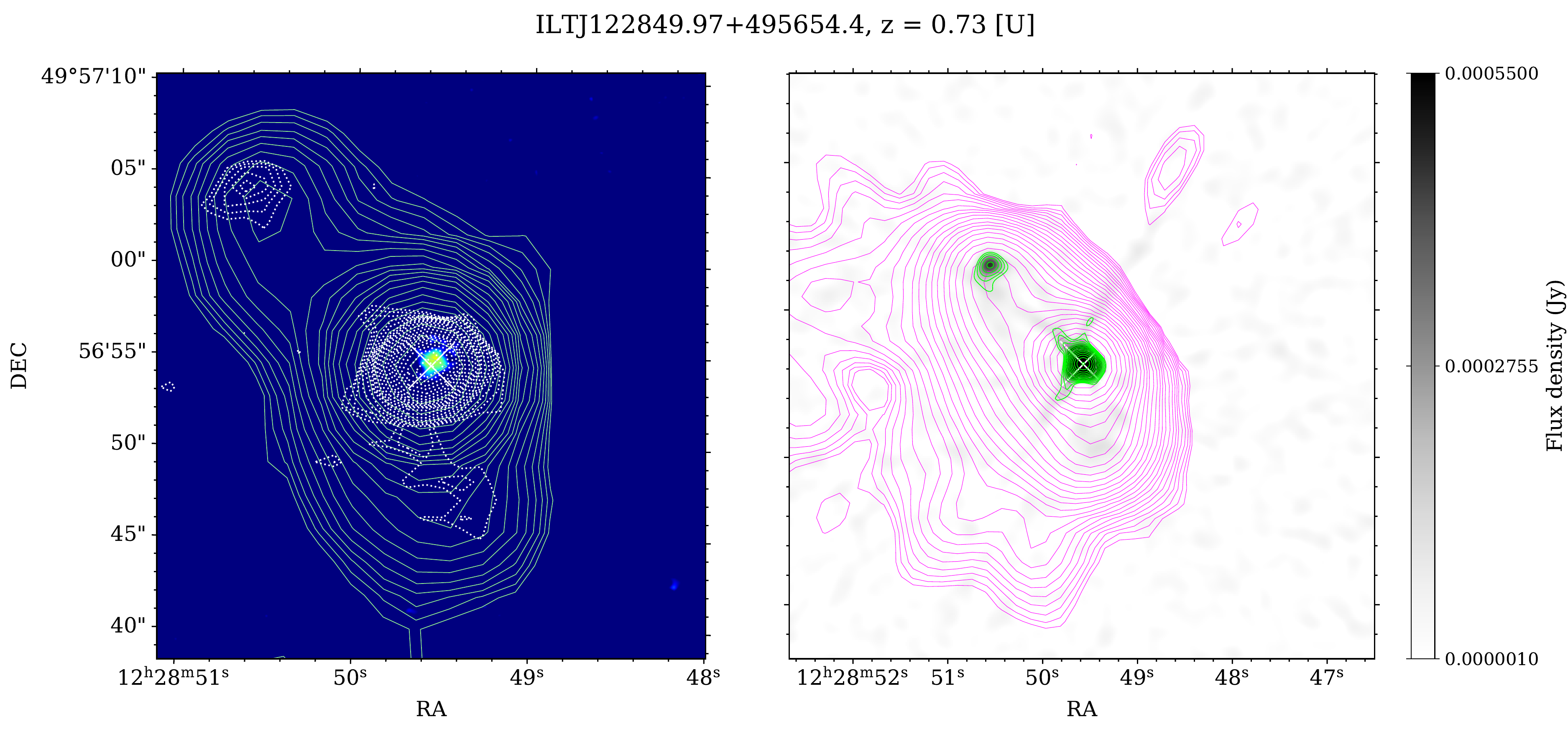}\\
\end{subfigure}
\end{adjustwidth}\caption{\textit{Cont}.}
\end{figure}
\begin{figure}[H]\ContinuedFloat

\begin{adjustwidth}{-\extralength}{0cm}
\centering
\begin{subfigure}
\centering
\includegraphics[scale=0.4]{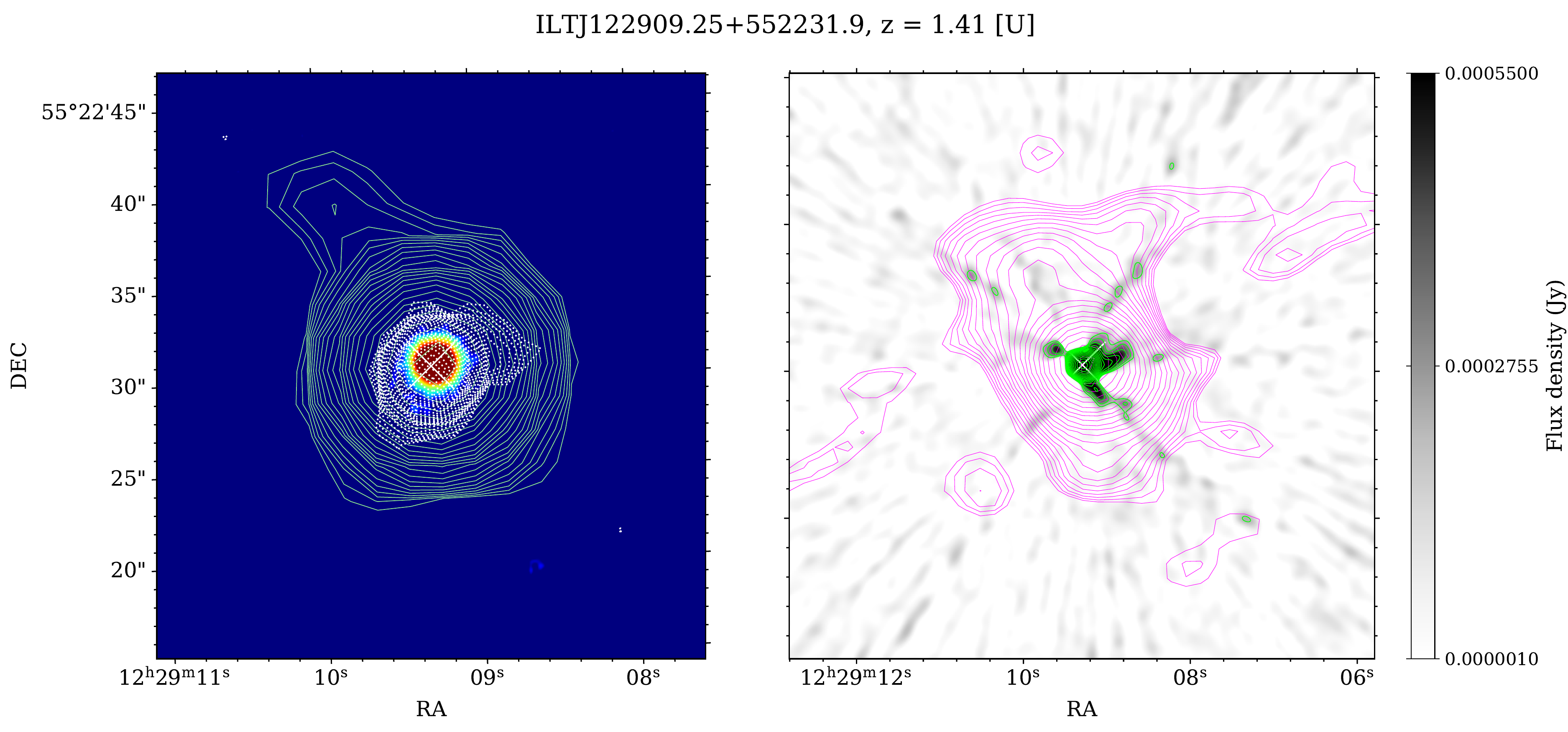}\\
\includegraphics[scale=0.4]{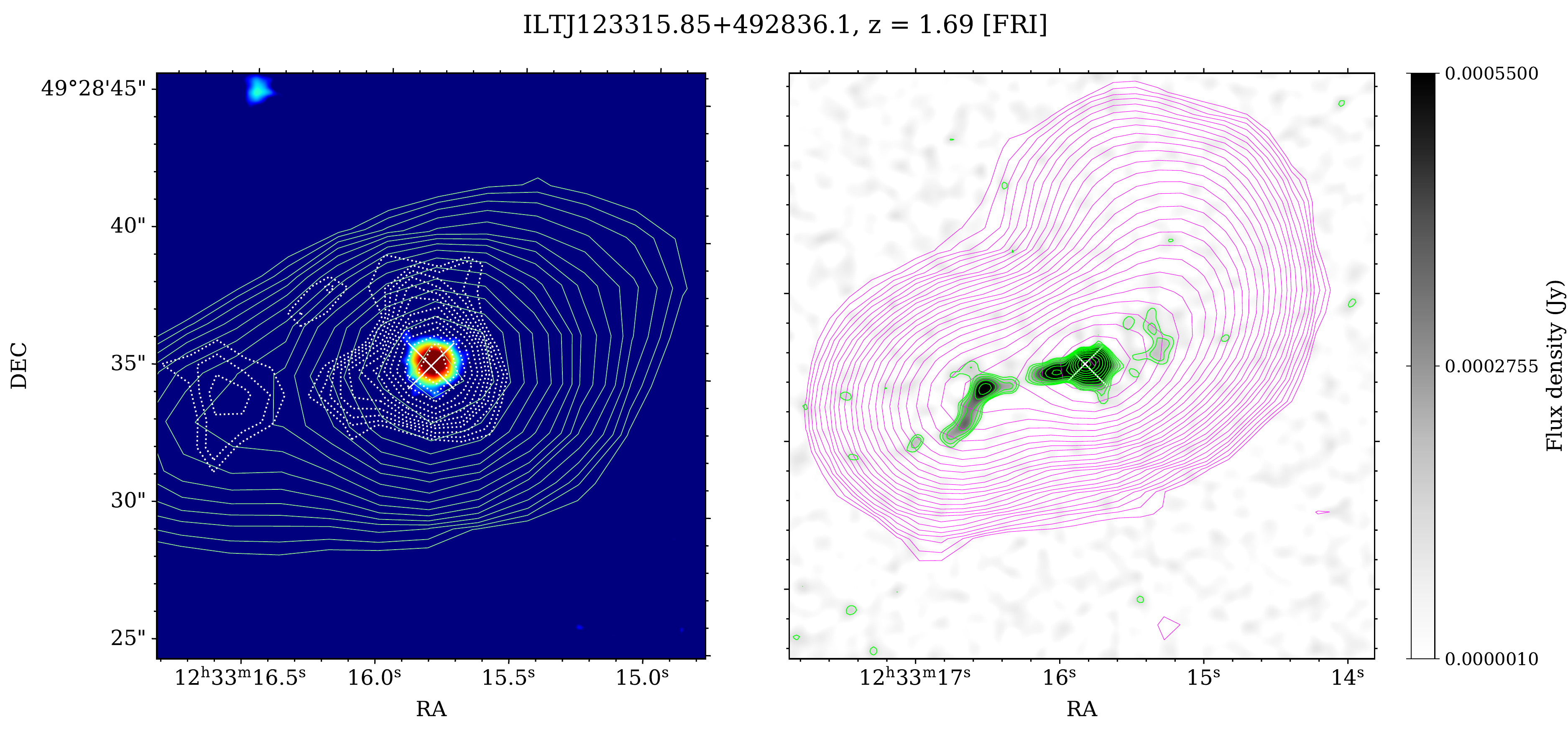}\\
\includegraphics[scale=0.4]{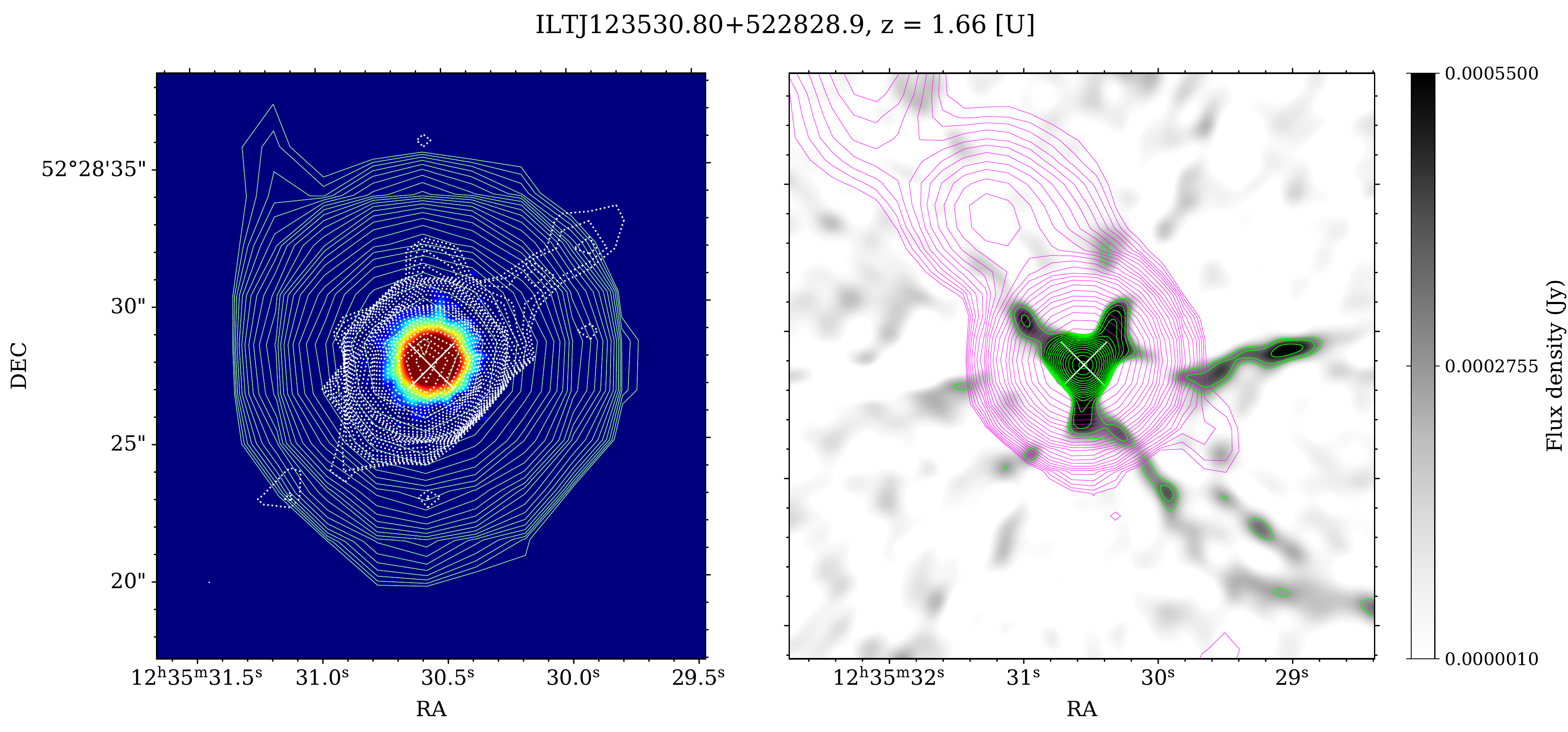}\\
\end{subfigure}
\end{adjustwidth}\caption{\textit{Cont}.}
\end{figure}
\begin{figure}[H]\ContinuedFloat

\begin{adjustwidth}{-\extralength}{0cm}
\centering
\begin{subfigure}
\centering
\includegraphics[scale=0.4]{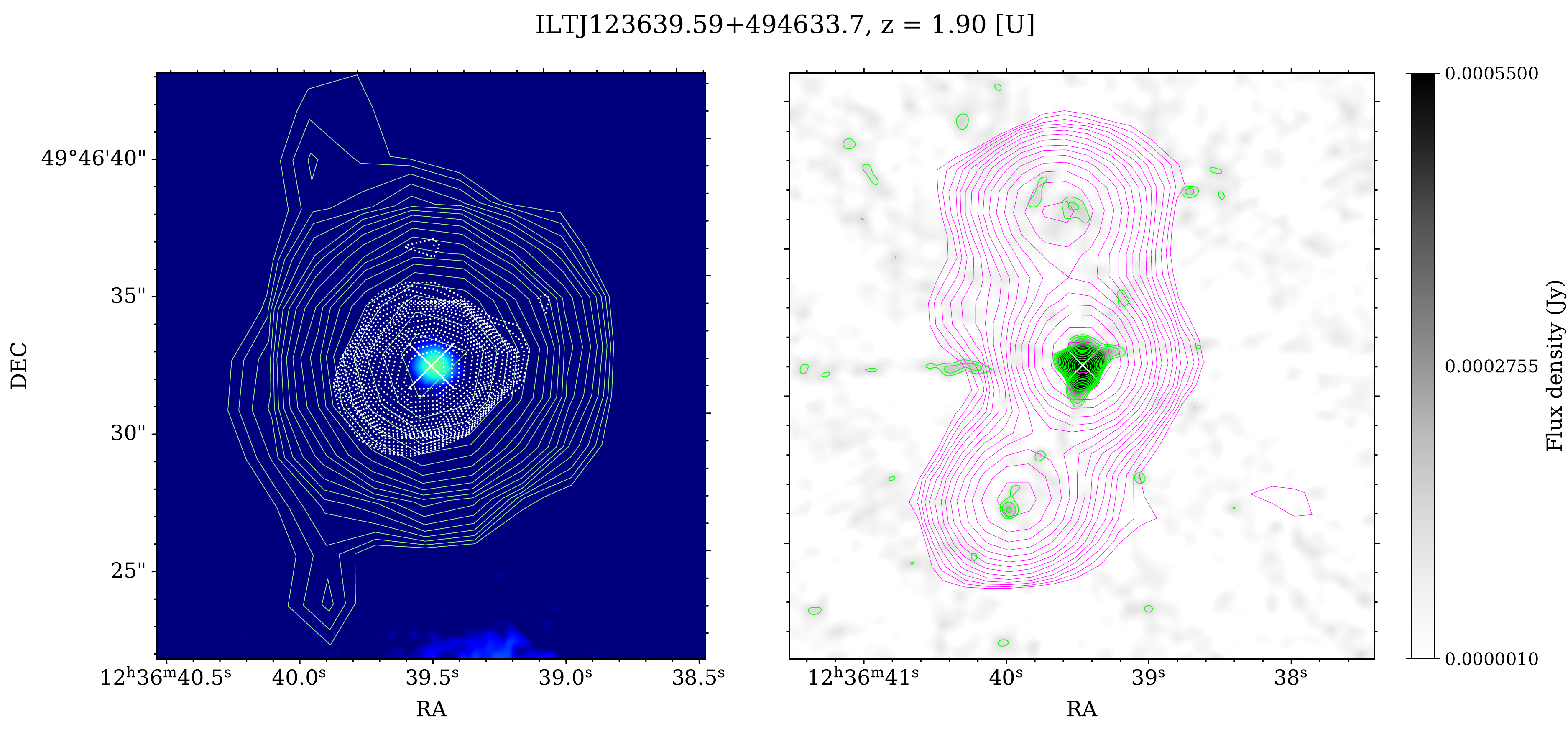}\\
\includegraphics[scale=0.4]{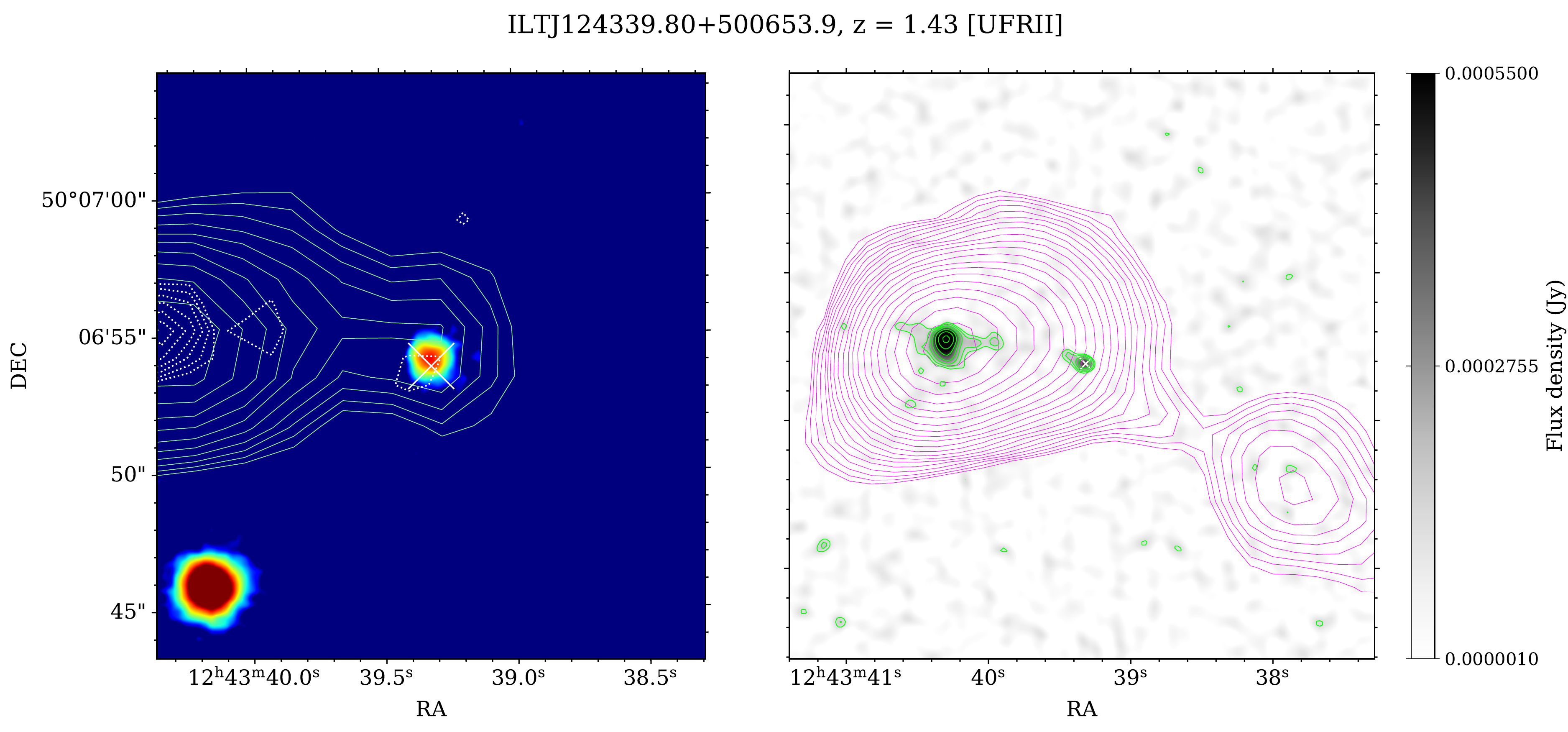}\\
\includegraphics[scale=0.4]{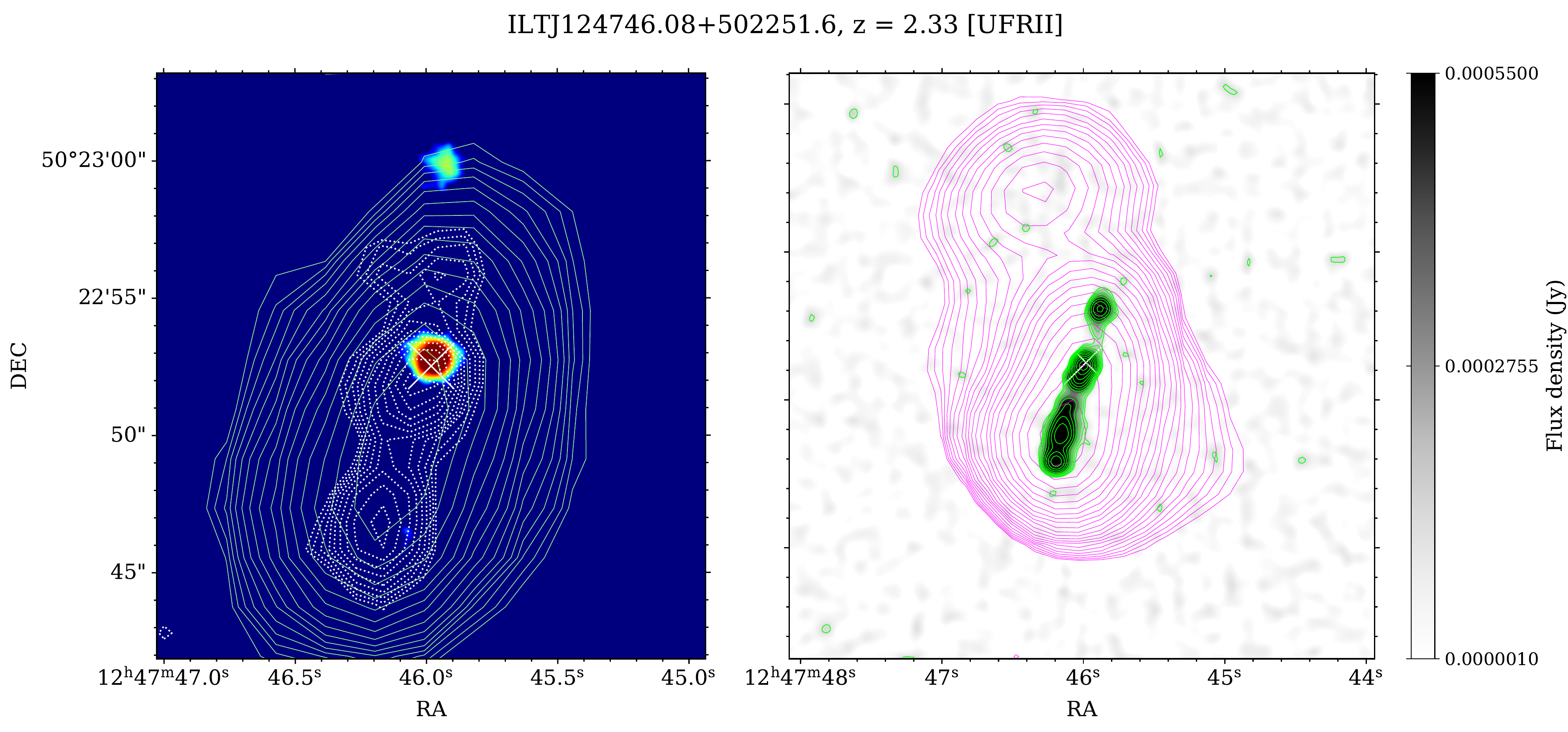}\\
\end{subfigure}
\end{adjustwidth}\caption{\textit{Cont}.}
\end{figure}
\begin{figure}[H]\ContinuedFloat

\begin{adjustwidth}{-\extralength}{0cm}
\centering
\begin{subfigure}
\centering
\includegraphics[scale=0.4]{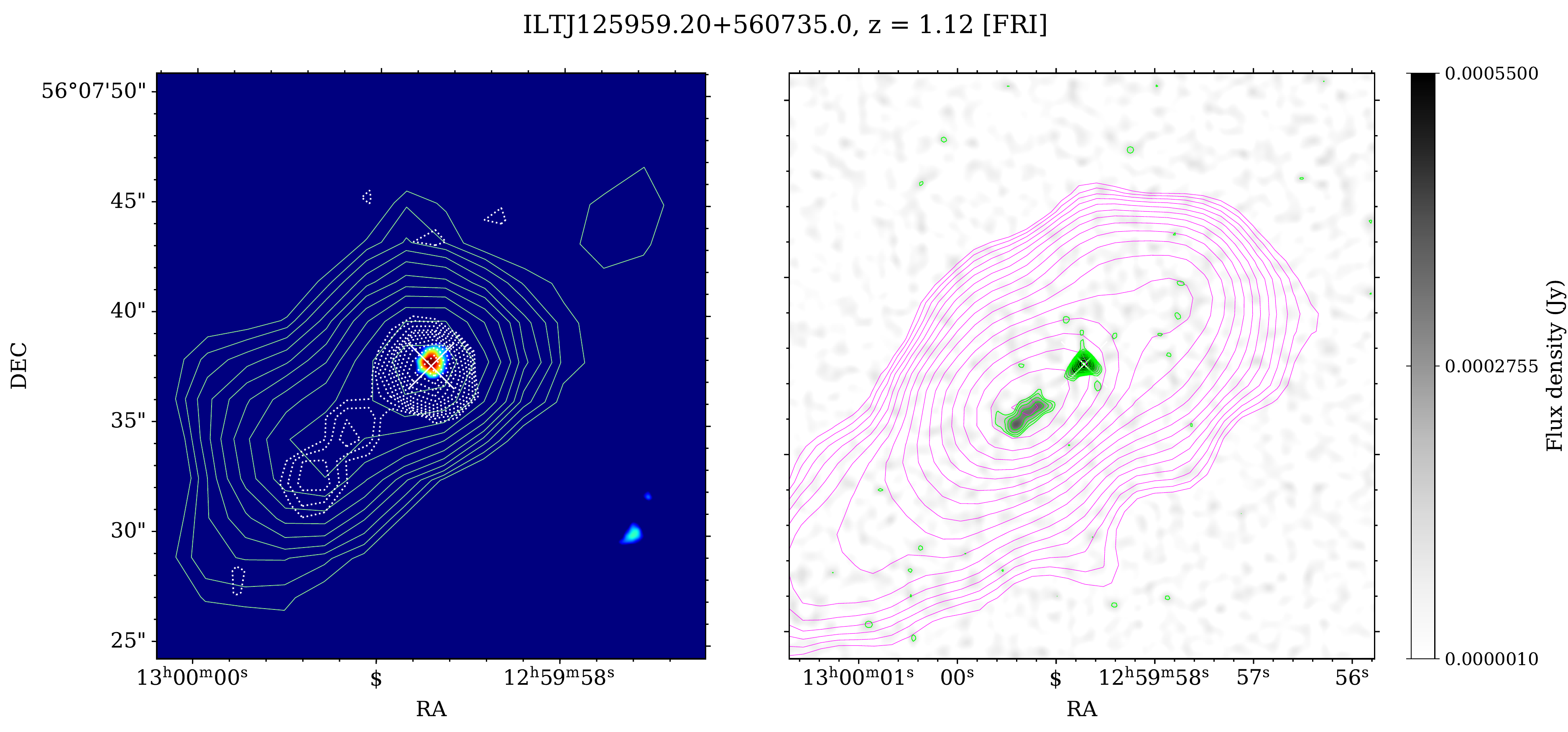}\\
\includegraphics[scale=0.4]{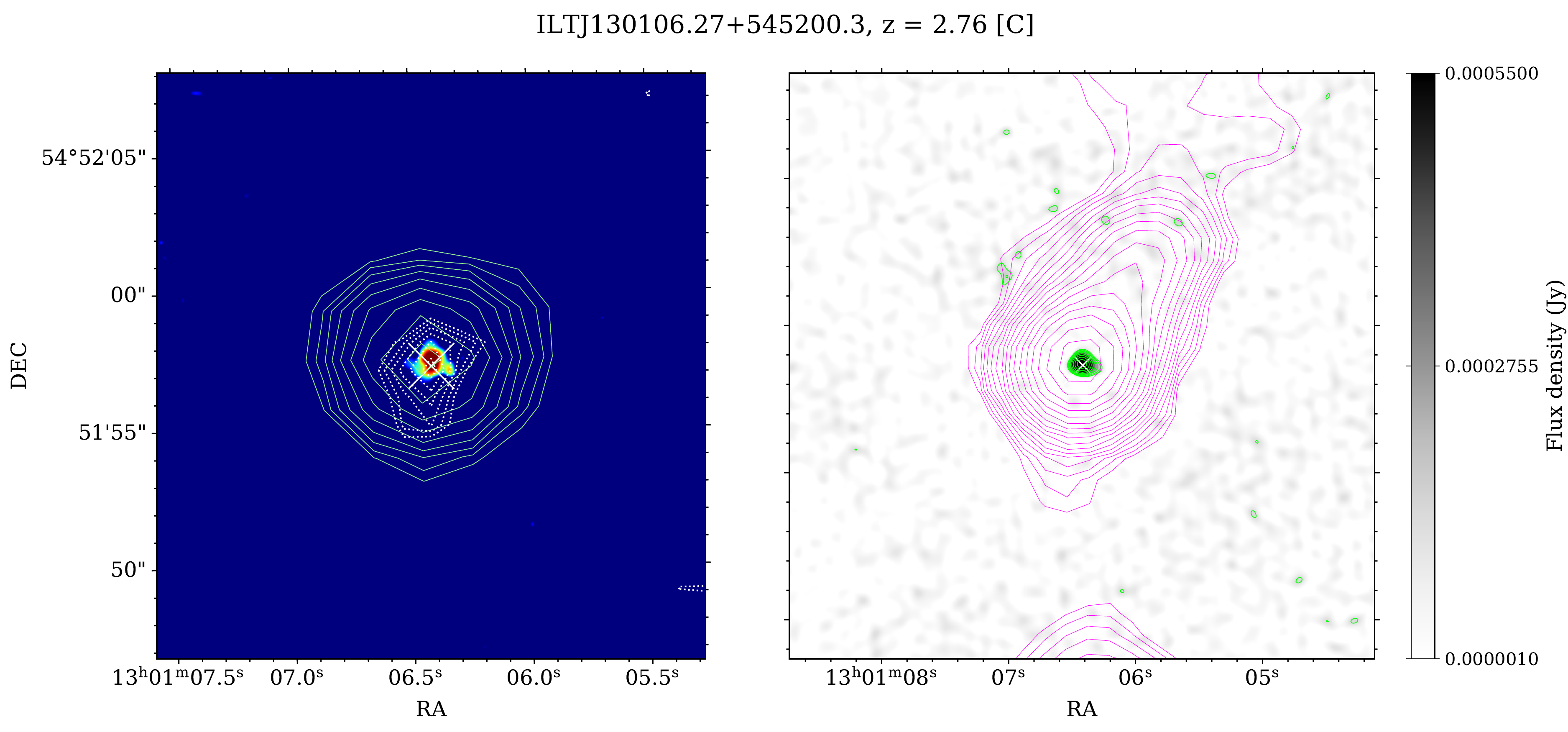}\\
\includegraphics[scale=0.4]{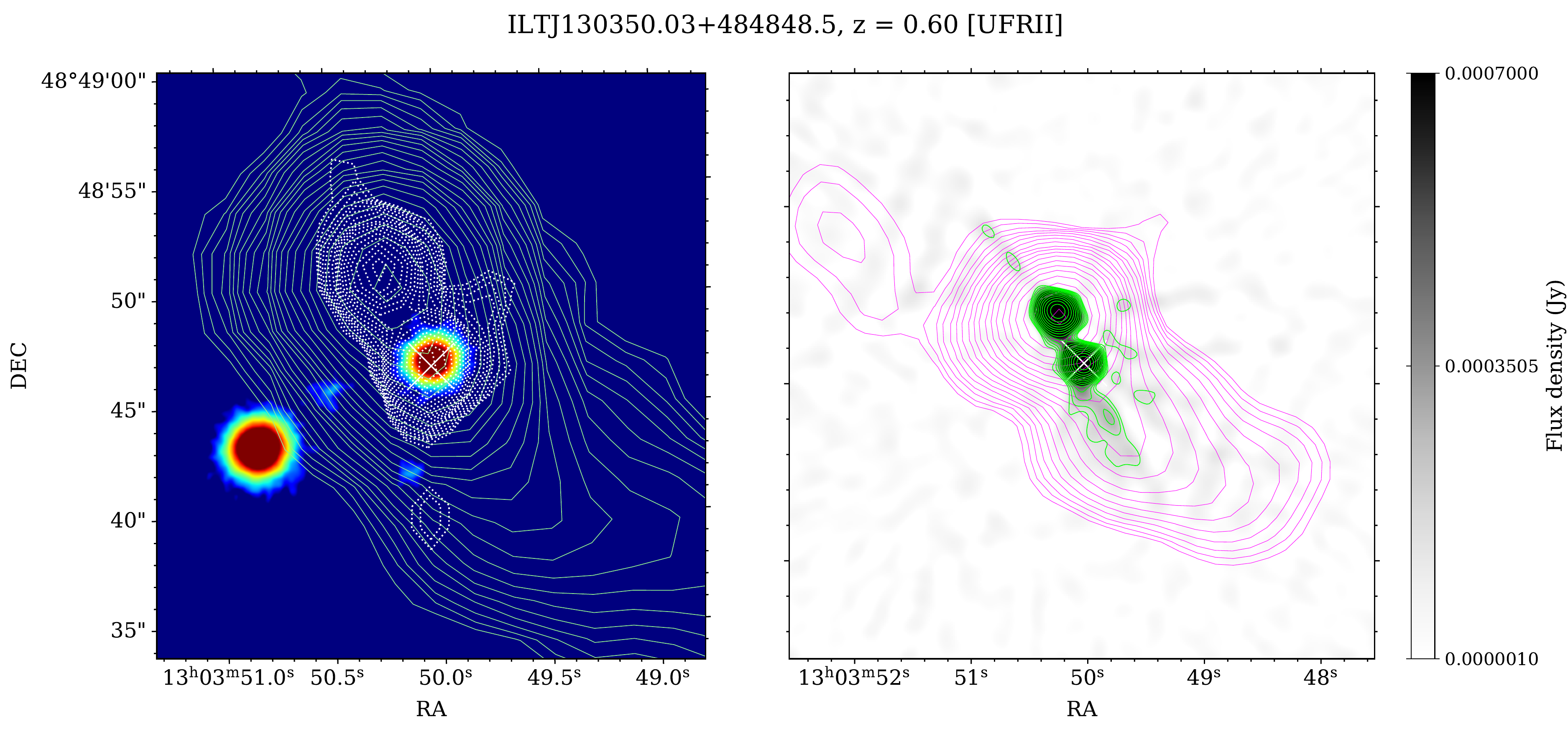}\\
\end{subfigure}
\end{adjustwidth}\caption{\textit{Cont}.}
\end{figure}
\begin{figure}[H]\ContinuedFloat

\begin{adjustwidth}{-\extralength}{0cm}
\centering
\begin{subfigure}
\centering
\includegraphics[scale=0.4]{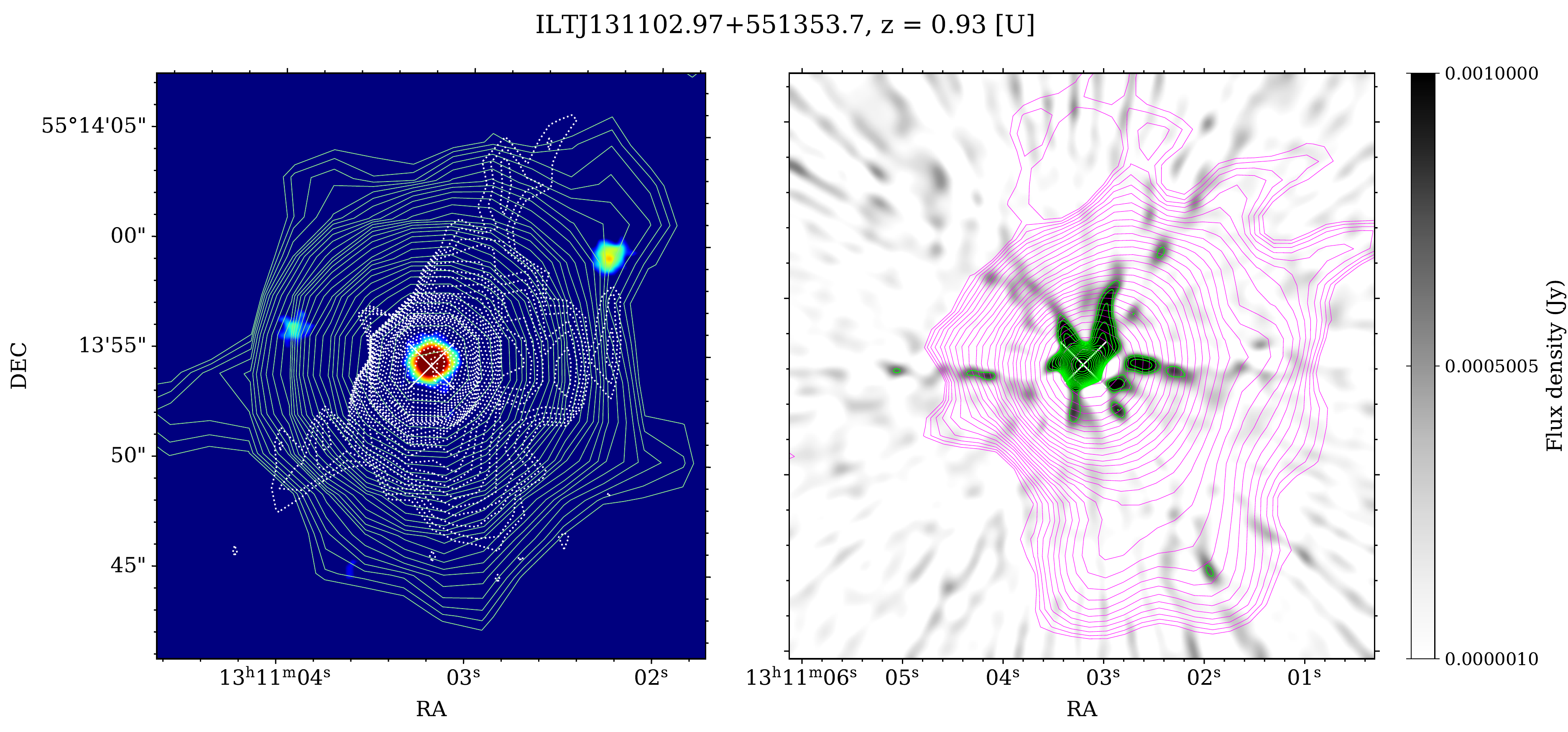}\\
\includegraphics[scale=0.4]{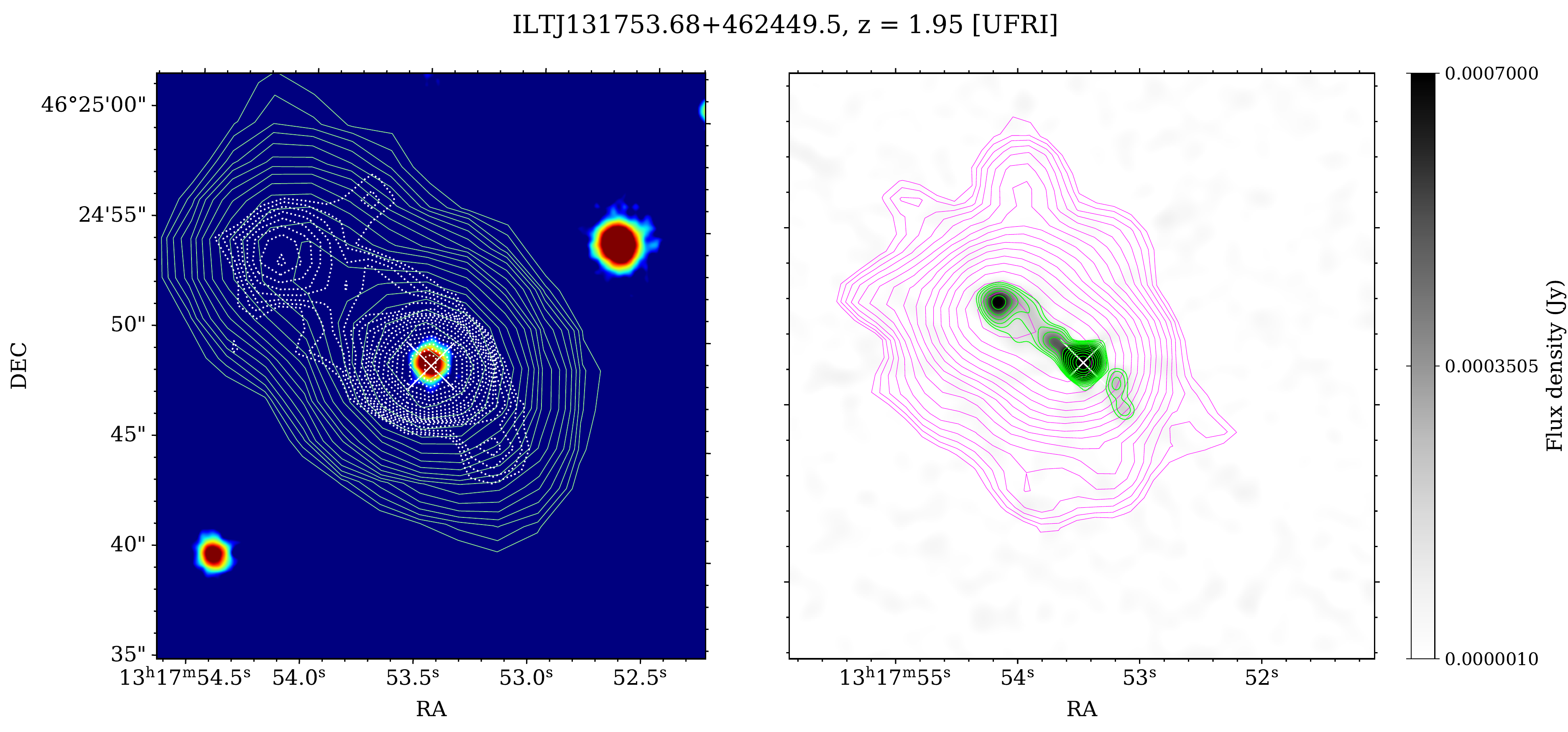}\\
\includegraphics[scale=0.4]{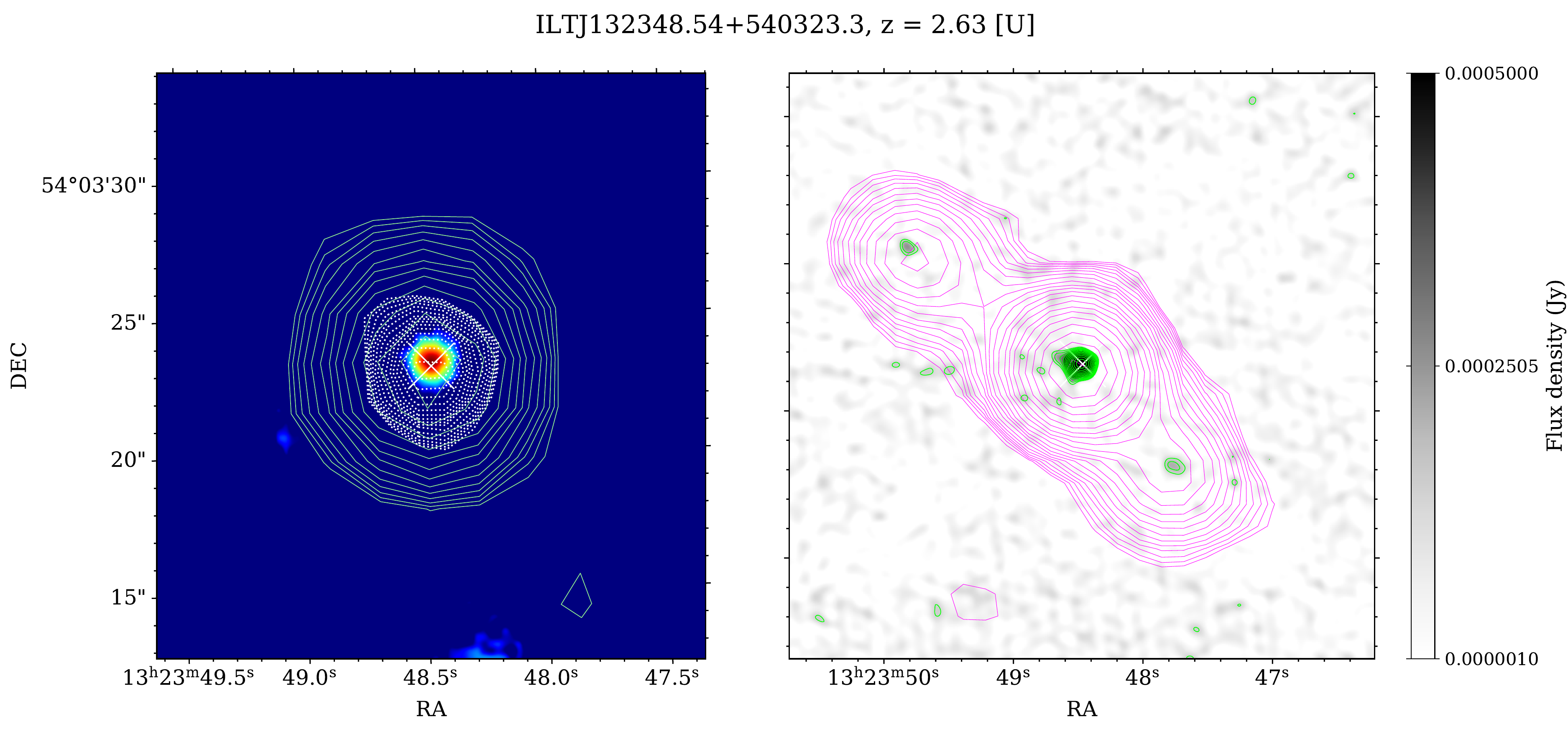}\\
\end{subfigure}
\end{adjustwidth}\caption{\textit{Cont}.}
\end{figure}
\begin{figure}[H]\ContinuedFloat

\begin{adjustwidth}{-\extralength}{0cm}
\centering
\begin{subfigure}
\centering
\includegraphics[scale=0.4]{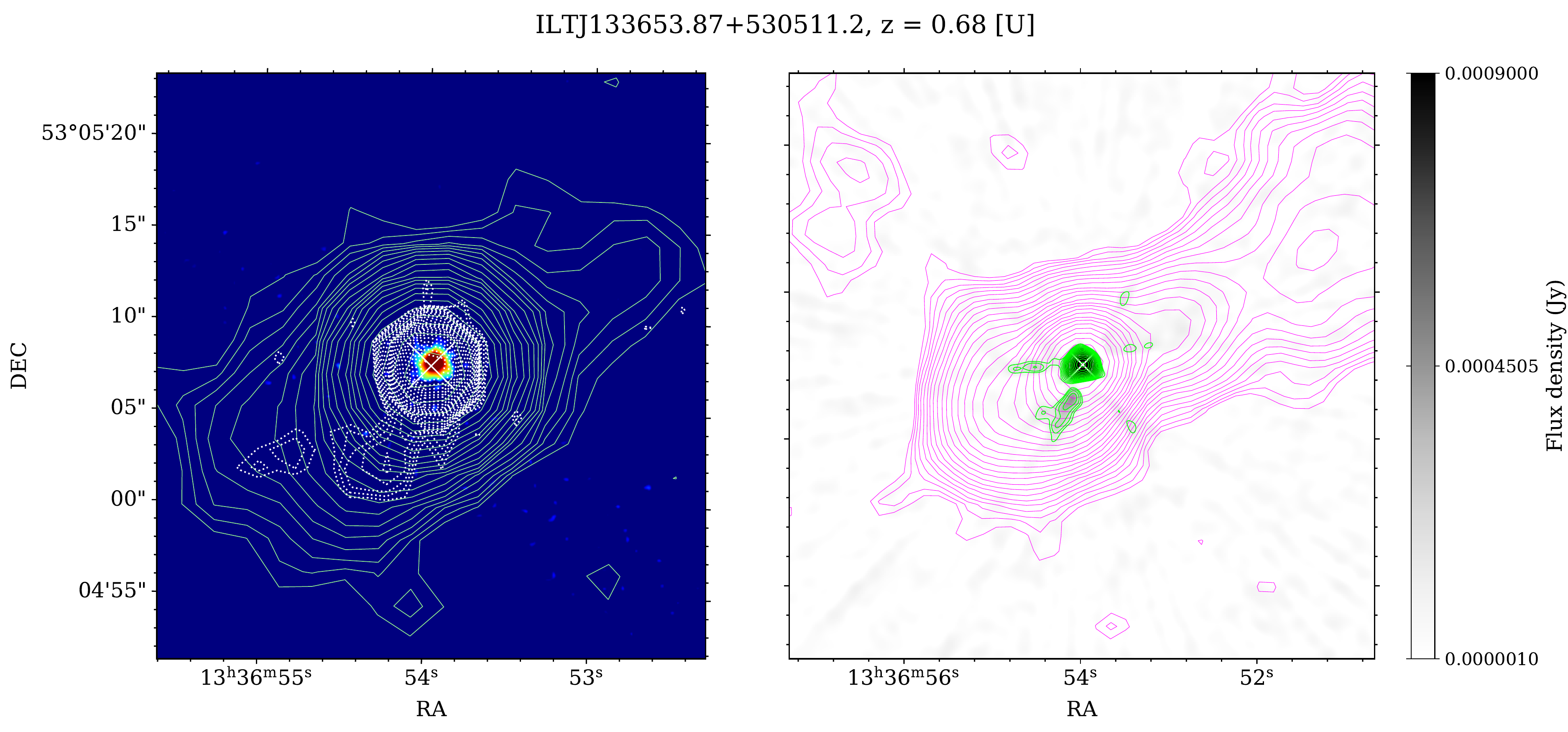}\\
\includegraphics[scale=0.4]{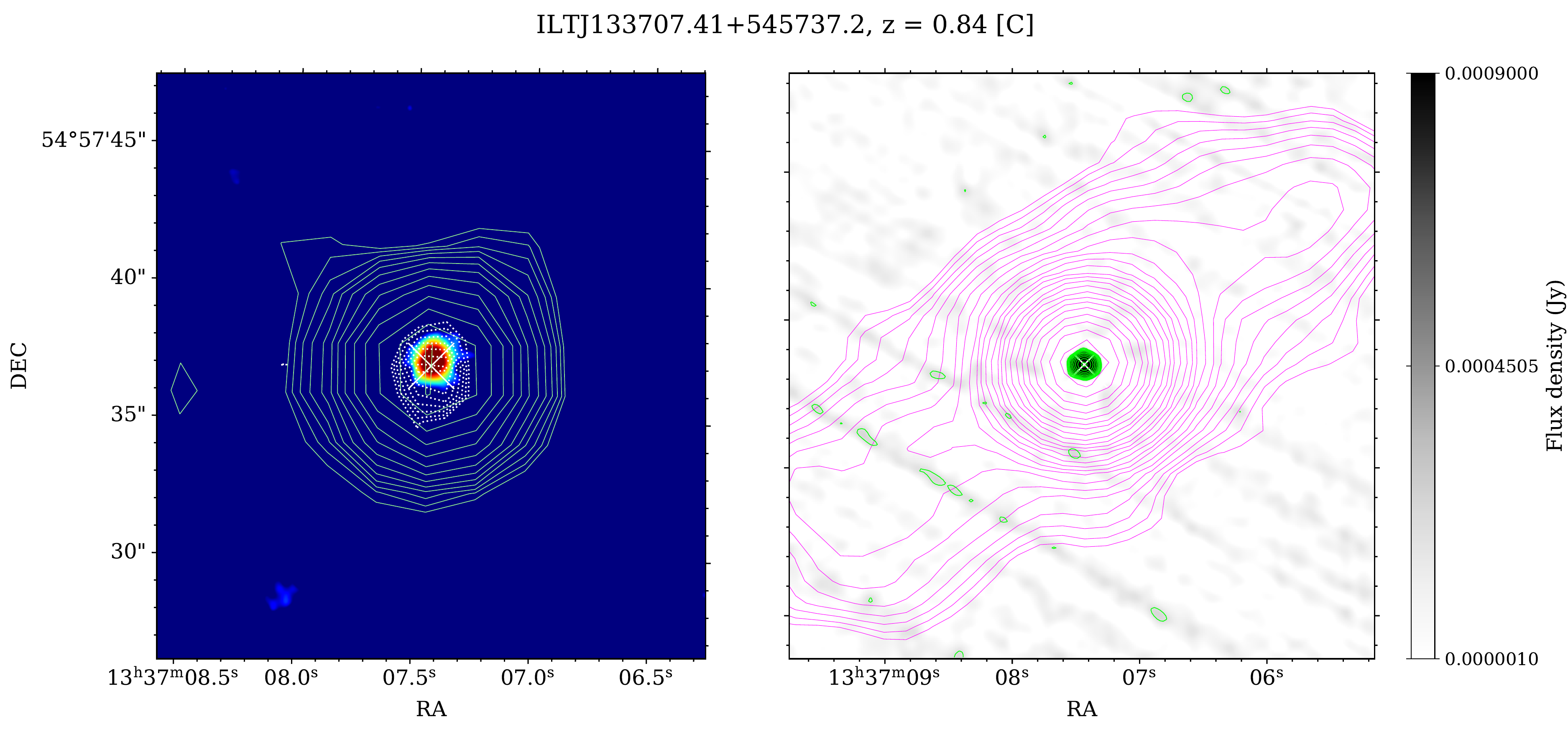}\\
\includegraphics[scale=0.4]{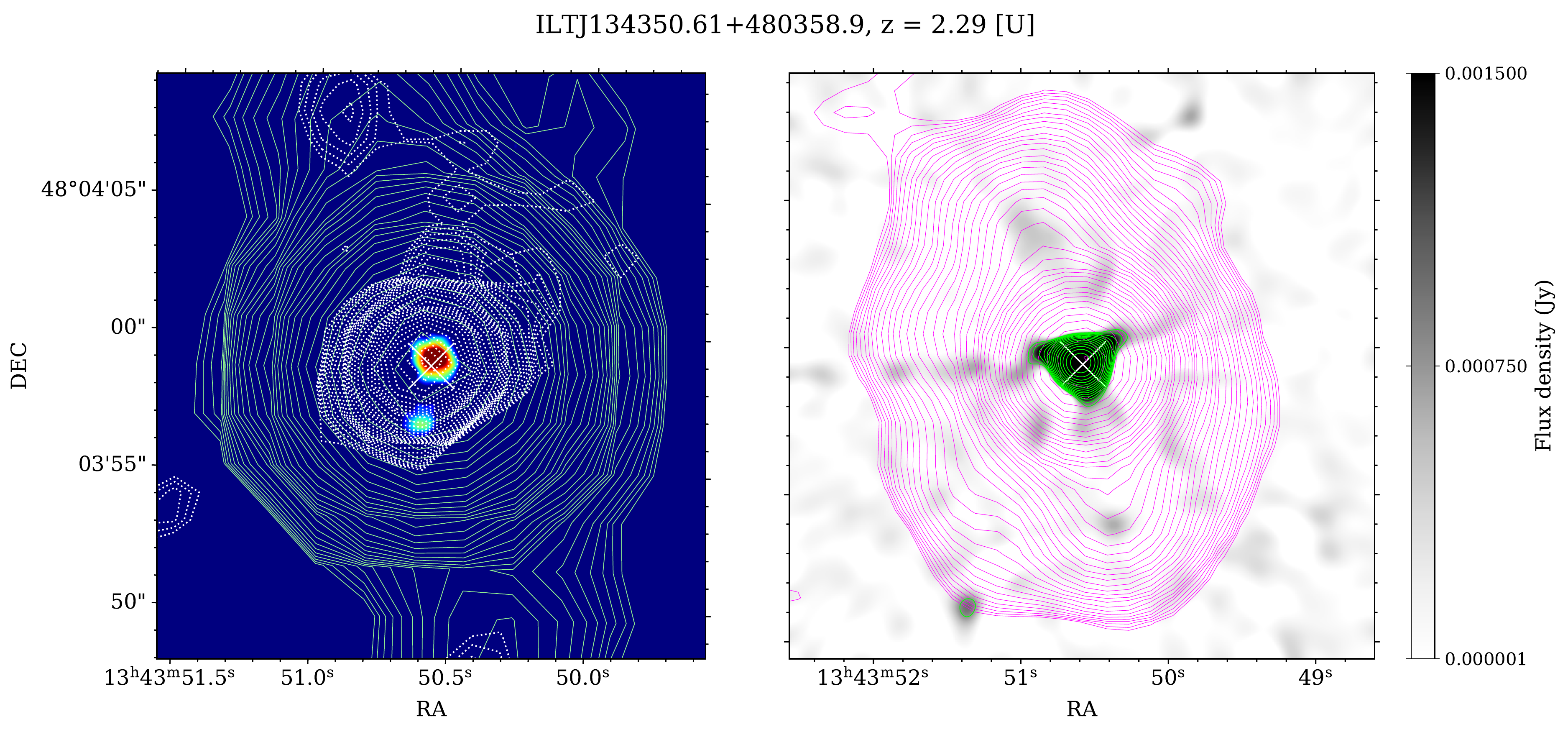}\\
\end{subfigure}
\end{adjustwidth}\caption{\textit{Cont}.}
\end{figure}

\begin{figure}[H]\ContinuedFloat

\begin{adjustwidth}{-\extralength}{0cm}
\centering
\begin{subfigure}
\centering
\includegraphics[scale=0.4]{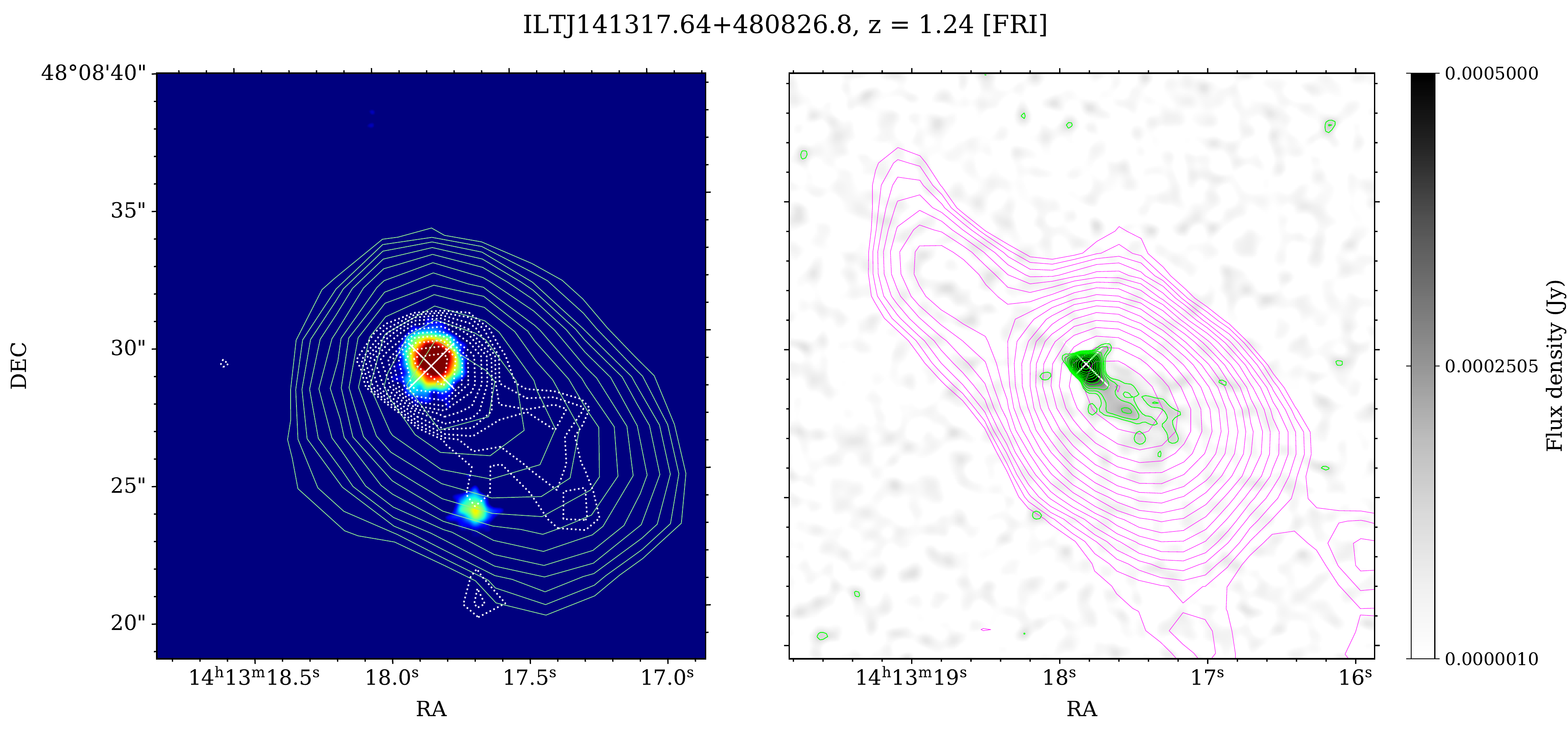}\\
\includegraphics[scale=0.4]{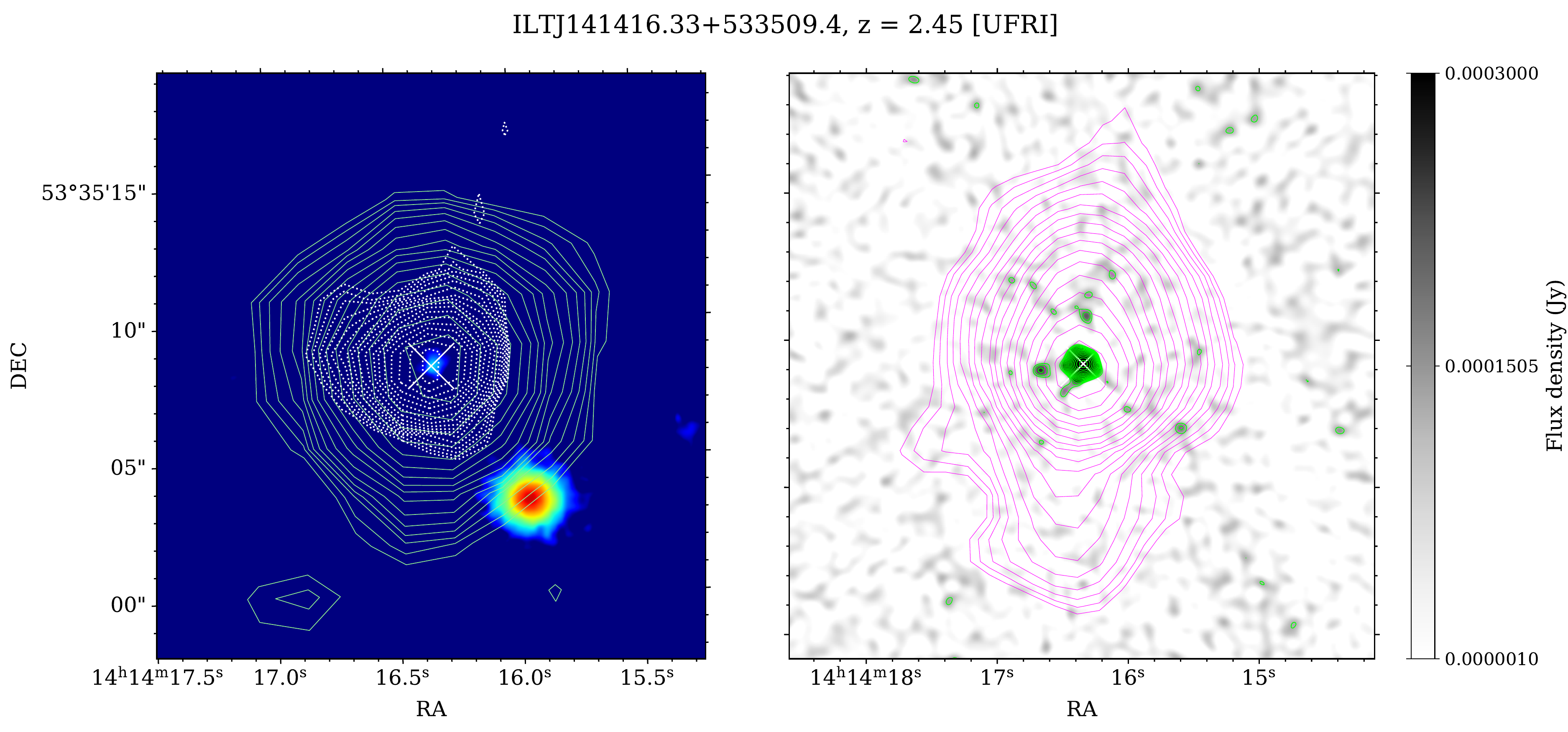}\\
\includegraphics[scale=0.4]{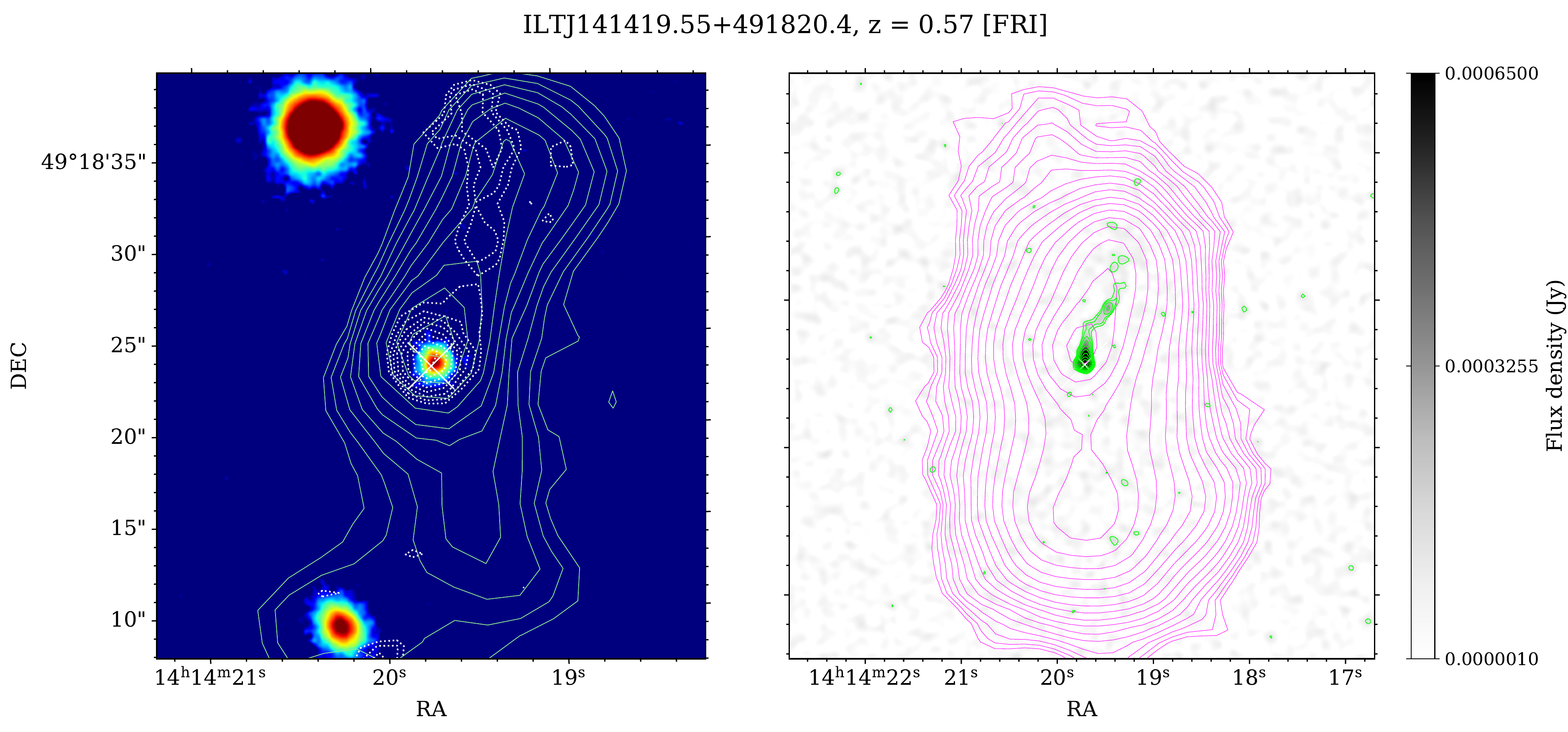}\\
\end{subfigure}
\end{adjustwidth}\caption{\textit{Cont}.}
\end{figure}
\begin{figure}[H]\ContinuedFloat

\begin{adjustwidth}{-\extralength}{0cm}
\centering
\begin{subfigure}
\centering
\includegraphics[scale=0.4]{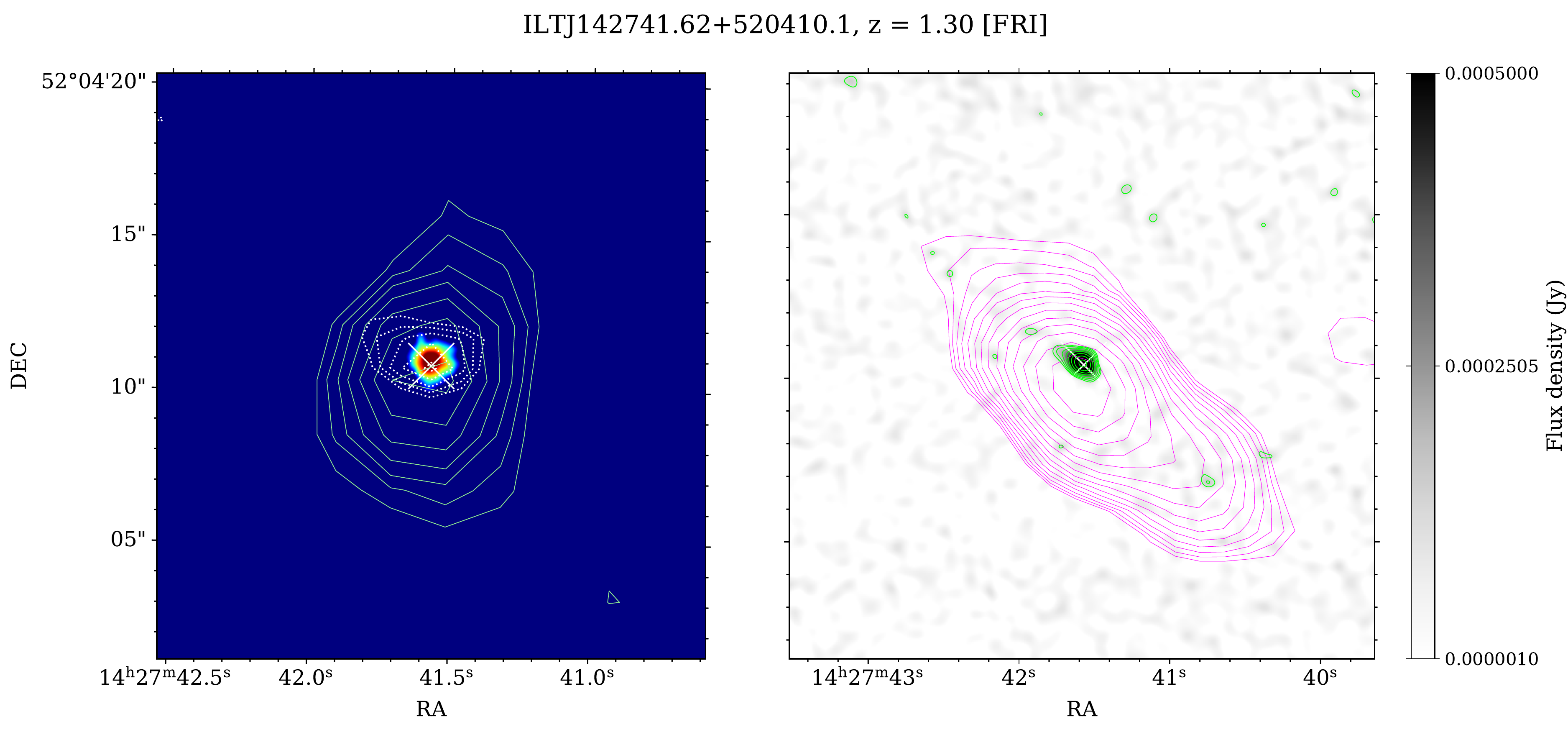}\\
\includegraphics[scale=0.4]{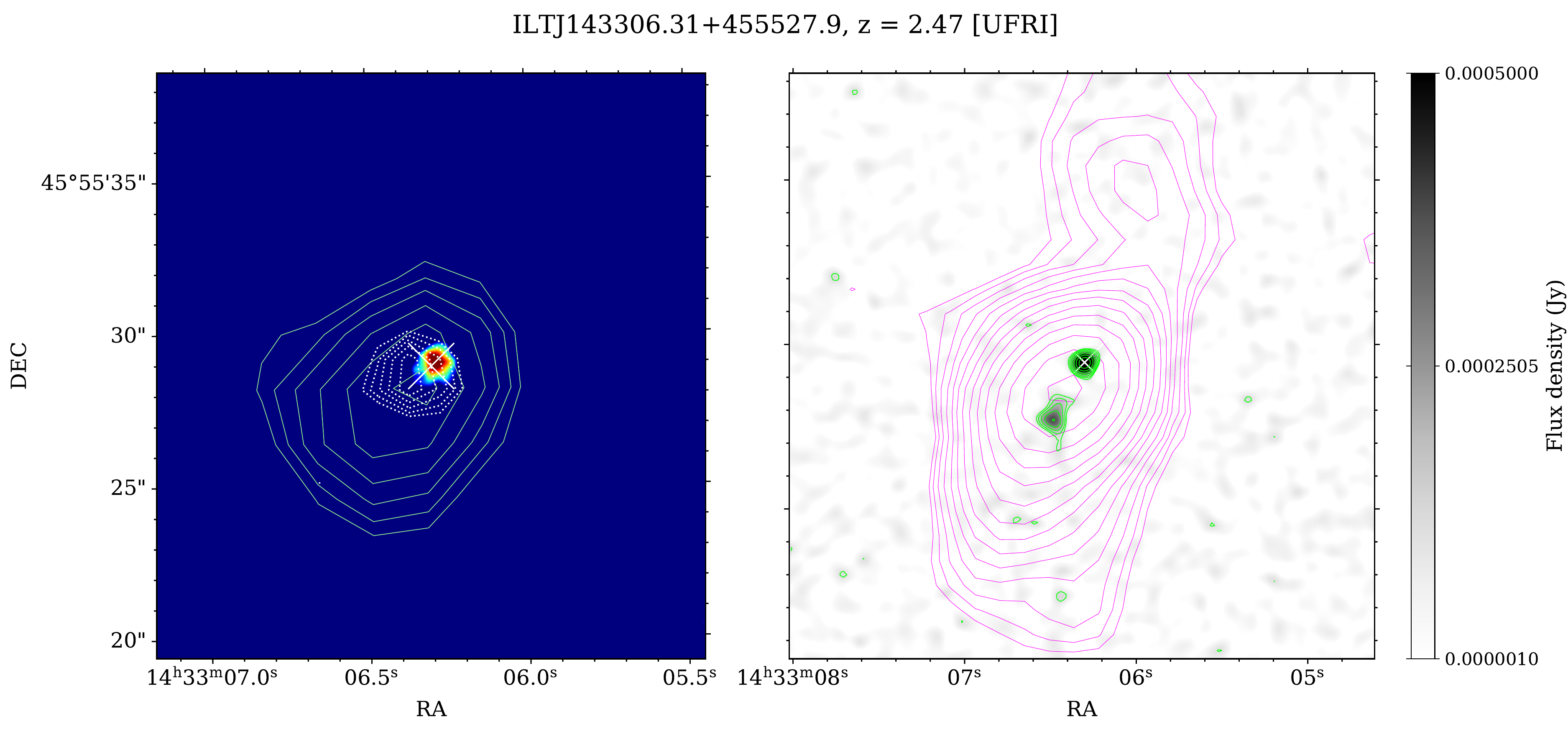}\\
\includegraphics[scale=0.4]{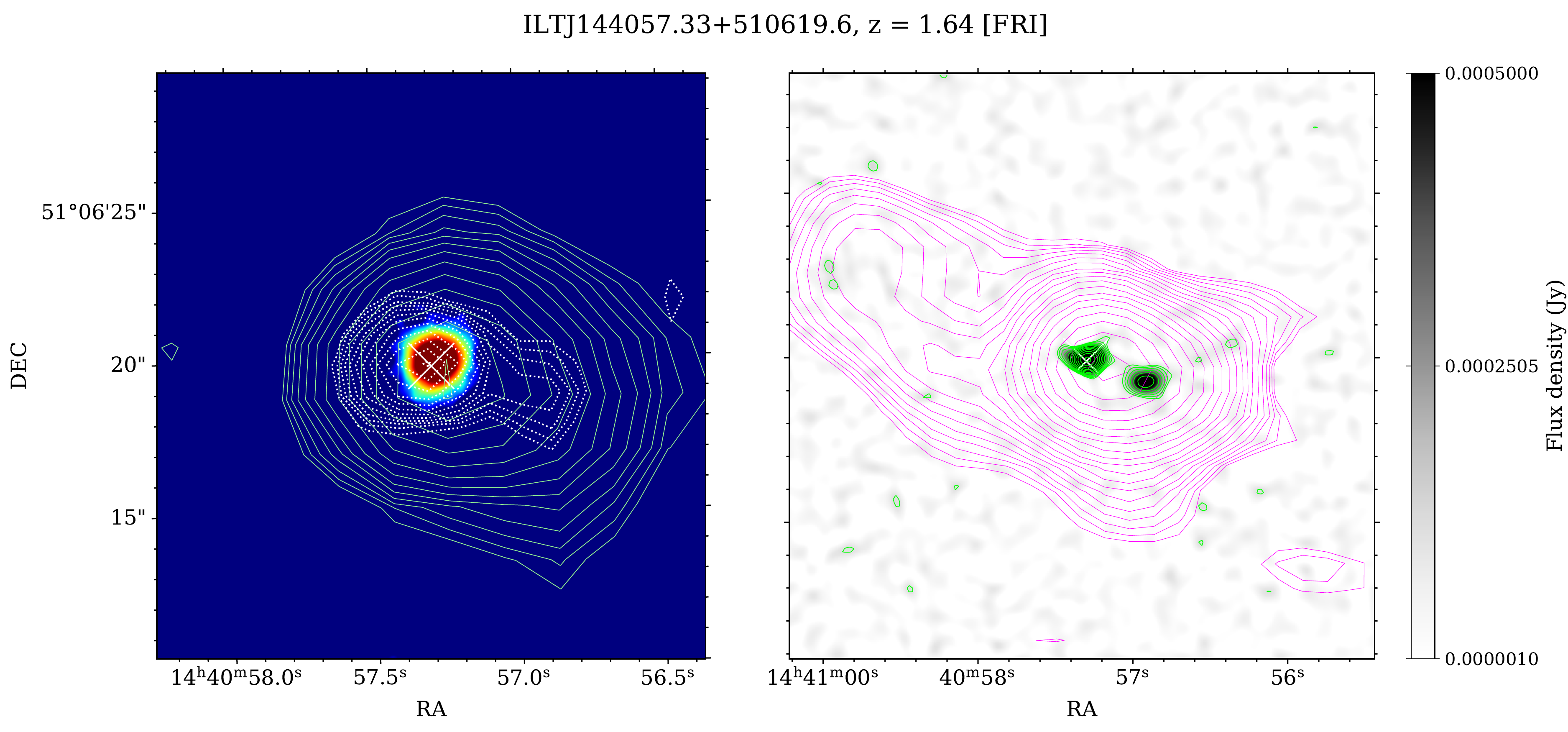}\\
\end{subfigure}
\end{adjustwidth}\caption{\textit{Cont}.}
\end{figure}
\begin{figure}[H]\ContinuedFloat

\begin{adjustwidth}{-\extralength}{0cm}
\centering
\begin{subfigure}
\centering
\includegraphics[scale=0.4]{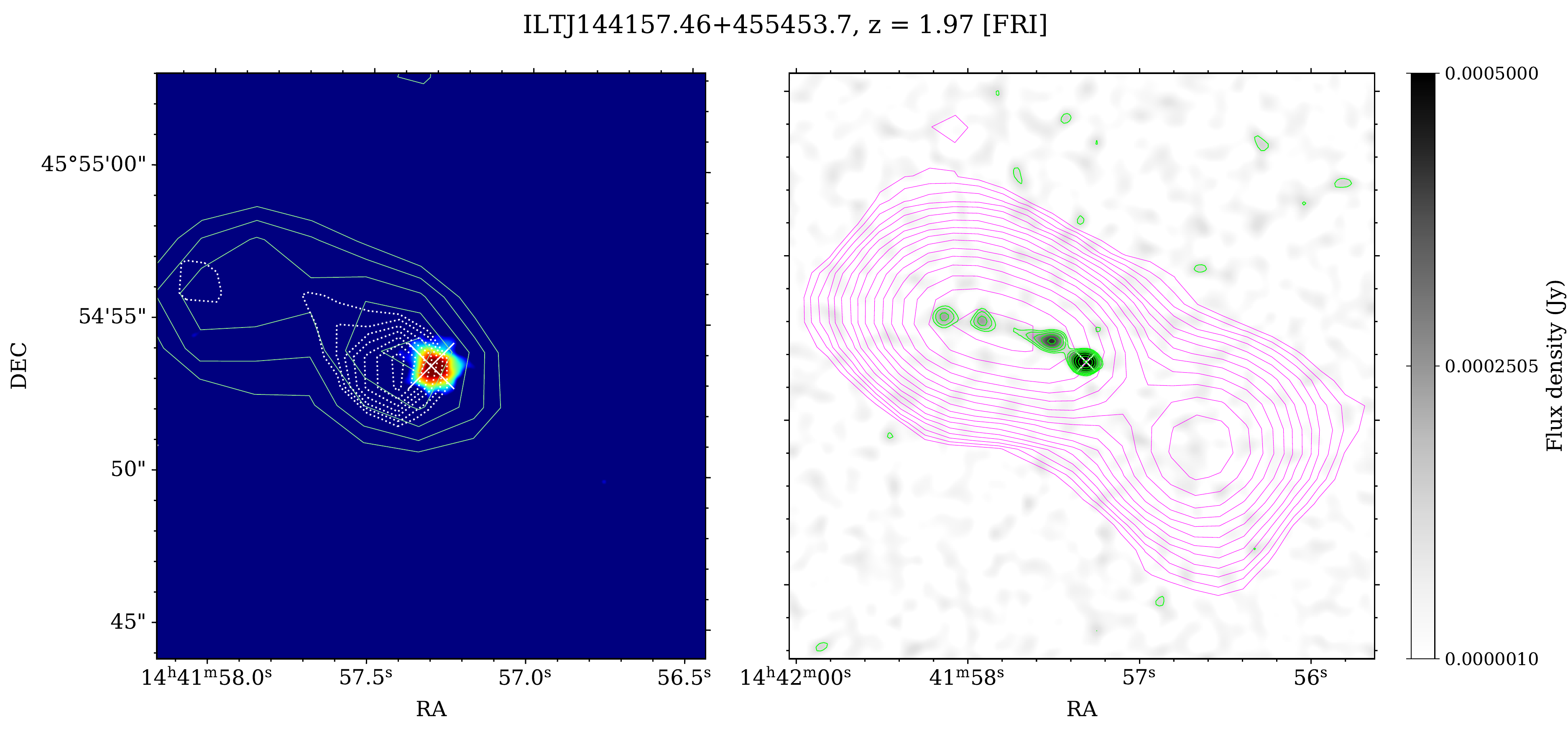}\\
\includegraphics[scale=0.4]{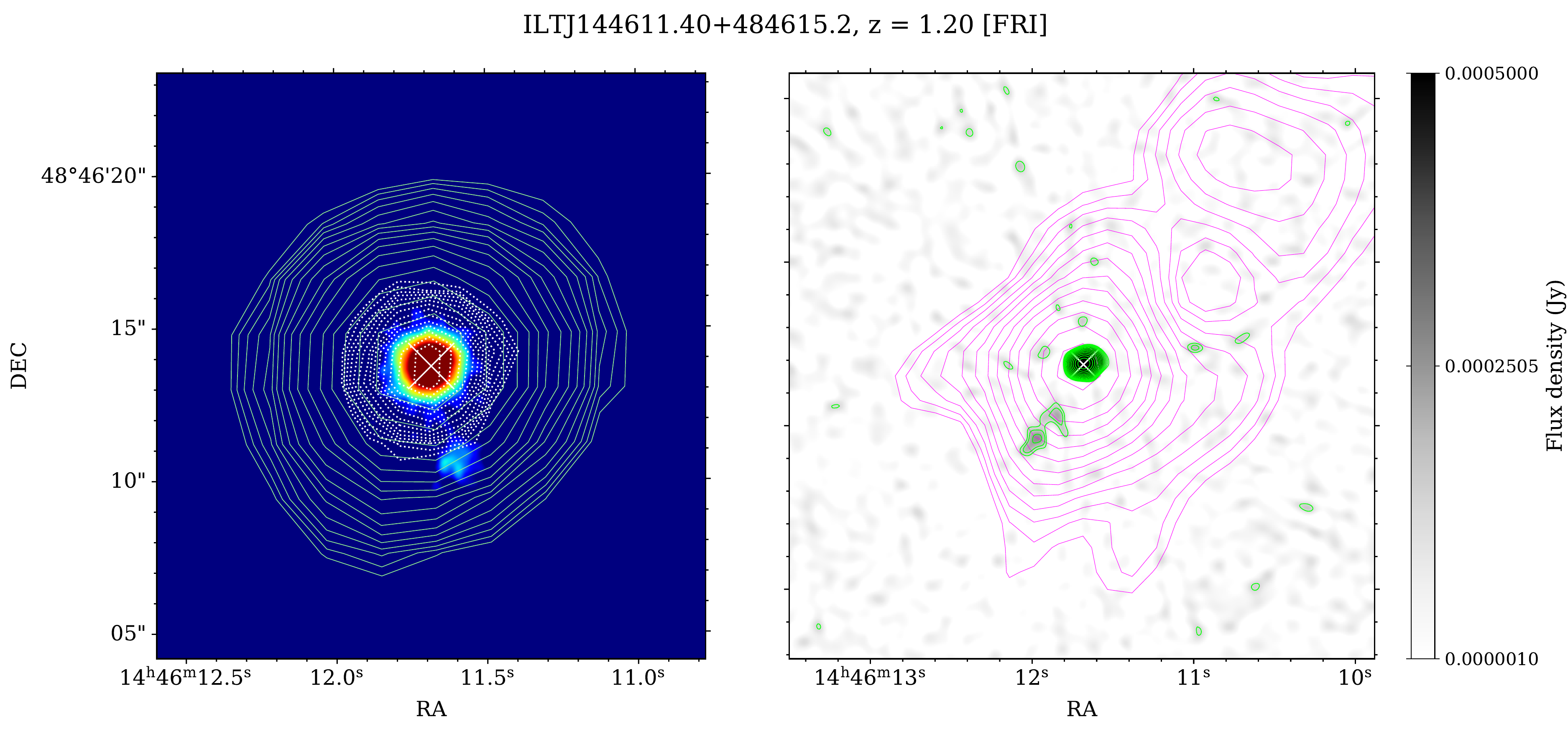}\\
\includegraphics[scale=0.4]{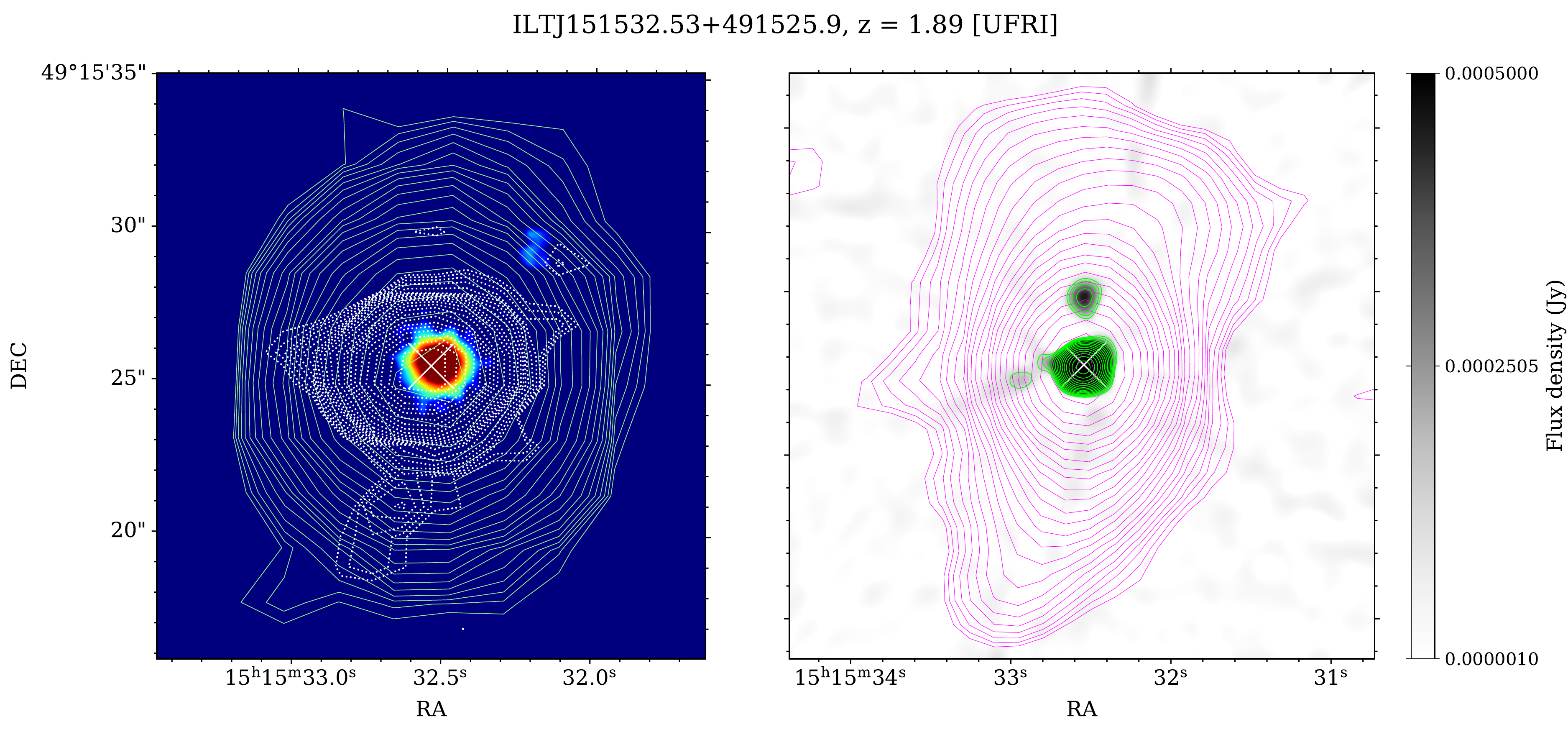}\\
\end{subfigure}
\end{adjustwidth}\caption{\textit{Cont}.}
\end{figure}



\begin{adjustwidth}{-\extralength}{0cm}
\reftitle{References}



\end{adjustwidth}

%


\end{document}